\documentclass[12pt]{article}
\usepackage{cancel}
\usepackage[normalem]{ulem}
\usepackage[dvipdfmx]{graphicx}
\usepackage{bmpsize}
\usepackage{bbding}
\usepackage{relsize}

\usepackage[utf8]{inputenc}

\usepackage[square,comma,numbers,sort&compress]{natbib}
\usepackage[left=20mm,right=20mm,top=20mm,bottom=20mm]{geometry}
\usepackage{bm}

\usepackage{epsf}
\usepackage{color}
\usepackage{xcolor}
\usepackage[normalem]{ulem}
\usepackage{cancel}
\usepackage{amsmath}
\usepackage{amssymb}
\usepackage{latexsym}
\usepackage{caption}
\usepackage{subcaption}
\usepackage{slashed}
\usepackage{float}
\usepackage{cases}
\usepackage{bm}
\usepackage{arydshln}
\usepackage{hyperref}
\hypersetup{ % http://www.biwako.shiga-u.ac.jp/sensei/kumazawa/tex/hyperref.html
 setpagesize=false,
 bookmarksnumbered=true,%
 bookmarksopen=true,%
 colorlinks=true,%
 linkcolor=blue,
 citecolor=red,
}
\usepackage{comment}

\usepackage{listings}
\usepackage{xcolor}
\usepackage{tikz}
\usepackage{tikz-feynman}

\usepackage[T1]{fontenc}
\usepackage{colortbl,xcolor,float}
\definecolor{orange}{rgb}{1,0.5,0}
\definecolor{codegreen}{rgb}{0,0.6,0}
\definecolor{codegray}{rgb}{0.5,0.5,0.5}
\definecolor{codepurple}{rgb}{0.58,0,0.82}
\definecolor{backcolour}{rgb}{0.95,0.95,0.92}

\lstdefinestyle{mystyle}{
    backgroundcolor=\color{backcolour},   
    commentstyle=\color{codegreen},
    keywordstyle=\color{magenta},
    numberstyle=\tiny\color{codegray},
    stringstyle=\color{codepurple},
    basicstyle=\ttfamily\footnotesize,
    breakatwhitespace=false,         
    breaklines=true,                 
    captionpos=b,                    
    keepspaces=true,                 
    numbers=left,                    
    numbersep=5pt,                  
    showspaces=false,                
    showstringspaces=false,
    showtabs=false,                  
    tabsize=2
}

\usepackage{mathtools}

\DeclarePairedDelimiterX\braket[2]{\langle}{\rangle}{#1 \delimsize\vert #2}

\usepackage{multicol}
\usepackage{lipsum}

\usepackage{braket}

\newcommand{\al}[1]{\begin{align}#1\end{align}}

\newcommand{\bp}{\begin{pmatrix}}
\newcommand{\ep}{\end{pmatrix}}
\newcommand{\bb}{\begin{bmatrix}}
\newcommand{\eb}{\end{bmatrix}}

\newcommand{\fn}[1]{\!\left(#1\right)}

\newcommand{\tx}[1]{\text{#1}}

\newcommand{\blue}{\textcolor{black}}

\definecolor{Orange}{cmyk}{0,0.50,1,0}

\newcommand{\beq}{\begin{equation}}
\newcommand{\eeq}{\end{equation}}
\newcommand{\bea}{\begin{eqnarray}}
\newcommand{\eea}{\end{eqnarray}}

\newcommand{\fulltoday}{\number\day\space \ifcase\month\or
    January\or February\or March\or April\or May\or June\or
    July\or August\or September\or October\or November\or December\fi
    \space\number\year}

 \makeatletter
    
    \@addtoreset{equation}{section}
  \makeatother

\usepackage{calc}
\newcounter{hours}\newcounter{minutes}
\makeatletter                   
\renewcommand*{\thehours}{\two@digits\c@hours}
\renewcommand*{\theminutes}{\two@digits\c@minutes}
\makeatother

\begin{document}

\allowdisplaybreaks[2]
\renewcommand{\thefootnote}{*}

\newlength{\mylength}

\title{
%a) Prospects of probing five-dimensional $L_\mu - L_\tau$ gauge interactions 
%in the light of elastic neutrino-electron scatterings with the DUNE near detector \\
%
% Prospects of a Five-Dimensional \texorpdfstring{$U(1)_{L_\mu - L_\tau}$}{U(1) Lmu-Ltau} Model at Future Muon Colliders:\\
%\texorpdfstring{$\mu$TRISTAN}{muTRISTAN} Scenarios and Bremsstrahlung-like Signatures
Aspects of a Five-Dimensional \texorpdfstring{$U(1)_{L_\mu - L_\tau}$}{U(1) Lmu-Ltau} Model \\ at Future Muon-Based Colliders
}

\author{
Dibyendu Chakraborty$^{1}$\thanks{E-mail: \tt dc282@snu.edu.in},
Arindam Chatterjee$^{1}$\thanks{E-mail: \tt arindam.chatterjee@snu.edu.in},
AseshKrishna Datta$^{2}$\thanks{E-mail: \tt asesh@hri.res.in},\\[5pt]
Ayushi Kaushik$^{1}$\thanks{E-mail: \tt ak356@snu.edu.in},
Kenji Nishiwaki$^{1}$\thanks{E-mail: \tt kenji.nishiwaki@snu.edu.in}
\bigskip\\
\it\normalsize
$^{1}$Department of Physics, School of Natural Sciences, \\ 
\it\normalsize
Shiv Nadar Institution of Eminence (Deemed to be University), \\
\it\normalsize
Tehsil Dadri, Gautam Buddha Nagar,
%Shiv Nadar Institution of Eminence, Tehsil Dadri, Gautam Buddha Nagar,\\
Uttar Pradesh, 201314, India
\bigskip\\
\it\normalsize
$^{2}$Harish-Chandra Research Institute, A CI of Homi Bhabha National Institute,\\
\it\normalsize
Chhatnag Road, Jhunsi, Prayagraj (Allahabad) 211019, India
}

\maketitle

\begin{abstract}
\noindent
We study a five-dimensional (5D) framework based on the 
\(U(1)_{L_\mu-L_\tau}\) gauge symmetry, where the associated gauge field \(V\) propagates in the bulk, giving rise to an infinite tower of Kaluza--Klein (KK) excitations \(V^{(n)}\) that couple selectively to the second- and third-generation leptons. Originally motivated by its potential to address the muon \(g-2\) anomaly, this framework remains of interest as a minimal, anomaly-free, phenomenologically well-motivated extension of the Standard Model (SM) of particle physics. We focus on high-energy muon-based colliders, which could directly probe the gauge structure without relying on the kinetic mixing between the SM hypercharge gauge boson and the 5D gauge boson  \(V\). We explore a set of complementary processes: the elastic $\mu^+\mu^+ \to \mu^+\mu^+$ scattering via off-shell exchange of KK (gauge) excitations \(V^{(n)}\); the bremsstrahlung production of \(V^{(n)}\) followed by their decays into neutrinos and into $\mu^-\mu^+$ at a future $\mu$TRISTAN collider. Further, we study the $\mu^-\mu^+ \to \mu^-\mu^+$ scattering via resonant KK excitation(s) at a future muon collider. Our results show that these future muon-based colliders could offer sensitive and complementary probes into regions in the parameter space of the scenario that are beyond the reach of low-energy experiments. In particular, such experiments would be able to probe both heavier KK gauge bosons with TeV-scale masses for relatively large gauge couplings, as well as the much lighter ones with masses in the MeV-scale for couplings as weak as \(g_D \sim \mathcal{O}(10^{-5})\), thereby offering a promising $2\sigma$ exclusion reach for such KK excitations, over an extensive range of masses, at these facilities. 

%\vspace{1em}
%\noindent
%\underline{$\square$ Color Scheme for comments:
%{\bf \color{red} Dibyendu},
%{\bf \color{black} Arindam},
%{\bf \color{blue} Asesh},
%{\bf \color{black} Ayushi},
%{\bf \color{Orange} Kenji}.
%}
%%%%%%%%%%%

\end{abstract}

\newpage

\renewcommand\thefootnote{\arabic{footnote}}
\setcounter{footnote}{0}

%\tableofcontents

%%%%%%%%%%%%%%%%%%%%%%%%%%%%%%%%%%%%%%%%%%%%%%%%%%%%%%%%
%%%%%%%%%%%%%%%%%%%%%%%%%%%%
\section{Introduction
\label{sec:intro}}
%%%%%%%%%%%%%%%%%%%%%%%%%%%%%%%%%%%%%
The Standard Model (SM) of particle physics \cite{Donoghue:1992dd} has witnessed enormous success 
in explaining various experimental results at the energies accessible today. However, there are several 
challenges, both theoretical and phenomenological, which hint towards the presence of physics beyond 
the SM (BSM). Notable ones, for example, are the generation of baryon asymmetry, the explanation 
of neutrino mass, the strong CP problem, and the nature of dark matter. Further, it offers no explanation for the large hierarchy between the electroweak and Planck energy scales. Significant efforts in search for the BSM physics at the high-energy frontier (see e.g., \cite{ATLAS:2025eii} for a reference), and complementary search for rather weakly coupled light particles at the intensity frontier \cite{Antel:2023hkf} are underway. 

Within a diverse set of BSM frameworks, various Abelian gauge extensions and their related phenomenologies have been extensively studied. The lepton sector provides a window to probe such extended gauge symmetries, as leptonic processes generally offer a clean environment that facilitates the search for new particles. Of particular interest are possible gauge extensions that discriminate between different flavours. In this context, a widely studied 
example is the  $U(1)_{L_\mu - L_\tau}$ gauge symmetry~\cite{Foot:1990mn,He:1990pn,He:1991qd,Foot:1994vd}. 
In this framework, the second and third-generation leptons carry charges $+1$ and $-1$, respectively, while 
all other SM fields remain uncharged under the \(U(1)_{L_{\mu}-L_{\tau}}\) gauge group. Consequently, the framework remains anomaly-free, without the necessity to 
include additional particles. In the absence of a tree-level gauge kinetic mixing term, the resulting neutral 
gauge boson ($Z'$) has no tree-level couplings to electrons or quarks due to their respective charge assignments, thus
suppressing its production in many conventional experiments.

Various phenomenological models based on such an extension have been widely studied in the context of neutrino mass models~\cite{Branco:1988ex,Heeck:2011wj,Asai:2017ryy,Asai:2018ocx,Asai:2019ciz,Joshipura:2019qxz,Araki:2019rmw,Fukuyama:2020swd,Bauer:2020itv,Majumdar:2020xws,Amaral:2021rzw}. In addition, this extension has also been explored in dark matter models~\cite{Baek:2008nz,Baek:2015fea,Patra:2016shz,Biswas:2016yan,Biswas:2016yjr,Asai:2017ryy,Arcadi:2018tly,Kamada:2018zxi,Foldenauer:2018zrz,Asai:2019ciz,Okada:2019sbb,Asai:2020qlp,Holst:2021lzm,Tapadar:2021kgw,Heeck:2022znj,Nagao:2022osm,KA:2023dyz,Figueroa:2024tmn} and in addressing the Hubble tension~\cite{Escudero:2019gzq,Araki:2021xdk,Carpio:2021jhu,Asai:2023ajh}.%
\footnote{It may be noted that one of the major motivations for considering this gauge symmetry was to provide an explanation for the discrepancy between the SM expectation and experimentally measured value of  the anomalous magnetic moment of the muon, $(g-2)_\mu$~\cite{Baek:2001kca,Ma:2001md,Harigaya:2013twa,Altmannshofer:2016brv,Hapitas:2021ilr}. However, in light of recent developments in lattice QCD estimations~\cite{Borsanyi:2020mff,Boccaletti:2024guq,Aliberti:2025beg}, this discrepancy is under scrutiny.} 
As this scenario gives rise to new gauge interactions involving the muon and tau sectors, both current and forthcoming muon-beam experiments offer promising opportunities to test these interactions. In the absence of gauge kinetic mixing with the hypercharge gauge boson, fixed-target experiments, notably, NA64$\mu$~\cite{NA64:2024nwj,Gninenko:2018tlp,Krnjaic:2019rsv,Sieber:2021fue,NA64:2022rme,NA64:2024klw,Andreev:2024lps}, offer strong constraints on such a framework through missing energy signatures using incident muon beams, restricting the gauge coupling $g' \lesssim {\cal O}(10^{-3})-{\cal O}(10^{-2})$ and the mass $m_{Z'}$ in the range of ${\cal O}(10^{-3})-{\cal O}(1)$ GeV of the extra gauge boson $Z'$. Low-energy $e^{-} e^{+}$ colliders, such as BaBar~\cite{BaBar:2016sci,Godang:2016gna,Filippi:2019lfq,Godang:2017nik}, also impose similar constraints for $m_{Z'} \lesssim 10$ GeV, by considering $Z'$ emissions from a muon pair in the final state, followed by its decay into a pair of muons~\cite{delAguila:2014soa,Nomura:2018yej,Jho:2019cxq,Belle-II:2019qfb}. 
Similar constraints have been derived from Kaon decay~\cite{Ibe:2016dir,Asai:2024pzx}, and neutrino trident 
experiments~\cite{Altmannshofer:2014pba,Altmannshofer:2019zhy,Ballett:2019xoj,Shimomura:2020tmg}. Further, supernovae 
cooling~\cite{Croon:2020lrf,Cerdeno:2023kqo,Manzari:2023gkt,Lai:2024mse}, and early Universe evolution (in particular, $\Delta N_{\rm eff}$)  provide additional constraints on the gauge coupling which are relevant for $m_{Z'} \lesssim \mathcal{O}(10)$ MeV~\cite{Escudero:2019gzq,Araki:2021xdk,Carpio:2021jhu,Asai:2023ajh}. In the energy frontier, the Large Hadron Collider (LHC) searches in the four-muon, as well as the three-muon plus missing transverse energy final states, impose a lower bound on the gauge coupling of $\mathcal{O}(10^{-2})$ for $m_{Z'}$ in the range of $\mathcal{O}(10-10^{2})$ GeV~\cite{ATLAS:2024uvu,ATLAS:2023vxg,CMS:2018yxg}. Thus, in general phenomenological grounds, the search for such an anomaly-free gauge extension of the SM remains well motivated.

Extra-dimensional frameworks have been widely investigated as possible explanations for the hierarchy 
problem and the observed weakness of the gravitational 
interaction~\cite{ArkaniHamed1998,Antoniadis1998,RandallSundrum1999a,RandallSundrum1999b}. A pioneering work on TeV-scale extra dimensions, originally motivated by string-theoretic constructions, was put forward in Ref.~\cite{Antoniadis:1990ew}, which laid the foundation for subsequent developments in higher-dimensional extensions of the SM. In a related context, the role of Kaluza--Klein (KK) excitations of leptophilic $U(1)$ gauge bosons propagating in extra dimensions has also been explored in reference to the muon anomalous magnetic moment within large extra-dimensional frameworks, including realisations with string-inspired embeddings~\cite{Anchordoqui:2021lmm,Antoniadis:2021mqz}. These further highlighted the phenomenological relevance of higher-dimensional gauge extensions, especially in relation to precision observables and related experimental constraints. An interesting possibility, in the form of extending the \(U(1)_{L_{\mu}-L_{\tau}}\) model in a 5-dimensional (5D) spacetime, both in the context of flat and warped extra dimensions, was first proposed in Ref.~\cite{Chakraborty:2024xxc}. In this model, all the SM fields are confined to the four-dimensional (4D) brane, while the 5D $U(1)_{L_\mu - L_\tau}$ gauge field $V$ propagates in a compactified extra spatial dimension, leading to a tower of KK excitations. Such a construction allows for a rather large extra-dimension, evading the stringent constraints from various experiments, in contrast to the case of the universal extra-dimensional frameworks 
\cite{Appelquist2001,Kakuda:2013kba,Deutschmann:2017bth,Flores:2021xwx}. A rather large compactification scale leads to light KK excitations of the 5D gauge boson $V$. Consequently, to evade the electroweak precision constraints, as well as various constraints from experiments in the intensity frontier, the respective $U(1)_{L_\mu - L_\tau}$ gauge coupling is assumed to be sufficiently weak. 

An important aspect of the scenario, in contrast to the four-dimensional $U(1)_{L_\mu - L_\tau}$ gauge extensions, 
is the presence of the compact extra-dimension, which is accessible to the gauge boson \(V\). From a four-dimensional perspective, the extra dimension manifests 
itself through the presence of an infinite tower of states. Therefore, it is essential to search at different energy scales to 
probe or constrain such a scenario. The relevant phenomenological 
constraints in this scenario have been studied in Ref.~\cite{Chakraborty:2025jbd}. Further, prospects for probing such a scenario in 
light of future experiments, such as MuSIC \cite{Acosta:2021qpx, Acosta:2022ejc}, $\rm M^3$ \cite{Kahn:2018cqs}, and a future high-energy muon beam-dump experiment \cite{Cesarotti:2022ttv}, have also been discussed.\footnote{While NA64$\mu$ can directly constrain the gauge coupling $g_D$ of the $U(1)_{L_{\mu}-L_{\tau}}$ gauge group without invoking any kinetic mixing, in the presence of a non-vanishing gauge kinetic mixing with the hypercharge gauge boson, constraints 
from NA64$e$ \cite{Andreev:2024lps} can also be effective. In Ref.~\cite{Chakraborty:2024xxc}, 
the low-energy constraints from elastic neutrino-electron scattering 
(E$\nu$ES), including BOREXINO \cite{Bellini:2011rx}, TEXONO 
\cite{Wong:2015kgl}, CHARM-II \cite{CHARM-II:1993phx,CHARM-II:1994dzw} 
and projected DUNE Near Detector sensitivities 
\cite{DUNE:2016hlj,DUNE:2020fgq}, were studied for this 
model, considering both flat and warped extra dimensions.} While various beam-dump experiments provide sensitivity to the light states in the KK 
tower, high-energy colliders will be able to shed light on the heavier KK 
excitations, complementing the low-energy probes. The lepton colliders offer a 
clean environment to search for such heavier excitations. In the present context, in the absence of any kinetic mixing between the hypercharge gauge boson and $V$, as the KK excitations of the 5D gauge boson \(V\) only to the second and the third generation leptons, the proposed muon colliders are most suitable for probing such scenarios.
%\footnote{\textcolor{blue}{The electron-positron colliders ~\cite{Nomura:2018yej,Zhang:2020fiu,Kaneta:2016uyt,Araki:2017wyg,Jho:2019cxq,Asai:2021wzx,Bandyopadhyay:2022klg,BaBar:2016sci,Filippi:2019lfq,Godang:2017nik,Heeck:2011wj}, high-energy Large Hadron Collider (LHC) \cite{ATLAS:2023vxg,ATLAS:2024uvu,CMS:2018yxg,Nomura:2020vnk,Galon:2019owl,delAguila:2014soa,Heeck:2011wj} constrain such scentarios. While resonant production of the respective gauge boson $Z^'$ is possible only in the presence of non-vanishing kinetic mixing, for a negligible kinetic mixing term constraints from cascade production of $Z^{'}$ have been studied. This imposes a lower limit of the gauge coupling  of ${\cal O}(10^{2})$ for the mass of $Z^{'}$ of ${\cal O}(10)-{\cal O}(10^{2})$ GeV \cite{ATLAS:2024uvu,ATLAS:2023vxg,CMS:2018yxg}.}} 
Several recent studies have demonstrated the scope of both \(\mu^+ \mu^+\)~\cite{Hamada:2022mua, Hamada:2022uyn} and \( \mu^- \mu^+\)~\cite{MuonCollider:2022xlm,Aime:2022flm,Black:2022cth,Huang:2021nkl,Franceschini:2022sxc, Abe:2019thb} 
colliders in probing the 4D $U(1)_{L_\mu - L_\tau}$ extensions.

In this article, we consider the high-energy frontier and investigate how future muon collider experiments could
probe this 5D $U(1)_{L_\mu - L_\tau}$ framework in the context of a flat extra-dimension. In particular, we 
consider the following proposed colliders: \(\mu\)TRISTAN, which is a \(\mu^+ \mu^+\) collider, and the 
\(\mu^- \mu^+\) collider, referred to as the muon collider. The direct coupling between the heavier KK excitations of the 5D gauge boson $V$ and the muons allows such colliders to probe the  $U(1)_{L_\mu - L_\tau}$ symmetric extension even in the absence of the gauge kinetic mixing effects. It is worth stressing that these experiments are sensitive to a parameter space with rather larger gauge couplings and heavier KK modes, which are typically inaccessible in fixed-target or low-energy experiments. Thus, the present study serves as a useful probe of the extra-dimensional realisation of the $U(1)_{L_\mu - L_\tau}$ gauge symmetry, providing an avenue to investigate the corresponding 
heavier excitations and complements our previous work, where only the low-energy KK states were accessible 
\cite{Chakraborty:2024xxc}.

The paper is organised as follows. In section~\ref{sec:ModelSetup}, the five-dimensional \( U(1)_{L_\mu - L_\tau} \) framework is reviewed, and the theoretical context is set up. In section~\ref{sec:Mutristan}, implications on $\mu^+ \mu^+$ scattering at the $\mu$TRISTAN collider are investigated, including the full KK tower with SM interference 
across an operational energy range of 2--20~TeV. In section~\ref{sec:Bremsstrahlung}, we analyse semi-visible signatures with missing transverse energy as well as all-visible four muon final state production, providing complementary probes of the model at a 2~TeV \(\mu\)TRISTAN collider. In section~\ref{sec:MuonC}, we focus on the resonant production of the KK gauge bosons ($V^{(n)}$) at a 3~TeV muon collider. A combined summary of results, including a few representative benchmark points, is presented in section~\ref{sec:Summary}. Finally, section~\ref{sec:conclusion} presents our conclusions and the outlook.
%Some technical details are collected in Appendices~\ref{appendix-a}, \ref{appendix-b}, \ref{appendix-c}, \ref{appendix-d}, and \ref{appendix-e}.
In Appendix~\ref{appendix-a}, formulae for the elastic scattering of $\mu^+ \mu^+$ are provided.
In Appendix~\ref{appendix-b}, technical details of the summation of the intermediate KK modes are discussed.
In Appendix~\ref{appendix-c}, we validate the significance approximation used for the analyses of the semi-visible final state.
In Appendix~\ref{appendix-d}, we share complementary plots for the semi-visible final state.
In Appendix~\ref{appendix-e}, formulae for the resonant production of $\mu^- \mu^+$ are provided.

\section{Theoretical framework
\label{sec:ModelSetup}}
%%%%%%%%%%%%%%%%%%%%%%%%%%%%%%%%%%%%%%%%%%%%%%%%%%%%%%%%%%%%%%%%%%%%%%%%%%%%%%%%%%%%%%%%%%%%%%%%%%%%%%%%%%%%%%%%%%%%
We consider a 5D extension of the SM with a \( U(1)_{L_\mu - L_\tau} \) gauge symmetry, where the corresponding gauge boson propagates in a flat compact extra dimension, while all SM fields are localised on a 4D brane positioned at \( y = y_{\text{SM}} \). This setup was originally developed and studied in detail in Ref.~\cite{Chakraborty:2024xxc}, where both flat and warped geometries were considered. In this work, we restrict our attention to the flat case, which serves as a minimal and representative benchmark for collider phenomenology. Although warped geometries can modify the KK spectrum and interaction profiles in important ways, they typically introduce additional model-dependent parameters and interpretational complexity. For the collider-centric analysis presented in this work, where the key phenomenological features are primarily governed by the KK mass scale and the couplings of the KK states to muons, the flat geometry provides a clean, controlled theoretical setting.

The extra spatial dimension is compactified on an interval \( y \in [0, \pi R] \), where \( R \) denotes the compactification radius. We impose the following twisted boundary conditions (BC) on the bulk \( U(1)_{L_\mu - L_\tau} \) gauge field, where \(V_\mu(x,y)\) stands for the respective 4D vector components, $\mu = 0,1,2,3$:
\begin{align}
V_\mu(x, y=0) \ \text{satisfies the Neumann BC}, \qquad
V_\mu(x, y=\pi R) \ \text{satisfies the Dirichlet BC}.
\end{align}
These boundary conditions eliminate the massless zero mode without invoking a Higgs mechanism. As a result, the effective 4D theory contains an infinite, equispaced tower of massive KK gauge bosons \( V^{(n)}_\mu \), with masses given by
\begin{align}
M_n = \left( 2n - 1 \right) m_{\text{KK}}, \qquad m_{\text{KK}} := \frac{1}{2R},
\label{eq:KK-mass-flat}
\end{align}
where \( m_{\text{KK}} \) sets the new physics scale, and \( n = 1, 2, 3, \ldots \) labels the KK excitation number.
\noindent
The 5D gauge coupling \( g_{5D} \) has mass dimension \(-1/2\), and the effective 4D coupling is defined as
\begin{align}
g_{D} = \frac{g_{5D}}{\sqrt{\pi R}},\footnotemark
\label{eq:gprime-definition}
\end{align}
\footnotetext{%
This parameter corresponds to \( g' \) in Ref.~\cite{Chakraborty:2024xxc}.}

\noindent
We also allow for kinetic mixing between the bulk \( U(1)_{L_\mu - L_\tau} \) and the brane-localized hypercharge \( U(1)_Y \) gauge bosons, characterized by the parameter
\begin{align}
\kappa_n = \kappa_D f_n,\footnotemark \qquad 
\kappa_D := \frac{\kappa_{5D}}{\sqrt{\pi R}}, \qquad 
f_n := f^{(n)}_V(y_{\text{SM}}),
\label{eq:kinetic-mixing}
\end{align}
\footnotetext{%
The parameter $\kappa_{5D}$, $\kappa_D$, $\kappa_n$ corresponds to the conventional kinetic mixing variable \(\epsilon_{5D}\), \(\epsilon_4\), \(\epsilon_n\) used in Ref.~\cite{Chakraborty:2024xxc}.}

\noindent
where \( f_n \) denotes the wavefunction profile of the \( n \)-th KK mode evaluated at the brane location, $y_{\mathrm{SM}}$. For convenience, we define the dimensionless parameter
\begin{align}
\tilde{y}_{\text{SM}} := \frac{y_{\text{SM}}}{R},
\end{align}
which denotes the position of the SM brane in units of the compactification radius. The effective Lagrangian for the neutral gauge bosons in the flat case, expressed in the mass-diagonal basis, takes the form:
\begin{align}
{\cal L}_{\text{free}}^{\text{eff}}
	&=
		-\frac{1}{4} \sum_n V^{(n)}_{\mu\nu} V^{(n)\mu\nu}
		-\frac{1}{4} Z_{\mu\nu} Z^{\mu\nu}
		-\frac{1}{4} A_{\mu\nu} A^{\mu\nu}
		+ \frac{1}{2} m_Z^2 Z_\mu Z^\mu
		+ \frac{1}{2} \sum_n M_n^2 V^{(n)}_\mu V^{(n)\mu}
		+ \mathcal{O}(\kappa_n^2),
\label{eq:Lagrangian-neutral-flat}
\end{align}
where \( A_\mu \), \( Z_\mu \), and \( V^{(n)}_\mu \) denote the photon, the \( Z \)-boson, and the \( n \)-th KK excitation of the \( U(1)_{L_\mu - L_\tau} \) gauge field, respectively, all expressed in the mass eigenbasis; and \( m_Z^2 = (g_1^2 + g_2^2)v^2/4 \) is the tree-level squared mass of the \( Z \)-boson as in the SM, $g_{1}$ and $g_{2}$ being the gauge couplings associated with the $U(1)_{Y}$ and $SU(2)_{L}$ gauge groups of the model, respectively. The effective interaction Lagrangian of the KK modes with SM leptons (in the mass basis) relevant for this study, including the effects of kinetic mixing, is given by
\begin{align}
{\cal L}_{\text{int}}^{\text{eff}}
	&= \mathlarger{\mathlarger{\sum_{a=e,\mu,\tau}}}
	\Bigg\{	\bar{l}^a_R \, i \slashed{\partial} \, l^a_R
	+ e\, \bar{l}^a_R \gamma^\mu l^a_R A_\mu
	+ g_1 s_W \, \bar{l}^a_R \gamma^\mu l^a_R \, Z_\mu \nonumber \\
	&\quad
	+ \bar{l}^a_R \gamma^\mu \Big\{
		g_1 \sum_n \left( \frac{\kappa_n}{c_W} + s_W t_W \frac{\kappa_n m_Z^2}{M_n^2 - m_Z^2} \right) V^{(n)}_\mu
		+ g_{D} Q_{L_\mu - L_\tau}^a \sum_n f_n \left( V^{(n)}_\mu - t_W \frac{\kappa_n m_Z^2}{M_n^2 - m_Z^2} Z_\mu \right)
	\Big\} l^a_R \nonumber \\
	&\quad
	+ \bar{l}^a_L \, i \slashed{\partial} \, l^a_L
	+ \bar{\nu}^a_L \, i \slashed{\partial} \, \nu^a_L
	+ \frac{g_2}{\sqrt{2}} \left( \bar{\nu}^a_L \gamma^\mu l^a_L W^+_\mu + \bar{l}^a_L \gamma^\mu \nu^a_L W^-_\mu \right) \nonumber \\
	&\quad
	+ \bar{l}^a_L \gamma^\mu \left[
		e A_\mu + \frac{1}{2} \left( -g_2 c_W + g_1 s_W \right) Z_\mu 
		+ \sum_n \big\{ \left( -g_2 c_W + g_1 s_W \right) t_W \frac{\kappa_n m_Z^2}{M_n^2 - m_Z^2} 
		+ g_1 \frac{\kappa_n}{c_W} \big\} V^{(n)}_\mu \right. \nonumber \\
	&\qquad\qquad\quad \left. +\, g_{D} Q_{L_\mu - L_\tau}^a \sum_n f_n \left( V^{(n)}_\mu - t_W \frac{\kappa_n m_Z^2}{M_n^2 - m_Z^2} Z_\mu \right)
	\right] l^a_L \nonumber \\
	&\quad
	+ \bar{\nu}^a_L \gamma^\mu \left[
		\frac{1}{2} \left( g_2 c_W + g_1 s_W \right) Z_\mu
		+ \sum_n \big\{ \left( g_2 c_W + g_1 s_W \right) t_W \frac{\kappa_n m_Z^2}{M_n^2 - m_Z^2} 
		+ g_1 \frac{\kappa_n}{c_W} \big\} V^{(n)}_\mu \right. \nonumber \\
	&\qquad\qquad\quad \left. +\, g_{D} Q_{L_\mu - L_\tau}^a \sum_n f_n \left( V^{(n)}_\mu - t_W \frac{\kappa_n m_Z^2}{M_n^2 - m_Z^2} Z_\mu \right)
	\right ] \nu^a_L
	\Bigg \} + \mathcal{O}(\kappa_n^2) .
\label{eq:Lagrangian-int-flat}
\end{align}
In the above expression, we define $c_{W} (s_W) := \cos\theta_{W} (\sin\theta_{W})$ and $t_{W} := \tan\theta_{W}$, where $\theta_{W}$ is the weak (Weinberg) mixing angle. The parameter $e \, (:= g_2 \sin\theta_{W})$ denotes the electromagnetic coupling. \footnote{
We adopt the SM electroweak conventions of Ref.~\cite{Denner:1991kt}.} For the sake of completeness, we also list the \( U(1)_{L_\mu - L_\tau} \) charges of the charged leptons, which are as follows:
\begin{align}
Q_{L_\mu - L_\tau}^e = 0, \qquad
Q_{L_\mu - L_\tau}^\mu = +1, \qquad
Q_{L_\mu - L_\tau}^\tau = -1.
\end{align}
\noindent
The effective interaction Lagrangian above provides the basis for the collider phenomenology discussed in the subsequent sections.

Throughout this work, we focus on the limit where kinetic mixing vanishes at tree-level, i.e., we set 
$$
\kappa_D= 0, \text{ and hence } \kappa_n = 0
$$
for all KK modes as an idealised benchmark to isolate the intrinsic sensitivity of a muon-based collider to the gauged $U(1)_{L_\mu-L_\tau}$ interaction.

In the minimal model, a finite kinetic mixing is radiatively generated through muon and tau loops~\cite{Araki:2017wyg}, with a typical loop-suppressed size $\kappa \sim \mathcal{O}(e g_D/16\pi^2)$, even if absent at tree level. The resulting non-vanishing kinetic mixing generates additional couplings of the $Z'$ boson to the electrically charged SM fermions, thereby leading to complementary constraints from electron- and hadron-based beam experiments. The impact of a non-vanishing kinetic mixing depends on the region of the involved parameter space and may become significant for certain experimental probes. For example, the phenomenology of muon beam-dump experiments and elastic neutrino--electron scattering can be significantly affected by kinetic mixing, as demonstrated in our earlier studies~\cite{Chakraborty:2024xxc,Chakraborty:2025jbd}. Since the purpose of the present work is to study the projected reach of a muon-based collider within the minimal 5D $U(1)_{L_\mu-L_\tau}$ framework, we neglect both the tree-level and the radiatively induced kinetic mixing throughout our phenomenological analysis. This approximation has also been adopted in previous phenomenological studies~\cite{Nomura:2020vnk}.\footnote{In the current $U(1)_{L_\mu - L_\tau}$ configuration, fermion loop corrections appear as finite terms due to the relative minus sign between $L_\mu$ and $L_\tau$. Furthermore, by fine-tuning the tree-level $\kappa_D$, it is possible to cancel the total effect at the one-loop level (at a specific renormalisation scale).} The general form of the Lagrangian is presented above for completeness and to highlight the potential interplay with kinetic mixing in broader contexts.

For our analysis, we have implemented the scenario in \texttt{FeynRules}~\cite{Alloul:2013bka} and generated the corresponding UFO model files~\cite{Degrande:2011ua,Darme:2023jdn}. For matrix element evaluation and event generation, those are interfaced with \texttt{MadGraph5\_aMC@NLO}~\cite{Alwall:2011uj}. On-shell decays of intermediate resonances are handled using \texttt{MadSpin}~\cite{Artoisenet:2012st}, while \texttt{MadAnalysis5}~\cite{Conte:2012fm} is used for cut-based analysis and visualisation of various kinematic distributions involving the signal and background events, and estimating their yields under the applied kinematic cuts. Analytical computations of matrix elements are carried out using  \texttt{FeynCalc}~\cite{Shtabovenko:2020gxv} in \texttt{Mathematica}~\cite{Mathematica}, and Feynman diagrams are generated with the help of the \texttt{FeynGame} package~\cite{Harlander:2020cyh,Harlander:2024qbn,Bundgen:2025utt}.

%%%%%%%%%%%%%%%%%%%%%%%%%%%%%%%%%%%%%%%%%%%%%%%%%%%%%%%%%%%%%%%%%%%%%%%%%%%%%%%%%%%%%%%%%%%%%%%%%%%%%%%%%%%%%%%%%%%%%%%%%%%%%%%
\section{Elastic scattering of $\mu^+ \mu^+$ at $\mu$TRISTAN %Collider \texorpdfstring{$\mu$TRISTAN}{muTRISTAN} Collider
\label{sec:Mutristan}}
%%%%%%%%%%%%%%%%%%%%%%%%%%%%%%%%%%%%%%%%%%%%%%%%%%%%%%%%%%%%%%%%%%%%%%%%%%%%%%%%%%%%%%%%%%%%%%%%%%%%%%%%%%%%%%%%%%%%%%%%%%%%%%%%
\begin{figure}[t]
    \centering
    \includegraphics[width=0.8\textwidth]{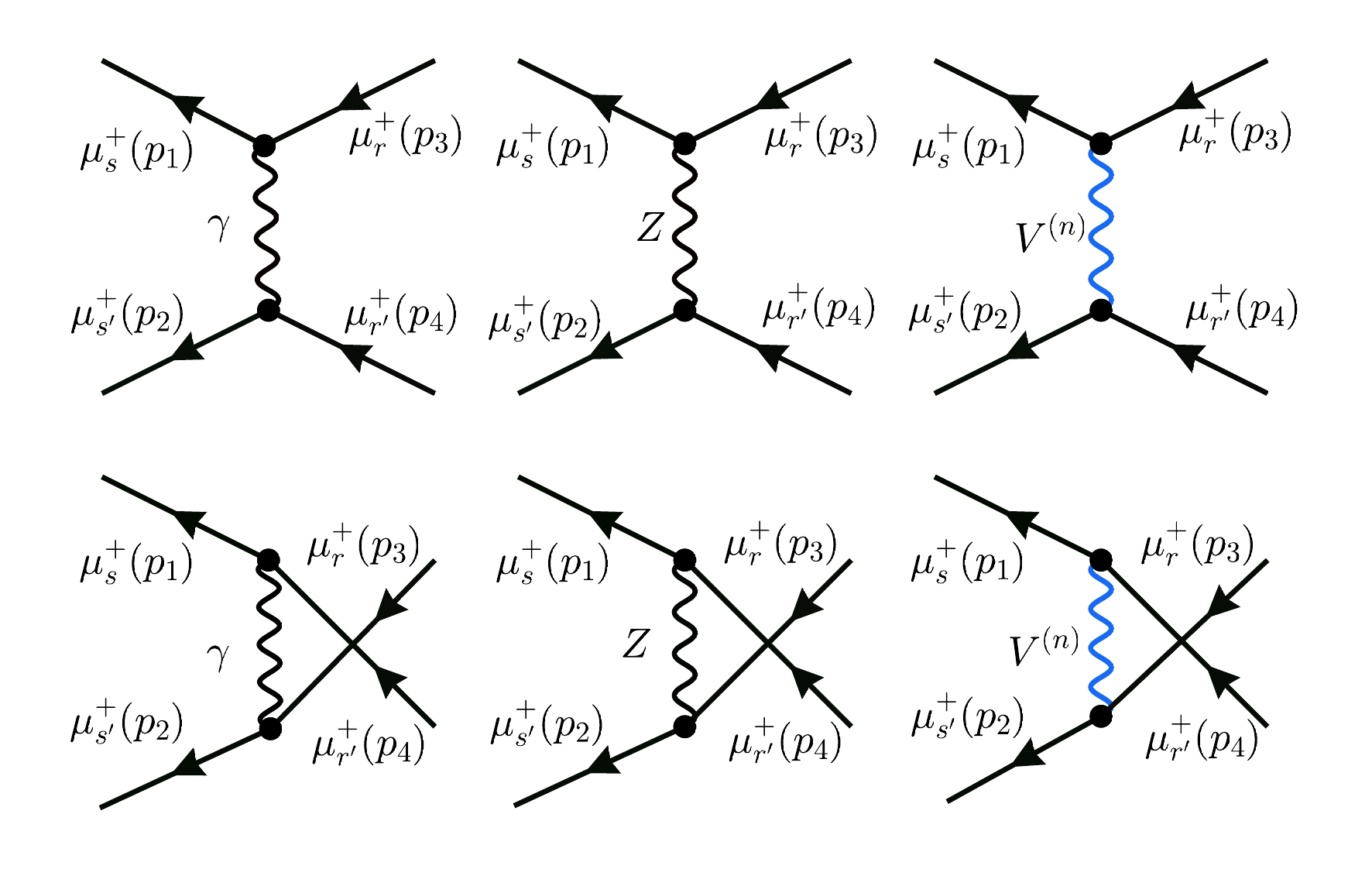}
    \caption{Representative Feynman diagrams for the process $\mu^+ \mu^+ \rightarrow \mu^+ \mu^+$ via $t$-channel (top) and $u$-channel (bottom) exchanges of photon, $Z$-boson and the KK gauge bosons, $V^{(n)}$.}
    \label{fig:mutrist_diag}
\end{figure}

The proposed $\mu$TRISTAN collider~\cite{Hamada:2022uyn} is envisioned as a high-luminosity $\mu^+ \mu^+$ machine dedicated to precision studies of the muon sector. Benefiting from the suppressed synchrotron radiation of muons (thanks to muons' heaviness, compared to electrons), it enables an efficient circular operation at high energies. It offers a unique platform for stringent electroweak tests and searches for new physics coupled to the second lepton generation, i.e., the muon sector.

In this section, we investigate the elastic scattering process
$\mu^+ \mu^+ \rightarrow \mu^+ \mu^+$
%\[
%\mu^+ \mu^+ \rightarrow \mu^+ \mu^+ \, ,
%\]
at the proposed $\mu$TRISTAN collider. The process proceeds via $t$- and $u$-channel exchanges of the SM gauge bosons (photon and $Z$), together with the entire KK tower of states associated with the 5D $U(1)_{L_\mu - L_\tau}$ gauge boson. Representative Feynman diagrams for the $t$- and $u$-channel topologies are shown in Fig.~\ref{fig:mutrist_diag}.
In the subsections that follow, we first derive the matrix element and cross-section(s) for the process. We then study how the angular distribution of the final-state muons, in the presence of KK contributions, deviates from the SM expectation. We follow this up by estimating the experimental sensitivity of such BSM effects, which, in turn, allows us to project out the region in the plane of $m_\tx{KK}$--$g_D$ that is excluded at a given confidence level, in the absence of a signal.

\subsection{Estimation of signal cross-section
\label{sec:sig-mutristan}}
The full matrix element for the elastic process $\mu^+\fn{p_1} \mu^+\fn{p_2} \rightarrow \mu^+\fn{p_3} \mu^+\fn{p_4}$ can be written in a compact form as
\begin{align}
\mathcal{M_{\text{Elastic}}} = \left( \mathcal{M}_t^{\gamma}  + \mathcal{M}_t^Z + \sum_{n=1}^{\infty}  \mathcal{M}_t^{(n)} \right) - \left( \mathcal{M}_u^{\gamma} + \mathcal{M}_u^Z + \sum_{n=1}^{\infty} \mathcal{M}_u^{(n)} \right) ,
\end{align}
where the superscripts $\gamma$, $Z$, and $(n)$ denote the photon, $Z$-boson, and the $n$-th KK excitation of the $U(1)_{L_\mu-L_\tau}$ gauge boson, respectively.
Each contribution can be expressed in the generic form
\begin{align}
 \mathcal{M_{\text{Elastic}}} &= 
\sum_{j=1,3,5} 
\bar{v}^{s'}(p_2)\, \gamma^\alpha (c_j + d_j \gamma^5)\, v^{r'}(p_4)\;
\bar{v}^{s}(p_1)\, \gamma_\alpha (\tilde{c}_j + \tilde{d}_j \gamma^5)\, v^{r}(p_3) \nonumber\\
&\quad
+ \sum_{j=2,4,6} 
\bar{v}^{s'}(p_2)\, \gamma^\alpha (c_j + d_j \gamma^5)\, v^{r}(p_3)\;
\bar{v}^{s}(p_1)\, \gamma_\alpha (\tilde{c}_j + \tilde{d}_j \gamma^5)\, v^{r'}(p_4).
\end{align}
Here \(s,s',r,r'\) denote the spin indices of the external fermions
associated with the momenta \(p_1,p_2,p_3,\) and \(p_4\), respectively.
Since all external particles in the present process are anti-muons,
their wavefunctions are represented by the Dirac spinors \(v(p)\).
Here \(c_j\) and \(d_j\) denote the vector and axial-vector couplings
at one vertex, while \(\tilde{c}_j\) and \(\tilde{d}_j\) denote the corresponding ones at the other vertex multiplied by the propagator
factor for the exchanged particle.
For the diagrams mediated by KK modes, the associated mode functions are
absorbed into the second (tilded) vertex together with the propagator factor,
so that the entire KK tower contribution is encoded in the effective vertex
structure. With this convention, the indices \(j=1,3,5\) correspond to the
\(t\)-channel exchanges of photon, \(Z\), and the KK modes, respectively,
while \(j=2,4,6\) denote the corresponding \(u\)-channel contributions.

The squared matrix element, averaged over the spins of the initial-state
fermions and summed over the spins of the final-state fermions, is defined as
\begin{align}
\overline{|\mathcal{M_{\text{Elastic}}}|^2}
=
\frac{1}{4}
\sum_{s,s',r,r'}
|\mathcal{M_{\text{Elastic}}}|^2 ,
\end{align}
which, in the high-energy limit,
(i.e., \(m_\mu \to 0\)) can be expressed in terms of the Mandelstam
variables \(s=(p_1+p_2)^2\), \(t=(p_1-p_3)^2\), and \(u=(p_1-p_4)^2\) as
\begin{align}
\overline{|\mathcal{M_{\text{Elastic}}}|^2}
= X\,s^2 + Y\,t^2 + Z\,u^2 \, ,
\end{align}
where the coefficient functions \( X \), \( Y \), and \( Z \) include the contributions from all interfering diagrams and are functions of \( c_j \) and \( d_j \). Explicit expressions for \( X \), \( Y \), and \( Z \) are collected in Appendix~\ref{appendix-a}.
The differential cross-section in the centre-of-mass (CM) frame then takes the form
\begin{align}
\left. \frac{d\sigma}{d\Omega} \right|_{\mathrm{CM}} = \frac{1}{64\pi^2 s} \left( \overline{|\mathcal{M_{\text{Elastic}}}|^2} \right),
\end{align}
where $\Omega$ is the solid angle. The definitions of the coefficients $c, d, \tilde{c}$ and $\tilde{d}$ are as follows:
\begin{itemize}
 \item {Photon Exchange:}
 \begin{align}
 c_1 &= e, & d_1 &= 0, & \tilde{c}_1 &= \frac{e}{t}, & \tilde{d}_1 &= 0, \nonumber\\
 c_2 &= e, & d_2 &= 0, & \tilde{c}_2 &= -\frac{e}{u}, & \tilde{d}_2 &= 0.
 \label{eq:Ph-coefficients}
 \end{align}
 \item {$Z$-boson Exchange:}
 \begin{align}
 c_3 &= \frac{3}{4} g_1 s_W - \frac{1}{4} g_2 c_W, & 
 d_3 &= \frac{1}{4} g_1 s_W + \frac{1}{4} g_2 c_W, & 
 \tilde{c}_3 &= \frac{c_3}{t - m_Z^2}, & 
 \tilde{d}_3 &= \frac{d_3}{t - m_Z^2}, \nonumber\\
 c_4 &= c_3, & 
 d_4 &= d_3, & 
 \tilde{c}_4 &= -\frac{c_3}{u - m_Z^2}, & 
 \tilde{d}_4 &= -\frac{d_3}{u - m_Z^2}.
 \label{eq:Z-coefficients}
 \end{align}
 \item {KK Tower Exchange (benchmark scenarios with $\tilde{y}_{\mathrm{SM}}=\pi/2$ and $0$}):

 For the brane configuration with $\tilde{y}_{\text{SM}} = {\pi}/{2}$:
 \begin{align}
 c_5 &=  g_D, & d_5 &= 0, &
 \tilde{c}_5 \Big|_{\tilde{y}_{\text{SM}} = \pi/2} &= -\frac{g_D \pi}{4 m_{\text{KK}} \sqrt{t}}  \tan\left( \frac{\pi \sqrt{t}}{2 m_{\text{KK}}} \right), &
 \tilde{d}_5 &= 0, \nonumber\\
 c_6 &= g_D, & d_6 &= 0, &
 \tilde{c}_6 \Big|_{\tilde{y}_{\text{SM}} = \pi/2} &= \frac{g_D \pi}{4 m_{\text{KK}} \sqrt{u}} \tan\left( \frac{\pi \sqrt{u}}{2 m_{\text{KK}}} \right), &
 \tilde{d}_6 &= 0.
 \label{eq:KK-coefficients-pi2}
 \end{align}

For the alternate brane configuration with $\tilde{y}_{\text{SM}} = 0$, all such coefficients are identical to the $\tilde{y}_{\text{SM}} = \pi/2$ case except for
\begin{align}
 \tilde{c}_5 \Big|_{\tilde{y}_{\text{SM}} = 0} &= -\frac{g_D \pi}{2 m_{\text{KK}} \sqrt{t}} \tan\left( \frac{\pi \sqrt{t}}{2 m_{\text{KK}}} \right), &
 \tilde{c}_6 \Big|_{\tilde{y}_{\text{SM}} = 0} &= \frac{g_D \pi}{2 m_{\text{KK}} \sqrt{u}} \tan\left( \frac{\pi \sqrt{u}}{2 m_{\text{KK}}} \right).
 \label{eq:KK-coefficients-0}
\end{align}
\end{itemize}
 A complete expression for the propagator sum valid for arbitrary $\tilde{y}_{\text{SM}}$ values is provided in Appendix~\ref{appendix-b},\footnote{
In Eqs.~\eqref{eq:KK-coefficients-pi2} and \eqref{eq:KK-coefficients-0}, the functions $\tan\left( \frac{\pi \sqrt{t}}{2 m_{\text{KK}}} \right)/\sqrt{t}$ and $\tan\left( \frac{\pi \sqrt{u}}{2 m_{\text{KK}}} \right)/\sqrt{u}$ are real in the physical region ($t,\,u < 0$) due to $\tan\fn{ix} = i \tanh\fn{x}$ for $x \in \mathbb{R}$.} where the effective-theory validity of the KK summation is also discussed.

\subsection{\texorpdfstring{{Deviations in muon angular distributions in elastic $\mu^+ \mu^+$ scattering}}{Angular Deviations in Elastic mu+ mu+ Scattering}}

\begin{figure}[t]
    \centering
    % Left panel (a)
    \begin{subfigure}[t]{0.35\textwidth}
        \centering
        \includegraphics[width=\textwidth]{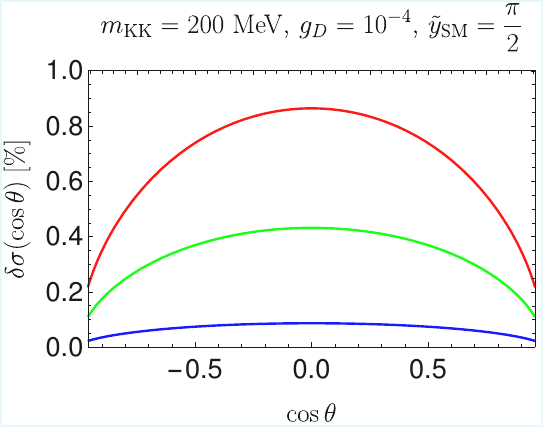}
        \caption{\hspace*{-2.5em}}
        \label{fig:2a}
    \end{subfigure}
    \hfill
    % Right panel (b)
    \begin{subfigure}[t]{0.35\textwidth}
        \centering
        \includegraphics[width=\textwidth]{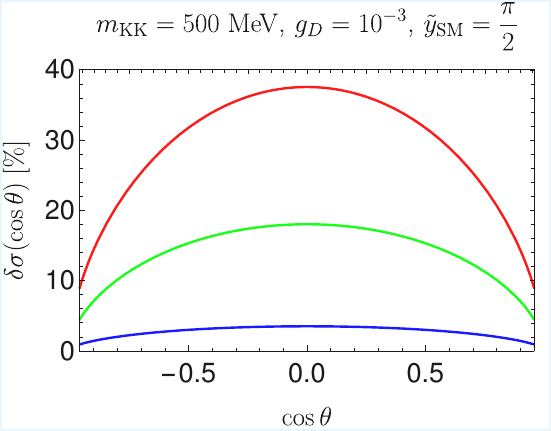}
        \caption{\hspace*{-2em}} 
        \label{fig:2b}
    \end{subfigure}
   \hfill
    % Legend 
    \begin{subfigure}[t]{0.18\textwidth}
       \centering
       \raisebox{3 em}{%
            \includegraphics[width=\textwidth]{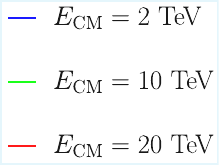}
        }
    \end{subfigure}
    \caption{Angular dependence of the deviation
    \( \delta\sigma(\cos\theta) \) of the total cross-section including 
    new physics from its SM-only prediction for three $\mu$TRISTAN CM energies of $E_{\mathrm{CM}}=$\, 2, 10 and 20 TeV and for
    (a) \( m_{\text{KK}} = 200~\text{MeV} \), \( g_D = 10^{-4} \), 
    and (b) \( m_{\text{KK}} = 500~\text{MeV} \), \( g_D = 10^{-3} \).}
    \label{fig:corrections}
\end{figure}

In this subsection, we analyse the deviations at the leading order in the angular distributions of the final state muons in the elastic scattering process \(\mu^+ \mu^+ \rightarrow \mu^+ \mu^+\) from their respective SM expectations, at the proposed \(\mu\)TRISTAN collider running at various CM energies. Such deviations are induced by processes involving exchange of a tower of KK modes (in the \( t \)- and \( u \)-channels) associated with a 5D \(U(1)_{L_\mu - L_\tau}\) gauge symmetry. The percentage deviation from the SM expectation, as a binned observable, is then given by
\begin{equation}
\delta\sigma(\cos\theta) \equiv \frac{\frac{d\sigma}{d\cos \theta} - \frac{d\sigma^{\text{SM}}}{d\cos \theta}}{\frac{d\sigma^{\text{SM}}}{d\cos \theta}} \times 100 \, ,
\end{equation}
where  $\frac{d\sigma}{d\cos\theta}$ and $\frac{d\sigma^{\text{SM}}}{d\cos\theta}$ are the differential cross-sections in our current 5D framework and the SM, respectively. Note that, in the present context, the leading contribution in $ \delta\sigma(\cos\theta)$ comes from the interference among the ($t$- and $u$-channels) processes mediated by photon, $ Z$, and the KK-excitations $V^{(n)}$. The pattern of variation of \( \delta\sigma(\cos\theta) \) across different angular bins can be used to indirectly infer the presence of new physics, even in the absence of on-shell production of massive KK states. Such deviations could serve as clean, sensitive probes of the underlying dynamics of the extra-dimensional framework. 

Figure~\ref{fig:corrections} illustrates such deviations in the angular differential cross-sections for two representative benchmark points involving 
the KK mass and coupling. The deviations are shown for three CM energies: \( E_\text{CM} = 2~\text{TeV}\), \(10~\text{TeV}\) and \(20~\text{TeV}\). In figure~\ref{fig:2a} we present the case of a lighter base KK mode having $m_{\mathrm{KK}}=200$ MeV with a relatively suppressed effective 4D gauge coupling of $g_D=10^{-4}$, which results in a correction of approximately \( 0.9 \% \) for the CM energy of 
\(20~\text{TeV}\). In contrast, figure \ref{fig:2b} 
illustrates the case with a heavier base 
KK mode of $m_{\mathrm{KK}}=500$ MeV and a stronger coupling of $g_D=10^{-3}$, leading to a sizable deviation of up to about \(40\% \) for the CM energy of \(20~\text{TeV}\). The \(t\)- and \(u\)-channel processes mediated by the KK gauge bosons, \( V^{(n)}\), alter the relative contributions from the vector and axial-vector types of interactions to the elastic process 
\(\mu^+ \mu^+ \rightarrow \mu^+ \mu^+\). Consequently, the angular (differential) distributions of the cross-section differ from the corresponding SM ones. Thus, attaining a wider angular coverage in the experimental setup, down to very forward and backward directions, would be important to fully characterise and discriminate such effects, a capability which is expected to be within technological reach in the future
\(\mu\)TRISTAN collider.
%
%%%%%%%%%%%%%%%%%%%%%%%%%%%%%%%%%%%%%%%%%%%%%%%%%%%%%%%%%%%%%%%%%%%%%%%%%%%%%%%%
%
\subsection{Projected \texorpdfstring{$2 \sigma$}{2 sigma} exclusion reach from elastic scattering
\label{sec:elastic-exclusion}}
\begin{figure}[t!]
    \centering 
%==================== Row 1 ====================%
\begin{subfigure}[t]{0.35\textwidth}
    \centering
    \includegraphics[width=\textwidth]{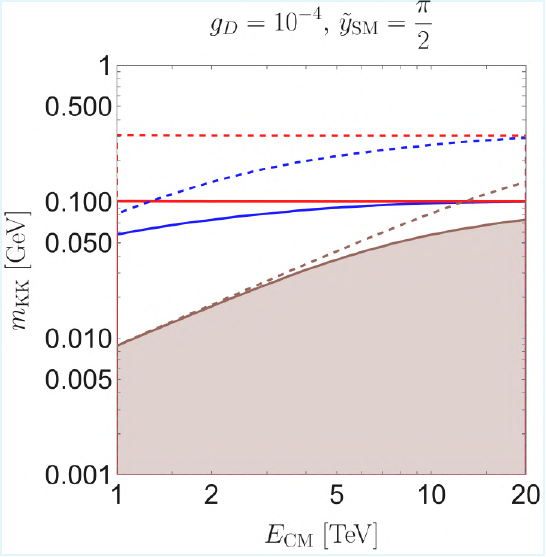}
    \caption{}
    \label{fig:plot3a}
\end{subfigure}
\hfill
\begin{subfigure}[t]{0.35\textwidth}
    \centering
    \includegraphics[width=\textwidth]{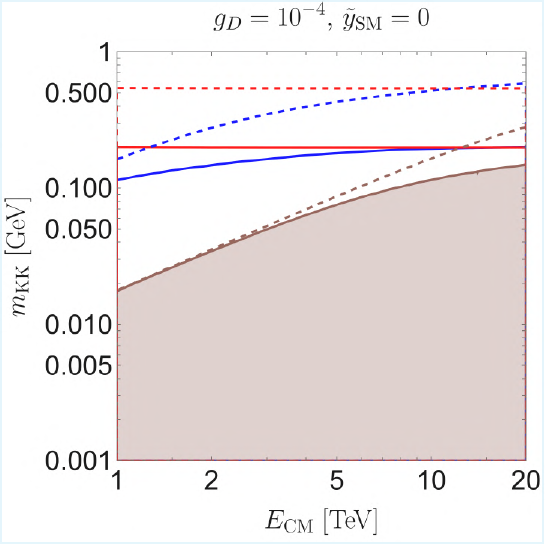}
    \caption{}
    \label{fig:plot3b}
\end{subfigure}
\begin{subfigure}[t]{0.25\textwidth}
    \centering
    \raisebox{3 em}{
    \includegraphics[width=\textwidth]{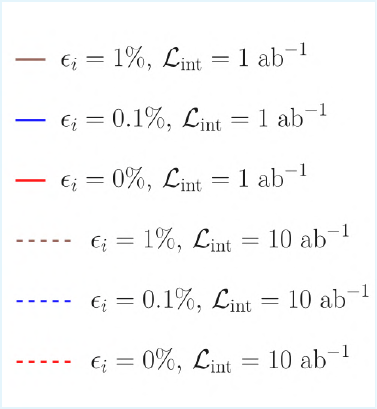}}
\end{subfigure}

\vspace{0.4cm}

%==================== Row 2 ====================%
\begin{subfigure}[t]{0.35\textwidth}
    \centering
    \includegraphics[width=\textwidth]{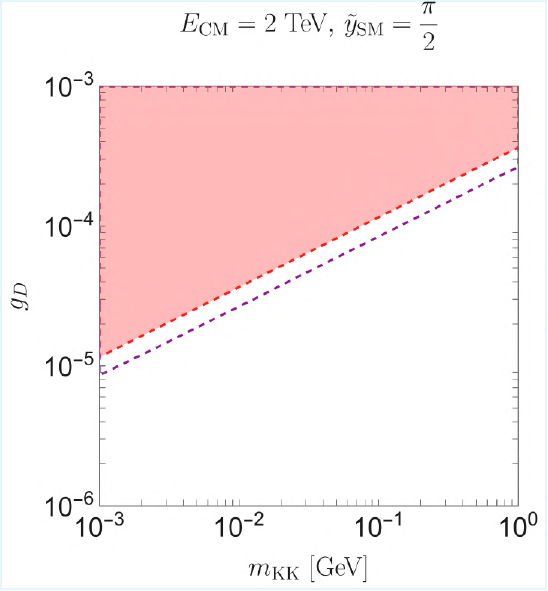}
    \caption{}
    \label{fig:plot3c}
\end{subfigure}
\hfill
\begin{subfigure}[t]{0.35\textwidth}
    \centering
    \includegraphics[width=\textwidth]{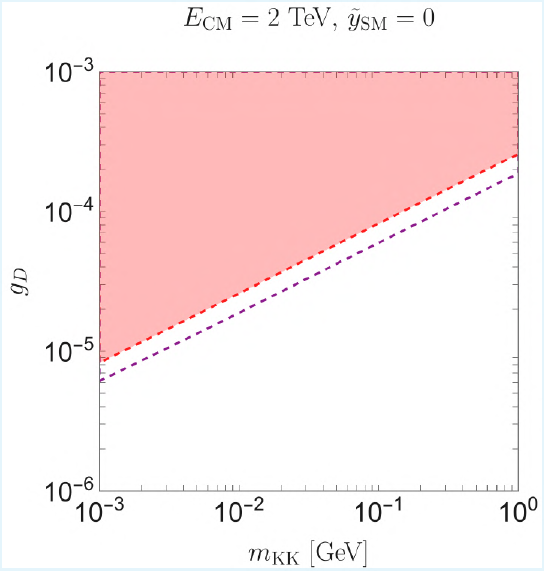}
    \caption{}
    \label{fig:plot3d}
\end{subfigure}
\begin{subfigure}[t]{0.25\textwidth}
    \centering
    \raisebox{6 em}{
    \includegraphics[width=\textwidth]{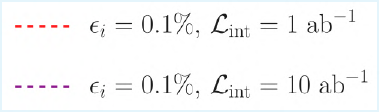}}
\end{subfigure}
   %%%%%%%%%%%%%%%
    \caption{Projected $2\sigma$ (exclusion) reach from elastic $\mu^+ \mu^+$ scattering at the $\mu$TRISTAN collider. 
The first row presents results in the $E_{\text{CM}}$--$m_{\text{KK}}$ plane, where parameter regions 
below each curve correspond to significances exceeding $2\sigma$. The second row shows the projected 
reach in the $m_{\text{KK}}$--$g_D$ plane.
In (a) and (b)/(c) and (d), for each (boundary) curve, the region below/above yields significance greater than $2\sigma$.
In each column, the left (right) panel corresponds to $\tilde{y}_{\text{SM}}={\pi}/{2}$ 
($\tilde{y}_{\text{SM}}=0$). The analysis includes the full KK tower with SM interference, assuming 
representative integrated luminosities and systematic uncertainties.
}
 \label{fig:elastic-exclusion}
\end{figure}
To evaluate the sensitivity to BSM effects, we adopt a binned likelihood approach based on the differential cross-section for the process \(\mu^+ \mu^+ \to \mu^+ \mu^+\). The statistical framework employs a $\chi^2$ test, defined over ten uniformly spaced bins in $\cos\theta$, following the methodology of Refs.~\cite{Okabe:2023esr,Hamada:2022uyn,Huang:2021nkl}. The angular range $\cos\theta \in [\cos\theta_1, \cos\theta_2]$, with $\theta_1 = 164^\circ$ and $\theta_2 = 16^\circ$,
where $\theta$ denotes the scattering angle and $\theta_1$ ($\theta_2$) stands for its maximum (minimum) value that we consider.

The total \(\chi^2\) is then defined as
\begin{align*}
\chi^2 = \sum_{i=1}^{10} \chi^2_i \, ,
\end{align*}
where the contribution from each bin is given by\footnote{
A bin-by-bin \(\chi^2\) analysis effectively captures angular distortions, including forward-backward asymmetries, thus offering enhanced sensitivity to BSM effects compared to inclusive measurements.
}
\begin{align}
\chi^2_i = \frac{\left[N^{\text{BSM}}_i\right]^2}{N^{\text{SM}}_i + \left(\epsilon_i N^{\text{SM}}_i\right)^2} \, .
\label{eq:chi-sqi}
\end{align}
\noindent
Here, the event counts, $N^{\text{BSM}}_i$ and $N^{\text{SM}}_i$, in the `$i$'-th bin are defined as
\begin{align}
N^{\text{BSM}}_i &= \mathcal{L}_{\text{int}} \int_{\text{i-th bin}} \left( \frac{d\sigma}{d\cos\theta} - \frac{d\sigma^{\text{SM}}}{d\cos\theta} \right) d\cos\theta \, , \\
N^{\text{SM}}_i  &= \mathcal{L}_{\text{int}} \int_{\text{i-th bin}} \frac{d\sigma^{\text{SM}}}{d\cos\theta} d\cos\theta \, ,
\end{align}
where $\mathcal{L}_{\mathrm{int}}$ denotes the integrated luminosity, and $\epsilon_i$ represents the fractional systematic uncertainty in the $i$-th bin, arising from experimental and theoretical sources. Such a parametrisation effectively captures the dominant uncertainties associated with the luminosity normalisation, beam energy and polarisation, detector acceptance, and uncertainties in the measured values of various fundamental parameters, following the standard practice in precision analyses at future lepton colliders, including the muon-based ones. For the present analysis, we consider two benchmark luminosity values, \(\mathcal{L}_{\text{int}} = 1\,\text{ab}^{-1}\) and \(10\,\text{ab}^{-1}\), and evaluate the impact of systematic uncertainties with \(\epsilon_i = 0\), \(10^{-3}\), and \(10^{-2}\), corresponding to a hypothetical vanishing uncertainty and its more realistic values like 0.1\%, and 1\%, respectively.

We now present the projected $2\sigma$ exclusion reach (henceforth to be simply called as `reach') in two
 regions of parameter space as derived from the elastic scattering process \( \mu^+ \mu^+ \to \mu^+ \mu^+ \) at the proposed \(\mu\)TRISTAN collider. For a projected 2$\sigma$ reach with two degrees of freedom (2~dof), we adopt the standard criterion \(\chi^2 > 6.18\).

Figure~\ref{fig:elastic-exclusion} illustrates the projected $2 \sigma$ reach in: Fig.~\ref{fig:plot3a} and \ref{fig:plot3b} for the lightest KK gauge boson mass \( m_{\text{KK}} \) versus the collider center-of-mass energy \( E_{\text{CM}} \), and  
Fig.~\ref{fig:plot3c} and \ref{fig:plot3d} for the effective 4D gauge coupling \( g_D \) (defined as \( g_{5D} / \sqrt{\pi R} \)) versus \( m_{\text{KK}} \), for benchmark values of the integrated luminosity. We show results for different values of the dimensionless brane position \( \tilde{y}_{\text{SM}} = y_{\text{SM}} / R \), introduced earlier. In all these cases, the contribution from the full KK tower is considered, including interference with the SM-$Z$ mediated process.

In Fig.~\ref{fig:plot3a}, we present the $2\sigma$ reach obtained for a fixed gauge coupling of $g_D = 10^{-4}$, considering two benchmark integrated luminosities, $1\,\text{ab}^{-1}$ and $10\,\text{ab}^{-1}$, and systematic uncertainties of $\epsilon_i = 0\%$, $0.1\%$, and $1\%$. As expected, the sensitivity to higher \( m_{\text{KK}} \) improves with increasing CM energy and luminosity, whereas the inclusion of systematic uncertainty weakens the $2 \sigma$ reach.

Figure~\ref{fig:plot3c} illustrates the corresponding projected $2\sigma$ reach in the 
$m_{\text{KK}}$--$g_D$ plane for a fixed
$E_{\text{CM}} = 2\,\text{TeV}$. In a scenario with systematic uncertainty, viz., $\epsilon_i = 0.1\%$, and 
$\mathcal{L}_{\text{int}} = 10\,\text{ab}^{-1}$, the effective 4D gauge coupling as small as $g_D \sim 9 \times 10^{-6}$ can be probed for sufficiently small values of the KK mass, reaching down to $m_{\mathrm{KK}} \sim 1\,\text{MeV}$. Other complementary versions of these plots, corresponding to the limiting case \(\tilde{y}_{\text{SM}} = 0\), are provided in Fig.~\ref{fig:plot3b} (analogous to Fig.~\ref{fig:plot3a}) and Fig.~\ref{fig:plot3d} (analogous to Fig.~\ref{fig:plot3c}). While the general features and trends of the contours remain qualitatively similar and in the expected lines across the \(\tilde{y}_{\text{SM}} = 0\) and \(\tilde{y}_{\text{SM}} = {\pi}/{2}\) cases, the derived $2 \sigma$ reach for \(\tilde{y}_{\text{SM}} = 0\) are slightly stronger. This enhancement arises because, at \(\tilde{y}_{\text{SM}} = 0\), the KK mode wavefunctions reach their maximal values, resulting in stronger couplings to SM fields. Consequently, their contribution to the scattering process is amplified, leading to a tighter $2 \sigma$ reach compared to the \(\tilde{y}_{\text{SM}} = {\pi}/{2}\) benchmark.

We note that these are future projections under idealised experimental conditions. The projected $2 \sigma$ reach presented here, therefore, illustrates the remarkable sensitivity of the elastic $\mu^+ \mu^+ \to \mu^+ \mu^+$ scattering at high energies to bulk $U(1)_{L_\mu - L_\tau}$ extensions, even in the absence of a sizable kinetic mixing.

%%%%%%%%%%%%%%%%%%%%%%%%%% Bremsstrahlung study %%%%%%%%%%%%%%%%%%%%%%%%%%%%%%%%%%%%%%%%%%%%%%%%%%%%%%%%%%%%%%%%%
%\section{\cmag{Same-sign dimuon + MET signatures at $\mu$TRISTAN}}
%
\section{Semi-visible and all-visible final states at $\mu$TRISTAN
\label{sec:Bremsstrahlung}}
%%%%%%%%%%%%%%%%%%%%%%%%%%%%%%%%%%%%%%%%%%%%%%%%%%%%%%%%%%%%%%%%%%%%%%%%%%%%%%%%%%%%%%%%%%%%%%%%%%%%%%%%
\begin{figure}[t]
    \centering
    \includegraphics[width=0.9\textwidth]{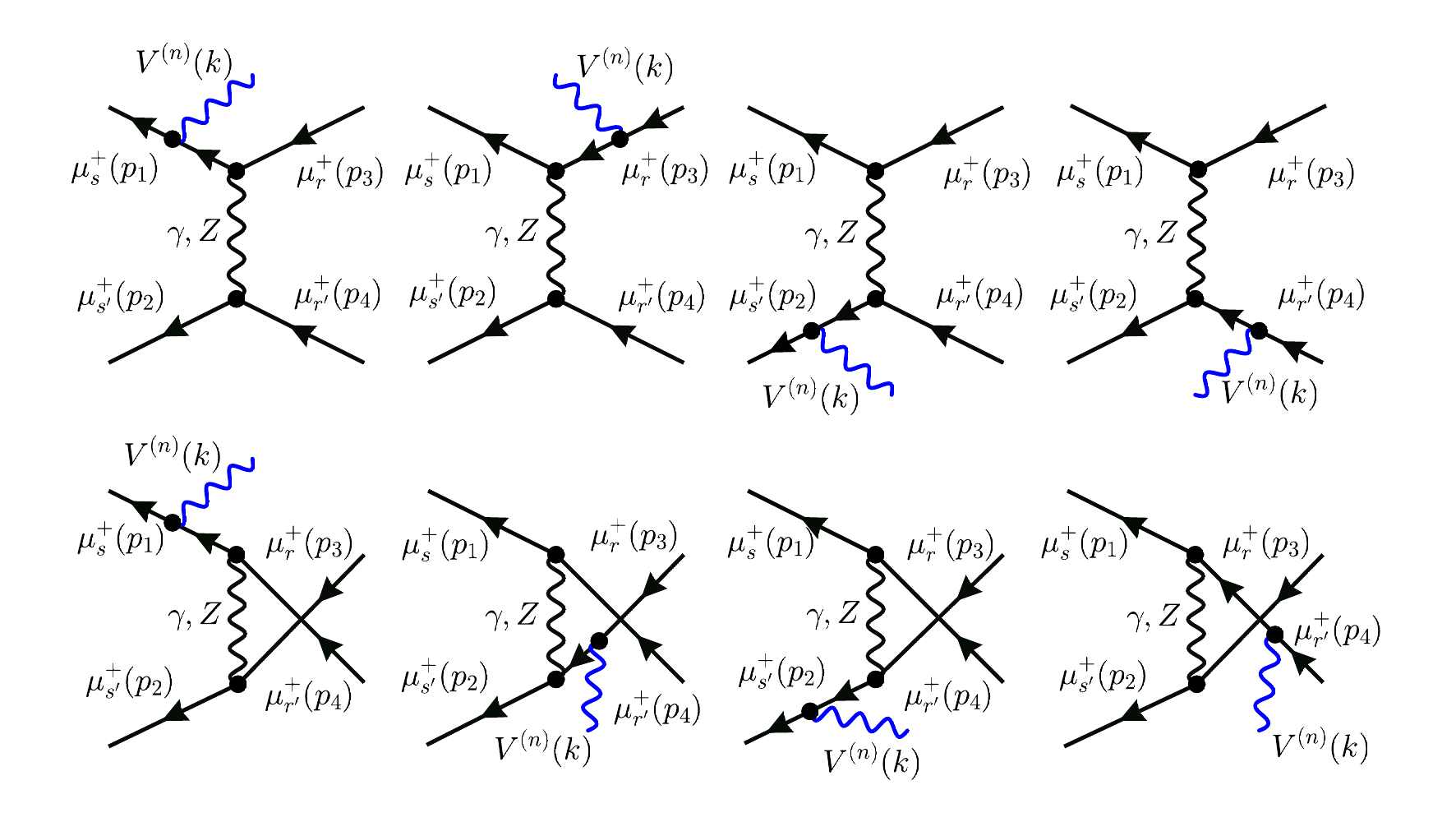}
   \caption{Tree-level Feynman diagrams for the bremsstrahlung-like process 
\(\mu^+ \mu^+ \to \mu^+ \mu^+ V^{(n)}\) (\(n = 1, 2, 3, \dots\)) in the \(t\)-channel (upper panel) and \(u\)-channel (lower panel).}
 \label{fig:threebody}
\end{figure}

We now turn to the study of semi-visible same-sign dimuon (SSDM) + missing transverse energy (MET; $\not\!\!E_T$) and all visible four muon final state with the right pair of opposite-sign dimuon (OSDM) reconstructed to a resonant signature arising from the production of on-shell KK excitations of the $U(1)_{L_\mu - L_\tau}$ gauge 
boson at the $\mu$TRISTAN collider. These KK excitations, $V^{(n)}$, are radiated off 
the external-state muons in scattering processes that proceed through $t$- and $u$-channel exchanges, in a bremsstrahlung-like topology, viz.,
\[ \mu^+ \mu^+ \to \mu^+ \mu^+ V^{(n)} , \]%
where, \(V^{(n)}\) denotes the \(n\)-th KK excitation of the \(U(1)_{L_\mu - L_\tau}\) gauge boson.
The tree-level Feynman diagrams for such processes are presented in Fig.~\ref{fig:threebody}. The subsequent decays of the on-shell KK gauge boson(s), $V^{(n)}$, determine the experimental final states of interest.

The first signal final state of interest is the
semi-visible one with SSDM+MET
\[
\mu^+ \mu^+ \rightarrow \mu^+ \mu^+ + \; \text{MET},
\]
where MET is carried away by neutrinos from the decay of an on-shell KK mode. In the narrow-width approximation (NWA), the signal can be factorised into productions of $V^{(n)}$ and their subsequent decays, i.e.,
\[
\mu^+ \mu^+ \to \mu^+ \mu^+ V^{(n)} , 
\quad \text{followed by} \quad 
V^{(n)} \to \nu_\alpha \bar{\nu}_\alpha, 
\]
thus giving rise to the semi-visible (SSDM+ MET) final state 
\[
\mu^+ \mu^+ \to \mu^+ \mu^+ V^{(n)} 
\to \mu^+ \mu^+ \nu_\alpha \bar{\nu}_\alpha, \quad \mathrm{with}
\quad n = 1, 2, 3, \, \dots \; \mathrm{, \; and} \;~\alpha = \mu, \tau \, ,
\]
where a same-sign muon pair recoils against the MET arising from the undetected neutrinos.

The second process of interest is an all-visible (four muon) final state, which is complementary to the semi-visible case. This emerges when the KK excitations decay to charged leptons, in particular, to an opposite-sign muon pair, giving rise to a striking all-visible signature with four muons in the final state, viz.,
%%%%
%
\[
\mu^+ \mu^+ \to \mu^+ \mu^+ V^{(n)} \to \mu^+ \mu^+ \mu^+ \mu^-, \qquad n=1,2,3, \, \dots \; .
\]
Here, the two same-sign primary muons in the final state are accompanied by an additional opposite-sign muon pair. The key experimental observable in the 4-body, all-visible final state is a narrow resonance peaking at the mass value of the produced KK gauge boson in the invariant mass distribution of the correct $\mu^+ \mu^-$ pair, which provides a clean probe to the KK spectrum and allows efficient separation of the coveted signal from SM backgrounds.
Thus, the above-mentioned two final states serve as complementary probes into the spectrum of the KK gauge bosons, which would enable us to draw a robust and multifaceted search strategy at the $\mu$TRISTAN collider.

%%%%%%%%%%%%%%%%%%%%%%%%%%%%%%%%%%%%%%%%%%%%%%%%%%%%%%%%%%%%%%%%%%%%%%%%%%%%%%%%%%%%%%%%%%%%%%%%%%%%%%%%%%%%%%%%%%%%%%

\subsection{Estimation of the signal cross-section
\label{sec:parameter scan}}
In this subsection, we describe how the total signal cross-section of a given KK tower is obtained by adding up contributions from the excited states, up to an appropriate cutoff below the collider energy. These summed cross-sections are then used in obtaining the $2 \sigma $ reach in the relevant parameter planes via scans of the parameter space. For clarity, the procedure is presented separately for the semi-visible and all-visible final states.

The signal cross-section incorporates the dominant tree-level electroweak contributions mediated by the photon and the \(Z\)-boson. Unlike in the case of the elastic process $\mu^+ \mu^+ \to \mu^+ \mu^+$ discussed in the last section, the subdominant effects from KK gauge-boson exchange in the propagators are neglected, as their contributions are suppressed by the small values of the effective 4D gauge coupling (\(g_D \lesssim 10^{-4}\)) relevant for our study. As the respective amplitude scales as \(g_D^{2}\), for  \(m_{\text{KK}} \sim \mathcal{O}(\text{1 MeV})\), summing over the KK modes up to \(\mathcal{O} (\text{TeV})\) (i.e., over six orders of magnitude) does not compensate for this suppression.

For both final states, the total signal cross-sections are obtained by summing over on-shell KK excitations of the 5D theory produced via a bremsstrahlung-like process. This procedure effectively yields the full 5D signal cross-section. In this analysis, we restrict ourselves to \(\sqrt{s} = 2~\text{TeV}\) as a representative benchmark for the proposed \(\mu\)TRISTAN collider.
As discussed later, increasing the CM energy does not seem to yield a commensurate increase in signal events in the present framework.
\subsubsection{Semi-visible final state ($\mu^+\mu^+ \to \mu^+\mu^+ \slashed{E}_T$)
\label{sec:semi-visible}}
To compute the total signal cross-section, we perform a summation over KK modes up to a cutoff of \( m_{\text{KK}}^{\text{max}} = 1\,\text{TeV} \), which lies below the collider energy \(\sqrt{s} = 2\,\text{TeV}\). The effective signal cross-section obtained after applying the MET cut of \(1\,\text{GeV} \leq \slashed{E}_T \leq 10\,\text{GeV}\), at each value of the lowest (base) KK mass \( m_{\text{KK}} \) (for which, there is a whole tower of excitations), is obtained by summing over all KK modes up to this cutoff, i.e.,
\al{
\sigma^{\text{semi-vis}}_{\text{signal}} = \sum_{n:\, M_n \leq 1\,\text{TeV}} \sigma^{\text{semi-vis}}_n \times \mathcal{E}_n \,,
}
where \( \sigma^{\text{semi-vis}}_n \) denotes the tree-level cross-section for the production of the \( n \)th KK mode, and \( \mathcal{E}_n \) represents the corresponding acceptance of the employed MET cut for the semi-visible final state. 

To avoid an explicit summation over many discrete KK states, we approximate the sum by a continuous integral. The procedure is as outlined below.
\begin{enumerate}
    \item We first compute the parton-level cross-sections, \( \sigma^{\text{semi-vis}}_{n} \), for a representative set of KK masses \( M_{n} \) spanning the range \( 1\,\text{MeV} \leq M_{n} \leq 1\,\text{TeV} \) (about 30 mass points covering the entire energy range which are equally spaced in logarithmic scale). Each cross-section is then multiplied by the corresponding MET cut efficiency \( \mathcal{E}_{n} \) to obtain the acceptance-weighted cross-section:
    \al{
    \sigma^{\text{semi-vis}}_{\text{eff}}(M_{n}) = \sigma^{\text{semi-vis}}_{n} \times \mathcal{E}_{n} \,.
    }
    The cross-section calculations and event generations are performed with \texttt{MadGraph5} and \texttt{MadSpin}, while the cut-based analyses are carried out with \texttt{MadAnalysis5}.
    \item The discrete set of points \( \{ M_{n}, \sigma^{\text{semi-vis}}_{\text{eff}}(M_{n}) \} \) is subsequently interpolated using a linear interpolation routine to construct a smooth, continuous function \( \sigma^{\text{semi-vis}}_{\text{eff}}(m) \), defined up to \( m = 1\,\text{TeV} \).
    
    \item Finally, the total signal cross-section is obtained by integrating over the KK tower, weighted by the KK mode density appropriate for a single flat extra dimension, as considered in our analysis:
    \al{
    \sigma^{\text{semi-vis}}_{\text{signal}} \simeq \int_{m = m_{\text{KK}}}^{1\,\text{TeV}} \frac{\sigma^{\text{semi-vis}}_{\text{eff}}(m)}{2 m_{\text{KK}}} \, dm \, .\footnotemark
    }
\footnotetext{
The continuum approximation implicit in the KK-mass integration is valid for sufficiently dense KK spectra. For larger KK scales ($m_\text{KK} \gtrsim 1$ GeV), where the KK mass spacing is non-negligible, the continuum description breaks down; we explicitly perform a discrete sum over individual KK states. Both contributions are combined to obtain the total signal cross-section.}
\end{enumerate}
Furthermore, the cross-section for each KK mode scales as
\[
\sigma^{\text{semi-vis}}_n \propto g_D^2 \, ,
\]
allowing us to efficiently obtain results for arbitrary values of \( g_D \) through rescaling, without requiring additional simulations.
\subsubsection{All-visible final state} 
In the all-visible final state, the signal arises predominantly from the KK modes whose masses fall within the selected invariant-mass window. The total signal cross-section is therefore obtained as  
\al{
\sigma^{\tx{all-vis}}_{\text{signal}} \;=\; \sum_{n:\, M_n \in \text{mass window}} \sigma^{\tx{all-vis}}_n \times \mathcal{E'}_n \, ,
}
where the summation runs over all KK modes whose masses 
$M_n$ lies inside the chosen invariant-mass window (hereafter referred to as the mass window). In other words, only those resonant states kinematically accessible within the applied di-muon mass cut contribute to the signal yield. The factor $\mathcal{E'}_n$ denotes the corresponding cut acceptance associated with the invariant-mass selection for the $n$-th KK mode within this window. For a narrow mass window, both the production cross-sections and acceptances of neighbouring KK states vary negligibly. In this regime, the discrete sum can be approximated by
\al{
\sigma^{\tx{all-vis}}_{\text{signal}} 
	\;\simeq\; 
		N_{\text{KK}}^{\text{window}} \times \sigma^{\tx{all-vis}}_{\text{eff}}(M_{n_\tx{rep}}) \, ,
}
where \( N_{\text{KK}}^{\text{window}}\) denotes the number of KK modes contained within the mass window, and 
$$
\sigma^{\tx{all-vis}}_{\text{eff}}(M_{n_\tx{rep}}) {:=} \sigma^{\tx{all-vis}}_{n_\tx{rep}} \times \mathcal{E'}_{n_\tx{rep}}
$$
represents the effective cross-section of any representative KK mode ($M_{n_\tx{rep}}$) whose mass lies inside the said mass window.
%%%%%%%%%%%%%%%%%%%%%%%%%%%%%%%%%%%%%%%%%%%%%%%%%%%%%%%%%%%%%%%%%%%%%%%%%%%%%%%%
\subsubsection{Simulation details} 
We implement a 4D vector boson \(Z'\), interpreted as a generic KK excitation ($V^{(n)}$) in \texttt{FeynRules} and generate the corresponding UFO model. By varying the \(Z'\) mass, we effectively probe the phenomenology associated with different KK states. Generating events for the signal and background proceeds as follows:
\begin{itemize}
    \item The \(2 \to 3\) production process \( \mu^+ \mu^+ \to \mu^+ \mu^+ Z^\prime \) is simulated using \texttt{MadGraph5}.
    \item Subsequent decays of \(Z^\prime\) are handled via \texttt{MadSpin} under the narrow-width approximation:  
    \( Z^\prime \to \nu_{\alpha} \bar{\nu}_{\alpha} \) (\( \alpha = \mu, \tau \)) for the
    semi-visible final state, and  
    \( Z^\prime \to \mu^+ \mu^- \) for the all-visible final state.  
    \item The SM backgrounds to the same final states are also simulated with \texttt{MadGraph5}, incorporating the full set of tree-level electroweak processes. For the semi-visible channel, the backgrounds receive contributions from photon, \(Z\), and \(W^\pm\) exchanges, whereas for the dimuon channel, they arise solely from photon- and \(Z\)-mediated processes.

\end{itemize}
%
%%%%%%%%%%%%%%%%%%%%%%%%%%%%%%%%%%%%%%%%%%%%%%%%%%%%%%%%%%%%%
\subsection{Kinematic cuts and projected \texorpdfstring{$2\sigma$}{2 sigma} reach
\label{sec:cut implementation}}
Kinematic variables provide the main experimental handles for distinguishing the signal from SM backgrounds in both the semi-visible and all-visible final states. In the semi-visible case, MET serves as the discriminator, while for the visible final state, the key observable is the invariant mass of the (right pair of) opposite-sign muons that originate from the decay of the on-shell KK boson produced in the primary $2 \to 3$ scattering process. For each of these final states, a cut-based analysis is performed using \texttt{MadAnalysis5} with an event-selection strategy based on relevant kinematic variables to optimise the signal-to-background ratio. 

The sensitivity of the proposed search is quantified using the Gaussian significance given by
\begin{equation}
S_{\rm Gauss} = \frac{s}{\sqrt{s + b}} \, ,
\label{eq:Gauss-met}
\end{equation}
where \(s\) and \(b\) are the expected signal and background yields after all analysis cuts~\cite{Biswas:2017lyg}.
The validity of this approximation is confirmed by comparison with the log-likelihood-based significance as described in Appendix~\ref{appendix-c}. In the asymptotic (i.e., for a large sample size) limit, the likelihood-ratio test statistic evaluated on the Asimov data set follows a $\chi^{2}$ distribution in accordance with Wilks’ theorem~\cite{wilks1938large}.
Accordingly, $S_{\text{Asimov}}$ provides the $\chi^{2}$-equivalent measure in Eq.~\eqref{eq:chi-sqi} for a single bin with vanishing systematic uncertainties. Since the hypothesis test involves scanning a single signal parameter $(m_{\text{KK}})$ with all other parameters held fixed, the corresponding profile-likelihood test statistic asymptotically obeys a $\chi^{2}$ distribution with one degree of freedom.

The event yields are obtained by multiplying the cross-sections by the integrated luminosity. The  $2\sigma$ reach is derived by requiring a minimum significance threshold of \(S_{\text{Gauss}} \ge 2\). The corresponding projected exclusion regions are presented in the \(m_{\text{KK}}\)--\(g_D\) plane, for a 
\(\sqrt{s} = 2\,\text{TeV}\) run of the \(\mu\)TRISTAN collider, assuming integrated luminosities of \(1\,\mathrm{ab}^{-1}\) and \(10\,\mathrm{ab}^{-1}\). This would highlight the capability of such a machine to probe the effective 4D gauge coupling and the lightest KK mass, as discussed below. The two final states provide complementary sensitivities: the semi-visible final state probes invisible decays via MET, while the all-visible final state exploits resonant peaks in the invariant-mass spectrum of an appropriate pair of opposite-sign muons. Together, these provide a comprehensive understanding of the extent of the projected $2 \sigma$ reach that could be achieved in the said parameter plane.
\subsubsection{Semi-visible final state
\label{sec:MET channel}}
\begin{figure}[t]
    \centering
    \includegraphics[width=0.7\textwidth]{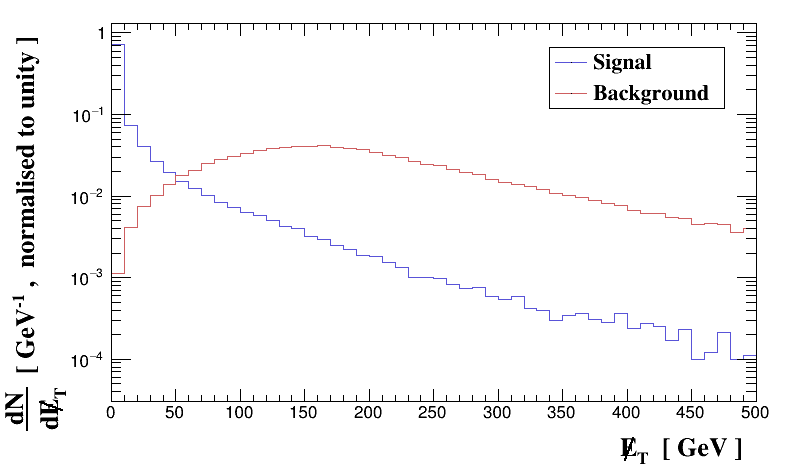}
    \caption{MET ($\slashed{E}_T$) distributions (normalised to unity) for the signal (in blue) with \(M_{Z^\prime} = 10\,\text{MeV} \) and \( g_D = 10^{-4} \), and for the SM background (in red).}
    \label{fig:met-spectrum}
\end{figure}
%%%
\begin{table}[t]
    \centering
    \begin{tabular}{|c|c|c|c|}
        \hline
        Cuts (GeV) & Signal ($s$) & Background ($b$) & $s/b$ \\
        \hline
        Initial (no cut)
        & $1.068130 \pm 0.000844$
        & $186901 \pm 180$
        & $5.71 \times 10^{-6} \pm 7.13 \times 10^{-9}$ \\
        \hline
        REJ: MET $>500.0$
        & $1.0659 \pm 0.0468$
        & $177206 \pm 196$
        & $6.02 \times 10^{-6} \pm 2.64 \times 10^{-7}$ \\
        \hline
        REJ: MET $>100.0$
        & $1.008 \pm 0.238$
        & $29808 \pm 160$
        & $3.38 \times 10^{-5} \pm 8.00 \times 10^{-6}$ \\
        \hline
        REJ: MET $>50.0$
        & $0.951 \pm 0.323$
        & $6872.3 \pm 81.6$
        & $1.38 \times 10^{-4} \pm 4.70 \times 10^{-5}$ \\
        \hline
        REJ: MET $>20.0$
        & $0.86 \pm 0.41$
        & $966.3 \pm 31.0$
        & $0.000889 \pm 0.000425$ \\
        \hline
        REJ: MET $>15.0$
        & $0.828 \pm 0.432$
        & $502.8 \pm 22.4$
        & $0.001646 \pm 0.000862$ \\
        \hline
        REJ: MET $>10.0$
        & $0.780 \pm 0.458$
        & $207.5 \pm 14.4$
        & $0.00376 \pm 0.00223$ \\
        \hline
        REJ: MET $>5.0$
        & $0.697 \pm 0.492$
        & $46.73 \pm 6.83$
        & $0.0149 \pm 0.0108$ \\
        \hline
        REJ: MET $>3.0$
        & $0.635 \pm 0.507$
        & $14.95 \pm 3.87$
        & $0.0425 \pm 0.0357$ \\
        \hline
        REJ: MET $>1.0$
        & $0.503 \pm 0.516$
        & $1.87 \pm 1.37$
        & $0.269 \pm 0.339$ \\
        \hline
    \end{tabular}
 \caption{Cut-flow table for the expected event yields (along with their associated Monte Carlo statistical uncertainties arising from the finite size of the simulated samples, as computed by \texttt{MadAnalysis5}) for the semi-visible signal process, and for a representative KK mode with $M_{Z^\prime}=10~\mathrm{MeV}$ and $g_D=10^{-4}$, as well as for the background. The event yields are normalised to an integrated luminosity of $1~\mathrm{ab}^{-1}$, with an initial signal cross-section of $1.07\times10^{-6}~\mathrm{pb}$ for the representative KK mode. Also shown are the corresponding signal-to-background ($s/b$) ratios with their Monte Carlo statistical uncertainties. The quoted event yields correspond to a single representative KK mode and do not represent the full 5D signal used to obtain the projected exclusion reach (see text for details).}

  \label{tab:MET-cutflow}
\end{table}

\begin{figure}[t]
    \centering
    % First plot: current yt = pi/2 case
    \begin{subfigure}[b]{0.35\textwidth}
        \centering
        \includegraphics[width=\textwidth]{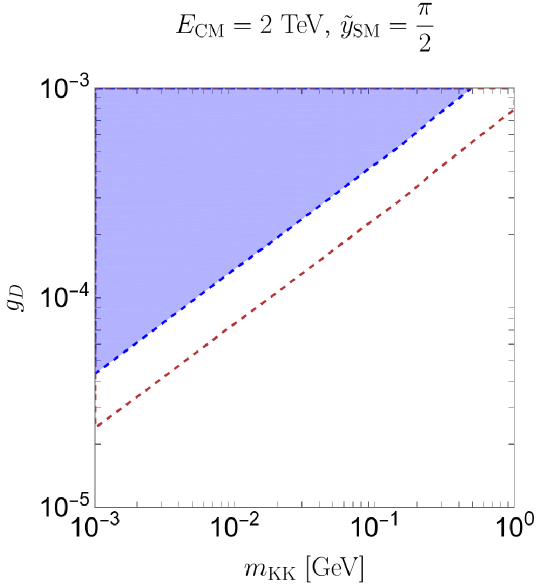}
        \caption{\hspace*{-3em}}
        \label{fig:met-yt-nonzero}
    \end{subfigure}
    \hfill
    % Second plot: yt = 0 case
    \begin{subfigure}[b]{0.35\textwidth}
        \centering
        \includegraphics[width=\textwidth]{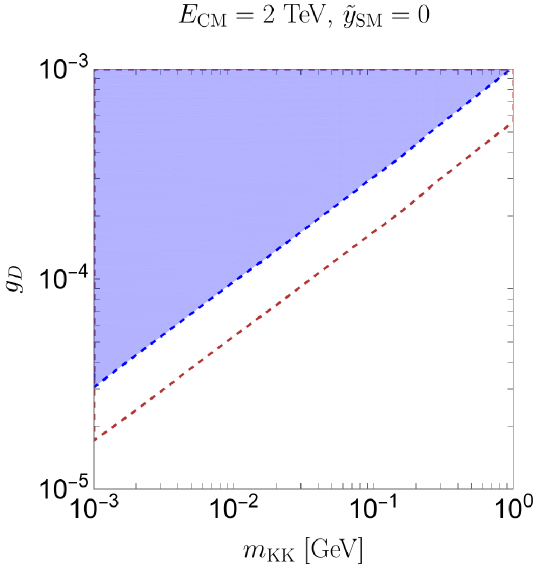}
       \caption{\hspace*{-3em}}
        \label{fig:met-yt-zero}
    \end{subfigure}
     \begin{subfigure}[b]{0.18\textwidth}
        \centering
        \raisebox{3cm}{
        \includegraphics[width=\textwidth]{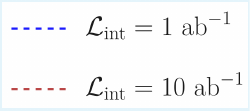}
        }
    \end{subfigure}
    \caption{Projected $2 \sigma$ reach in the 
    \(m_{\text{KK}} \)-\( g_D \) plane for the semi-visible process 
\( \mu^+ \mu^+ \rightarrow \mu^+ \mu^+ \nu_\alpha \bar{\nu}_\alpha \) (\(\alpha = \mu, \tau\)) at a \(\sqrt{s} = 2\,\text{TeV}\) \(\mu\)TRISTAN collider. The parameter region above each curve corresponds to the significance more than $2 \sigma$ with integrated luminosities of \(1~\text{ab}^{-1}\) and \(10~\text{ab}^{-1}\), respectively, for \(\tilde{y}_{\text{SM}} = {\pi}/{2} \) (plot (a)) and \(\tilde{y}_{\text{SM}} = 0\) (plot (b)).}
\label{fig:6}
\end{figure}

In both signal and background events, MET arises from the undetected neutrinos. However, the kinematic features of the MET for the signal and background differ significantly. In the signal process \( \mu^+ \mu^+ \rightarrow \mu^+ \mu^+ V^{(n)} \rightarrow \mu^+ \mu^+ \nu_\alpha \bar{\nu}_\alpha \) (with \(n = 1,2,3,\ldots\) and \(\alpha = \mu, \tau\)), the neutrinos originate from the decay of a resonant KK gauge boson, resulting in MET distributions that are typically confined to the low-energy region as is expected for not so heavy KK states.
In contrast, the SM backgrounds in $\mu^+ \mu^+$ collisions generate neutrinos through non-resonant electroweak processes involving off-shell gauge boson exchange and multi-body final states, where neutrinos can be produced at different vertices and are not constrained by a common parent resonance.
Consequently, the resulting MET distributions are broader and extend to significantly higher values.
These features are clearly observed in the MET distributions for the signal and the background that we present in Fig.~\ref{fig:met-spectrum}, which are evaluated at a CM energy of \(\sqrt{s} = 2\,\text{TeV}\), for a representative scenario having $M_{Z'}=10\, \tx{MeV}$ and \(g_D = 10^{-4}\).
The signal cross-section for the above-mentioned representative scenario with a single $Z'$ state corresponding to an KK excitation is approximately \(10^{-6}\,\text{pb}\), and after summing over the full KK tower, the total signal cross-section increases to \(\sim 10^{-4}\,\text{pb}\). In contrast, the total SM background cross-section is approximately \(0.18\,\text{pb}\). This calls for a carefully chosen set of kinematic cuts to improve/maximise the signal significance.

To this end, a comprehensive cut-flow analysis is performed to evaluate the impact of the MET cut choice on the signal and background strengths, and hence on the power to discriminate the signal from the background. The results are summarised in Table~\ref{tab:MET-cutflow}, illustrating the effect of each selection criterion on event yields. Signal-to-background ratios (\(s/b\)) are presented for a representative KK mode with $M_{Z^\prime}=10~\mathrm{MeV}$ and $g_D=10^{-4}$, rather than considering the full KK-tower contribution, alongside labels indicating the rejection (`REJ') steps. The table is intended to demonstrate the impact of the MET selection windows on the signal efficiency and background rejection for an individual KK mode. The quoted $s$ and $b$ values, therefore, are not the actual numbers entering into our analysis~\eqref{eq:Gauss-met} that lead to the projected exclusion regions. Instead, the latter are obtained by multiplying the initial number of events by the cut-acceptance for each KK mode, and then summing over the full tower. The analysis reveals that more stringent MET cuts generally enhance the signal significance, as evidenced by the increasing \(s/b\) ratio with tighter upper MET cuts for the representative KK mode. For example, rejecting events with \(\slashed{E}_T > 5\,\mathrm{GeV}\) yields \blue{a} higher significance than using a threshold of \(\slashed{E}_T > 10\,\mathrm{GeV}\).

At this point, in our search for an optimal MET cut, we note that a dedicated study of MET resolution at the \(\mu\)TRISTAN collider is not yet available in the literature. However, existing detector-performance analyses~\cite{Marshall:2012ry,Bartosik:2020xwr} indicate that particle-flow techniques could achieve a precision of a few per cent in measuring missing momenta at future lepton colliders. Guided by these expectations and considering a possible presence of additional low-energy backgrounds at the \(\mu\)TRISTAN collider, we adopt the following MET window:
\al{
1~\text{GeV} \leq \slashed{E}_T \leq 10~\text{GeV}, %%
	\label{eq:toneutrino_METcut}
}
which is expected to fall within the detector's realistic capabilities. This window preserves most of the signal-enhancing advantages of narrower intervals, such as \(1\,\mathrm{GeV} \leq \slashed{E}_T \leq 5\,\mathrm{GeV}\).
The upper cut on MET efficiently suppresses background events characterised by high MET, while preserving a substantial fraction of signal events localised in the low-MET region. To mitigate theoretical and experimental ambiguities associated with near-vanishing MET, a lower cut of \(1\,\mathrm{GeV}\) is imposed on it. Events with vanishingly small MET may originate from soft final states or detector effects not systematically accounted for in our simulation. The MET window mentioned above is found to optimise the signal-to-background ratio while respecting realistic detector resolution thresholds. All  plots for projected $2 \sigma$ reach are presented in Fig.~\ref{fig:6}

It is noteworthy that the signal events analysed here correspond to a representative parameter choice of \(g_D = 10^{-4}\) and \(M_{Z^\prime} = 10\,\mathrm{MeV}\). In this regime, the signal receives contributions from a large number of closely spaced KK modes, resulting in a smooth, stable event distribution. Therefore, the projected $2 \sigma$ reach shows negligible sensitivity to moderate changes in the MET window, thus affirming the robustness of the derived constraints.

The projected $2 \sigma$ reach is obtained by evaluating the statistical significance of the signal after applying MET-based selection criteria, assuming negligible systematic uncertainties. The resulting projected $2\sigma$ reach are displayed in Figs.~\ref{fig:met-yt-nonzero} and~\ref{fig:met-yt-zero} for the cases \(\tilde{y}_{\text{SM}} = {\pi}/{2}  \) and \(\tilde{y}_{\text{SM}} = 0\), respectively.
For example, for the benchmark point having \(g_D = 10^{-4}\), and \(\tilde{y}_{\text{SM}} = {\pi}/{2}\), with $\sqrt{s}=2$ TeV, the expected $2\sigma$ reach can be as low as \(g_D \sim 2 \times 10^{-5}\), for an integrated luminosity of \(10~\mathrm{ab}^{-1}\). The sensitivity depends on the localisation of the SM brane in the extra dimension: when the brane is located at \(\tilde{y}_{\text{SM}} = 0\), where the bulk gauge boson profile attains its maximum, the effective coupling to SM fields is enhanced. Consequently, the projected $2\sigma$ reach is moderately strengthened relative to the \(\tilde{y}_{\text{SM}} ={\pi}/{2} \) configuration, thus reflecting an enhanced sensitivity. A similar behaviour has been observed in the elastic scattering process as discussed in section~\ref{sec:Mutristan}. This demonstrates the robustness of the semi-visible final state as a probe to extra-dimensional gauge interactions across configurations with different brane localisations.
Further sophistication of such an analysis, by incorporating a full detector simulation and treating the systematic uncertainties in detail, would refine and strengthen these projections. While increasing the CM energy generally improves the collider sensitivity by extending its kinematic reach, the situation in the present framework is more subtle. For relatively small KK scales, the total signal events are already dominated by the lowest-lying modes, such that increasing $\sqrt{s}$ leads to, at best, a modest (or, not so sizable) improvement. Even for larger KK scales (e.g., $m_{\mathrm{KK}} \sim \text{a few hundred GeV}$), where additional KK excitations become kinematically accessible at higher energies, their individual contributions are increasingly suppressed. This results in a rapidly convergent overall contribution from the excitations in the KK tower, thus keeping the cumulative enhancement modest. 

In addition, the present analysis focuses on a low MET window, $\mathrm{MET} \in [1,10]\,\mathrm{GeV}$, which preferentially selects contributions from the lighter KK modes. In contrast, the heavier excitations tend to produce harder MET spectra and are therefore further suppressed by our event selection criteria. Consequently, no significant increase in the total signal cross-section is expected beyond the benchmark CM energy of $\sqrt{s} = 2\,\mathrm{TeV}$. Since the actual MET response at a muon-based collider will depend on the detector performance and beam-induced backgrounds, which are not included in our parton-level analysis, we additionally consider the following broader MET windows,
\begin{align}
1~\mathrm{GeV}\leq \slashed{E}_T \leq 50~\mathrm{GeV},
\qquad
1~\mathrm{GeV}\leq \slashed{E}_T \leq 100~\mathrm{GeV}.
\end{align}
The corresponding projected $2\sigma$ reaches are presented in Appendix~\ref{appendix-d}. These provide a quantitative assessment of the dependence of the projected reach on the selection of MET windows. As expected, extending the MET window increases the surviving SM background and weakens the projected reach.

\subsubsection{All-visible final state
\label{sec:Dimuon Channel}}
\begin{figure}[t]
    \centering
    \includegraphics[width=0.7\textwidth]{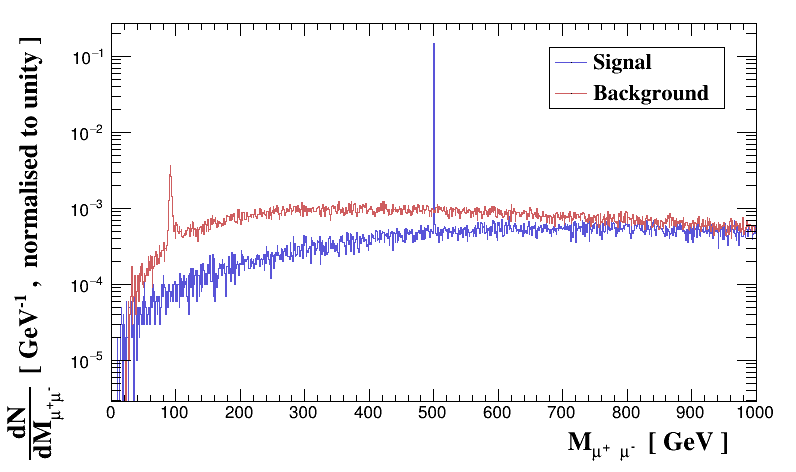}
    \caption{Invariant mass distributions of opposite-sign muon pairs for the signal with \( M_{Z'}=500\,\text{GeV}\) and \(g_D = 10^{-2}\) (in blue) and the SM background (in red), normalised to unity. In events containing three $\mu^+$'s and one $\mu^-$, the distributions shown are obtained by pairing the lone $\mu^-$ with the highest-energy $\mu^+$.}
    \label{fig:invmass500}
\end{figure}

\begin{figure}[t]
  \centering
  % --- Row 1 ---
  \begin{subfigure}{0.31\textwidth}
    \includegraphics[width=\linewidth]{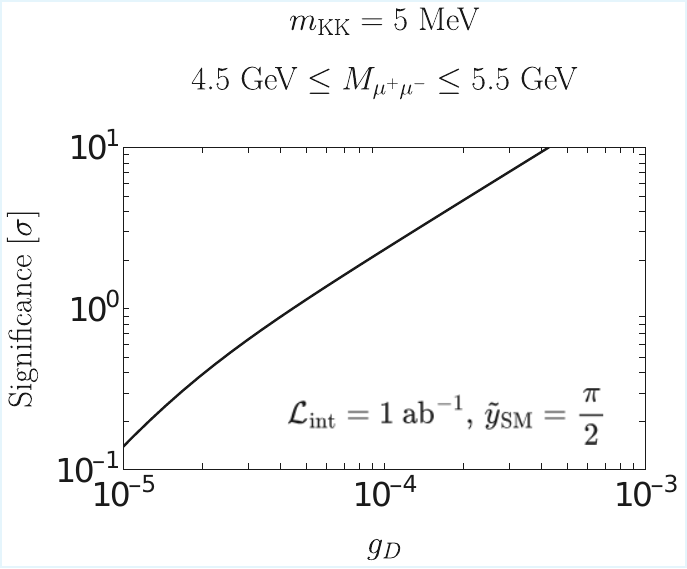}
    \caption{\hspace*{-2em}}
    \label{fig:8a}
  \end{subfigure}\hfill
  \begin{subfigure}{0.31\textwidth}
    \includegraphics[width=\linewidth]{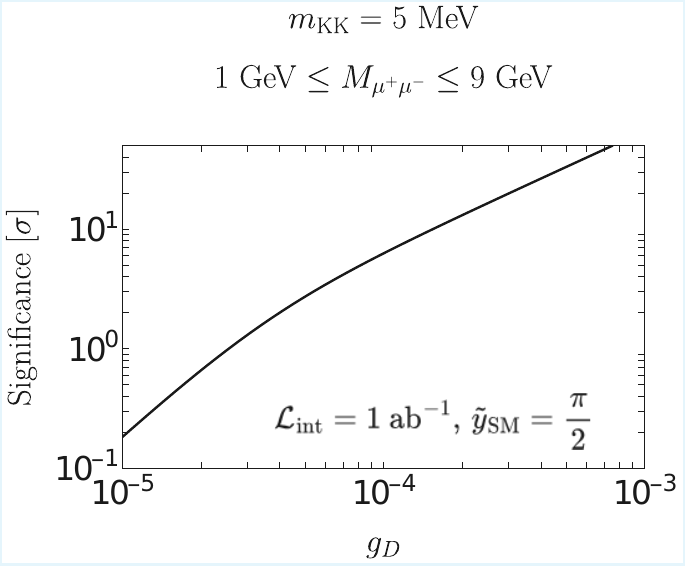}
    \caption{\hspace*{-2em}}
    \label{fig:8b}
  \end{subfigure}\hfill
  \begin{subfigure}{0.31\textwidth}
    \includegraphics[width=\linewidth]{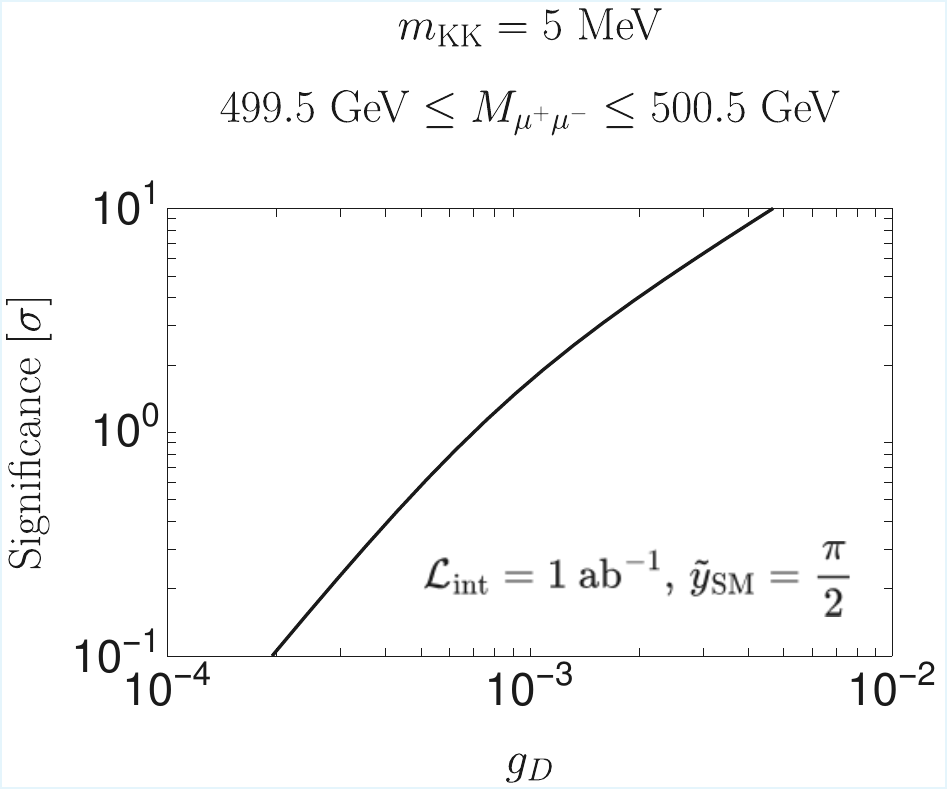}
    \caption{\hspace*{-2em}}
    \label{fig:8c}
  \end{subfigure}

  \vspace{0.4cm}
  % --- Row 2 ---
  \begin{subfigure}{0.31\textwidth}
    \includegraphics[width=\linewidth]{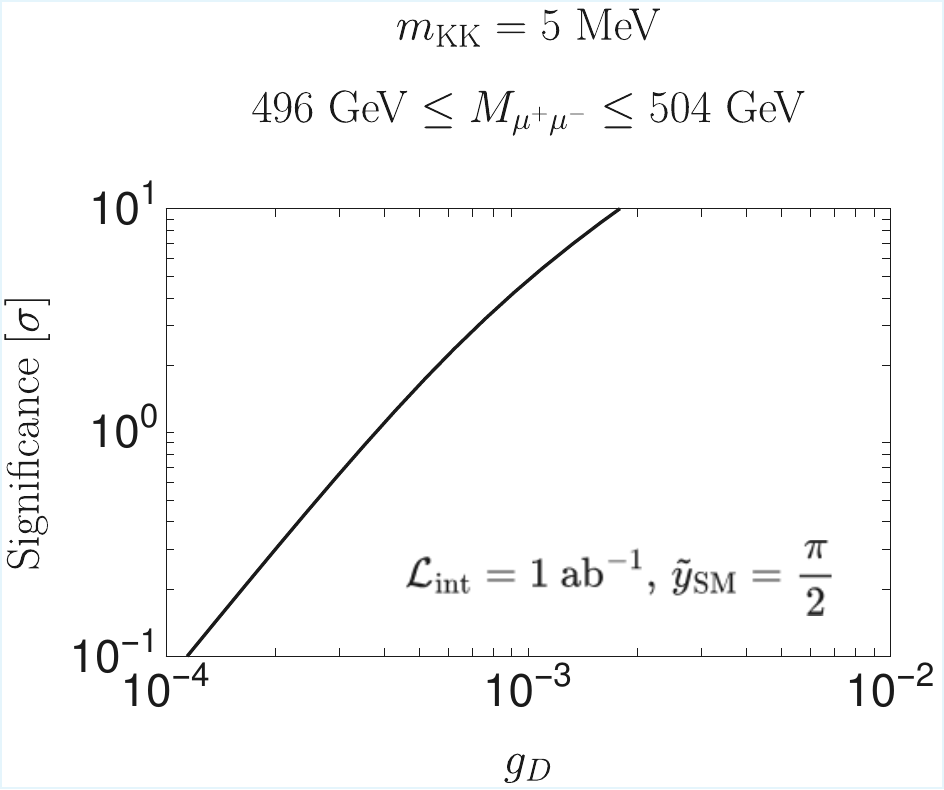}
   \caption{\hspace*{-2em}}
   \label{fig:8d}
  \end{subfigure}\hfill
  \begin{subfigure}{0.31\textwidth}
    \includegraphics[width=\linewidth]{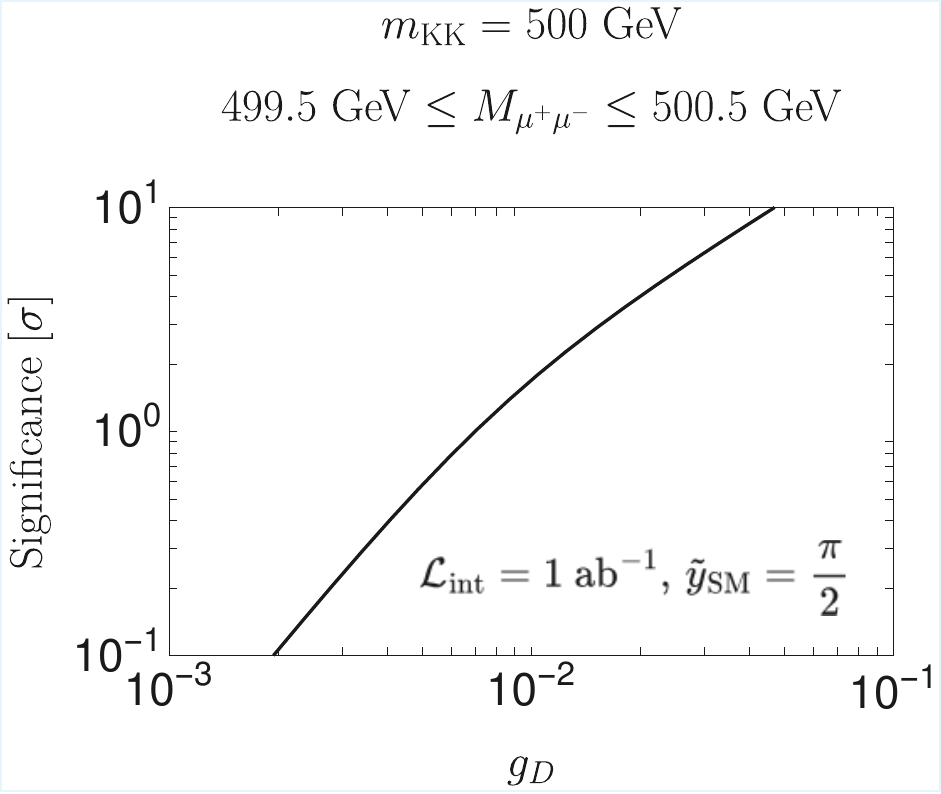}
    \caption{\hspace*{-2em}}
    \label{fig:8e}
  \end{subfigure}\hfill
  \begin{subfigure}{0.31\textwidth}
    \includegraphics[width=\linewidth]{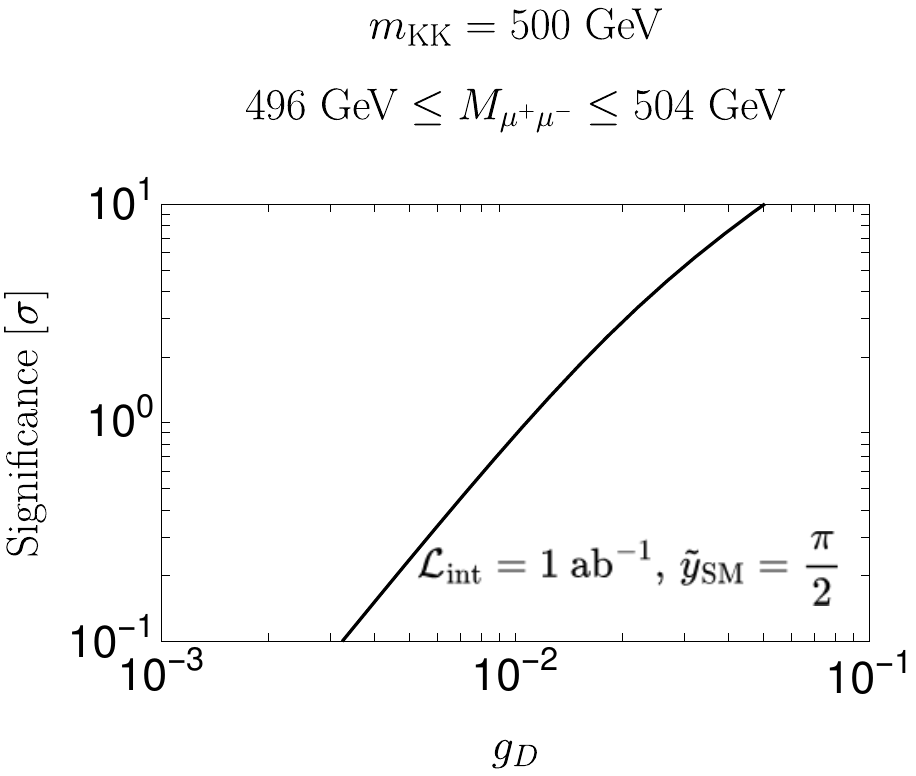}
    \caption{\hspace*{-2em}}
    \label{fig:8f}
  \end{subfigure}
  \caption{Statistical significance for a representative set of parameters, evaluated at $\mathcal{L}_{\text{int}} = 1~\text{ab}^{-1}$ with SM brane position 
  $\tilde{y}_{\text{SM}} = {\pi}/{2}$, for the all-visible four muon final state.}
  \label{fig:eightpanel}
\end{figure}

In the all-visible four-muon final state, the key observable is the invariant mass of the opposite-sign muon pair, denoted by \(M_{\mu^+\mu^-}\), originating from the decay of a resonant KK gauge boson ($Z'$) radiated from an initial or final state muon, with the maximum mass accessible for it being limited by the beam energy. Consequently, unlike in the semi-visible case, where a universal MET window can be applied across the KK spectrum in a given KK tower, in the present case, the choice of the invariant-mass window involving both its central value and its width must be adapted to the specific KK mass under consideration. In events containing three positively charged muons and one negatively charged muon, three opposite-sign muon pairings are possible: $(\mu^+_1, \mu^-)$, $(\mu^+_2, \mu^-)$, and $(\mu^+_3, \mu^-)$. 
The $M_{\mu^+\mu^-}$ distributions for the signal (blue) and background (red) events for a representative point with $M_{Z^\prime} = 500~\mathrm{GeV}$ and $g_D = 10^{-2}$ are shown in Fig.~\ref{fig:invmass500}, corresponding to the pairing of the lone $\mu^-$ with the highest-energy $\mu^+$. An event is retained if at least one opposite-sign muon pair satisfies the invariant-mass selection criterion of
\[
M_{\min} \le M_{\mu^+\mu^-} \le M_{\max} \, ,
\]
where $M_{\min}$ and $M_{\max}$ denote the lower and upper limits of the kinematic window, for the specific KK mode under consideration. 
This requirement suppresses the continuum-like combinatorial background arising from multiple wrong (non-resonant) pairings, while retaining events containing the truly resonant muon pair.

In Fig.~\ref{fig:eightpanel} we present the statistical significance of the signal, expressed in units of standard deviation, as a function of the gauge coupling $g_D$, for various representative mass values of the KK modes and choices of the invariant-mass windows. This allows us to identify the regions of the highest sensitivity for different mass regimes, and to determine the optimal choice of the mass windows for both MeV- and GeV-scale KK modes, assuming the SM brane position at $\tilde{y}_{\text{SM}} = \pi/2$, at the $\mu$TRISTAN collider operating with $\sqrt{s} = 2~\text{TeV}$, and an integrated luminosity of ${\cal L}_{\text{int}}= 1~\text{ab}^{-1}$.

From Figs.~\ref{fig:8a} and \ref{fig:8b} it is evident that the significance is enhanced when a wider mass window is adopted. A comparison between Figs.~\ref{fig:8c} and \ref{fig:8d} reveals the same trend. Furthermore, a comparison of Figs.~\ref{fig:8b} and \ref{fig:8d} shows that allowing for a smaller value of the lower cut on the invariant mass results in substantially larger significances. This is expected; for, in the case of a MeV-scale KK mass, a wider window encompasses multiple KK resonances, thereby boosting the signal rate, while, at the same time, the resulting access to a region with lower invariant mass cashes in on a relatively larger associated cross-section as compared to that for the region having higher invariant mass. We therefore adopt the configuration of Fig.~\ref{fig:8b} as the optimal mass window for MeV-scale KK modes. Accordingly, the benchmark point BP1 in Table~\ref{tab:dimuon} is analysed using that window and presented alongside the parameter values in Table~\ref{tab:dimuon}.
\begin{table}[t]
\centering
\begin{tabular}{|c|c|c|c|}
\hline
BP & $m_{\text{KK}}$ [GeV] & $g_D$ & Mass window [GeV] \\
\hline
BP1 & $5 \times 10^{-3}$ & $4.01222 \times 10^{-5}$ & $M_{\mu^+ \mu^-} \in [1, 9]$ \\
BP2 & 5 & $8.6474 \times 10^{-4}$ & $M_{\mu^+ \mu^-} \in [4.5, 5.5]$ \\
BP3 & 500 & $ 1.1306 \times 10^{-2}$ & $M_{\mu^+ \mu^-} \in [499.5, 500.5]$ \\
\hline
\end{tabular}
\caption{Benchmark points (BP) for the all-visible final state analysis. 
The couplings $g_D$ are chosen such that a $2\sigma$ statistical significance is achieved for the corresponding $m_{\text{KK}}$ values.
}
\label{tab:dimuon}
\end{table}

The situation is reversed for the case involving GeV-scale masses. As shown in Figs.~\ref{fig:8e} and \ref{fig:8f}, the significance improves when an appropriate narrower mass window is employed. Here, the spectrum is sparse, with only a single KK mode contributing within the window. Narrowing of the window suppresses the background without sacrificing the signal, thereby maximising the sensitivity. Consequently, we select the kinematic configuration of Fig.~\ref{fig:8f} as the optimal choice for GeV-scale KK modes. The benchmark points BP2 and BP3 in Table~\ref{tab:dimuon} are therefore evaluated using the mass windows indicated in Fig.~\ref{fig:8f}, and presented alongside the respective parameter entries in Table~\ref{tab:dimuon}.
The all-visible final state constitutes a particularly clean and robust probe for new physics searches at lepton colliders. At muon colliders, this channel benefits from the high reconstruction efficiency and precise momentum resolution achievable for muon tracks, enabling sensitive studies of potential deviations from the SM predictions.
%

%%%%%%%%%%%%%%%%%%%%%%%%%%%%%%%%%%%%%%%%%%%%%
\section{Resonant production at muon collider
\label{sec:MuonC}}
%%%%%%%%%%%%%%%%%%%%%%%%%%%%%%%%%%%%%%%%%%%%%
\begin{figure}[t]
    \centering
    \includegraphics[width=0.9\textwidth]{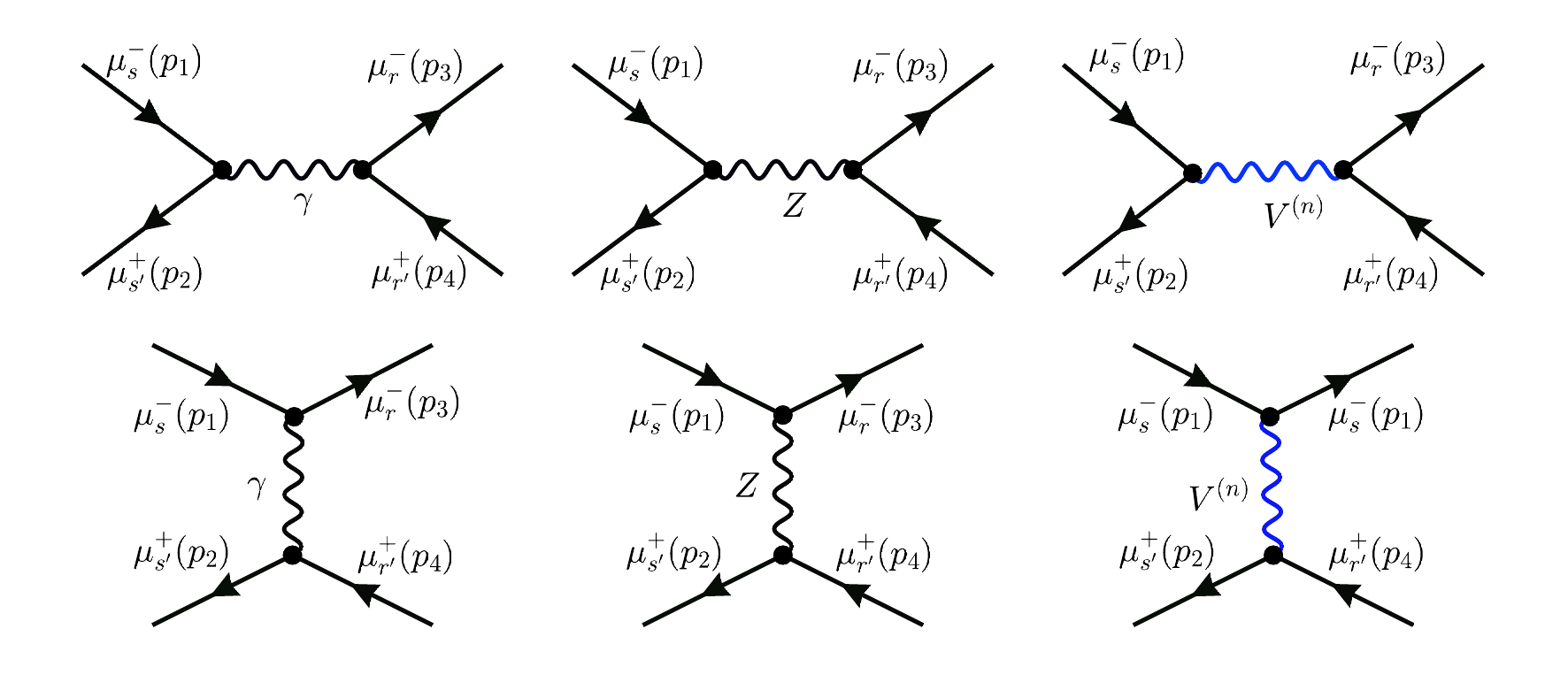}
   \caption{Feynman diagrams for the process $\mu^- \mu^+ \rightarrow \mu^- \mu^+$ in
   \(s\)-channel (upper panel) and \(t\)-channel (lower panel).}
  \label{fig:MuonC_diag}
\end{figure}
Muon colliders~\cite{Hamada:2022mua,Hamada:2022uyn,Okabe:2023esr} combine the clean experimental environment of lepton colliders with an extended energy reach into the multi-TeV regime. The ability to probe precision SM processes and new physics scenarios makes a high-energy $\mu^- \mu^+$ collider an ideal setting for studying resonant scattering phenomena.

In this section, we focus on the process \( \mu^- \mu^+ \to \mu^- \mu^+ \) involving resonant productions of KK $Z'$-bosons from a given KK tower, at a future high-energy muon collider. The scattering amplitude receives contributions from both \(s\)- and \(t\)-channel diagrams mediated by the photon, \(Z\)-boson, and the KK tower of the additional gauge bosons, as illustrated in Fig.~\ref{fig:MuonC_diag}. In what follows, we first derive the analytical expression for the matrix element by including the full interference between the SM and KK gauge boson-mediated contributions, and therefore work out the expression for the differential cross-section in the CM frame, the frame in which the muon collider is planned to operate. We then proceed to present our approach to estimating the collective contribution of the KK modes from a given KK tower to the cross-section, in the presence of finite spreads of the muon beam energies, which cannot be ignored in a precision search of a resonant state at lepton colliders.

%%%%%%%%%%%%%%%%%%%%%%%%%%%%%%%%%%%%%%%%%%%%%%%%%%%%%%%%%%%%%%%%%%%%%%%%%%
\subsection{Estimation of signal cross-section
\label{subsec:matrix-res}}
%%%%%%%%%%%%%%%%%%%%%%%%%%%%%%%%%%%%%%%%%%%%%%%%%%%%%%%%%%%%%%%%%%%%%%%%%%%%%%%%%%%%%%%
We now analyse the elastic scattering process,
$\mu^-\fn{p_1} \mu^+\fn{p_2} \to \mu^-\fn{p_3} \mu^+\fn{p_4}$, which includes resonant 
contributions from the SM $Z$-boson and the KK gauge boson(s), along with the non-resonant ones, which proceed via \(s\)- and \(t\)-channels, as shown in Fig.~\ref{fig:MuonC_diag}. The complete matrix element receives six distinct contributions given by
\begin{align}
\mathcal{M_{\text{Res}}} = -\left(\mathcal{M}_s^{\gamma} +  \mathcal{M}_s^Z + \sum_{n=1}^{\infty} \mathcal{M}_s^{(n)} \right)+ \left(\mathcal{M}_t^{\gamma} + \mathcal{M}_t^Z + \sum_{n=1}^{\infty} \mathcal{M}_t^{(n)} \right)\, ,
\label{eqn:totmat1}
\end{align}
which can be rendered in a compact form involving Dirac spinors and $\gamma$ matrices as follows:
\begin{align}
 \mathcal{M_{\text{Res}}} &= 
\sum_{j=1,3,5} 
\bar{u}^r(p_3)\, \gamma^\alpha (c_j + d_j \gamma^5)\, v^{r'}(p_4)\;
\bar{v}^{s'}(p_2)\, \gamma_\alpha (\hat{c}_j + \hat{d}_j \gamma^5)\, u^{s}(p_1) \nonumber\\
&\quad
+ \sum_{j=2,4,6} 
\bar{u}^r(p_3)\, \gamma^\alpha (c_j + d_j \gamma^5)\, u^{s}(p_1) \;
\bar{v}^{s'}(p_2)\, \gamma_\alpha (\hat{c}_j + \hat{d}_j \gamma^5)\, v^{r'}(p_4).
\end{align}
Here \(s,s',r,r'\) denote the spin indices of the external fermions associated with the momenta \(p_1,p_2,p_3,p_4\), respectively. The symbols \(u(p)\) and \(v(p)\) denote the standard Dirac spinors for muons and anti-muons, respectively. Further, \(c_j, d_j, \hat{c}_j, \hat{d}_j\) represent the coefficients for the vector and axial-vector couplings, where the coefficients $c_j$ and $d_j$ have already appeared in section~\ref{sec:Mutristan}, and we will soon define the coefficients \(\hat{c}_j\) and 
\(\hat{d}_j\).
With the above convention, indices $j=1,3,5$ denote the $s$-channel exchanges of photon, $Z$-boson, and the KK modes, respectively, while indices $j=2,4,6$ denote the corresponding $t$-channel contributions. The squared matrix element, averaged over the spins of the initial-state fermions and summed over the spins of the final-state fermions, in the high-energy limit, can be expressed as
\begin{align}
\overline{|\mathcal{M}_{\text{Res}}|^2}
= P\,s^2 + Q\,t^2 + R\,u^2 \, ,
\end{align}
where the coefficient functions \( P \), \( Q\), and \( R \) include the contributions from all interfering diagrams and are functions of \( c_j \) and \( d_j \). Explicit expressions for \( P \), \( Q \), and \( R \) are collected in Appendix~\ref{appendix-e}. In the CM frame, the differential cross-section can be written down as
\begin{align}
\left. \frac{d\sigma}{d\Omega} \right|_{\mathrm{CM}} = \frac{1}{64\pi^2 s} \left( \overline{|\mathcal{M_{\text{Res}}}|^2} \right).
\end{align}
where $\Omega$ is the solid angle and  \(s\), \(t\), and \(u\) are the usual Mandelstam variables defined with the same momentum convention as in section~\ref{sec:sig-mutristan}.

For completeness, we list below the explicit expressions for the vertex coefficients \(c_j\), \(d_j\), \(\hat{c}_j\), and \(\hat{d}_j\) associated with each channel:
\begin{itemize}
    \item {Photon exchange (\(s\)- and \(t\)-channels):}
    \begin{align}
        c_1 &= e \, , & d_1 &= 0 \, , & \hat{c}_1 &= -\frac{e}{s} \, , & \hat{d}_1 &= 0 \, , \nonumber \\
        c_2 &= e \, , & d_2 &= 0 \, , & \hat{c}_2 &= \frac{e}{t} \, , & \hat{d}_2 &= 0 \, .
    \end{align}

    \item {\(Z\)-boson exchange (\(s\)- and \(t\)-channels):}
    
\begin{align}
    c_3 &=  \frac{3}{4} g_1 s_W - \frac{1}{4} g_2 c_W \, , & 
    d_3 &= \frac{1}{4} g_1 s_W + \frac{1}{4} g_2 c_W \, , \nonumber \\
    \hat{c}_3 &= - \frac{c_3}{s - m_Z^2 + i \Gamma_Z m_Z} \, , & 
    \hat{d}_3 &= - \frac{d_3}{s - m_Z^2 + i \Gamma_Z m_Z} \, , \nonumber\\
    c_4 &= c_3 \, , & 
    d_4 &= d_3 \, , \nonumber \\
    \hat{c}_4 &= \frac{c_3}{t - m_Z^2} \, , & 
    \hat{d}_4 &= \frac{d_3}{t - m_Z^2} \, .
\end{align}
   \item {KK tower vector boson exchange (\(s\)- and \(t\)-channels):}
\begin{align}
    c_5 &= g_D \, , & d_5 &= 0 \, , & 
    \hat{c}_5 &= - \sum_{n=1}^{\infty} \frac{g_D [f_V^{(n)}(\tilde{y}_{\text{SM}})]^2}{s - M_n^2 + i \Gamma_n M_n} \, , & \hat{d}_5 &= 0 \, , \nonumber\\
    c_6 &= g_D \, , & d_6 &= 0 \, , & 
    \hat{c}_6 &= \sum_{n=1}^{\infty} \frac{g_D [f_V^{(n)}(\tilde{y}_{\text{SM}})]^2}{t - M_n^2} \, , & \hat{d}_6 &= 0 \,. 
\end{align}
\end{itemize}

\vspace{0.5em}
\noindent
Contributions from the gauge bosons of the KK towers are captured in the coefficients \(\hat{c}_5\) and \(\hat{c}_6\), which encode the full KK summation via propagators and wavefunctions evaluated at the SM brane position \( \tilde{y}_{\text{SM}}\). Here, \(\Gamma_n\) denotes the total decay width of the \(n\)-th KK mode. Additional analytical details of these two terms are provided in subsection~\ref{sec:mupmtomupm-technical}.
%%%%%%%%%%%%%%%%%%%%%%%%%%%%%%

In collider experiments, the energies of the incoming particles are not exactly fixed.
The analytical formulation of summing over the KK states in a given tower is intricately connected to experimental issues such as the finite mass resolution (a detector issue) and the beam energy spread (BES; an accelerator issue). In particular, these issues are crucial for the low-lying KK states with sub-GeV masses, which are comparable to the expected energy (mass) resolutions of such machines. That the inherent spreads in the beam energies also fall in the same energy window makes the interplay all the more significant.
%%%%%
For our current purposes, at the systematic level, we assume the CM energy to possess a Gaussian profile~\cite{Cerri:2016bew} around its central  (nominal, or operational) value,
$\sqrt{s}_c$, ensuring the proper normalisation over the physical domain \(\sqrt{s}\ge 0\),\footnote{
The denominator in
Eq.~\eqref{eq:BESnorm} can be evaluated analytically as
\(
\sqrt{\frac{\pi}{2}}\,
\sigma_{\sqrt{s}}
\left[
1+\operatorname{erf}
\!\left(
\frac{\sqrt{s}_c}
{\sqrt{2}\sigma_{\sqrt{s}}}
\right)
\right]
\) in terms of the Gauss error function.
For the BES considered in this work,
\(
\sqrt{s}_c/\sigma_{\sqrt{s}}\gg1
\),
so that
\(
\operatorname{erf}
\!\left(
\frac{\sqrt{s}_c}
{\sqrt{2}\sigma_{\sqrt{s}}}
\right)\simeq1
\).
Consequently, the normalised distribution is numerically indistinguishable from a standard Gaussian defined on the full real line.} and is given by
%%%%%%%%%%
\begin{equation}
f(\sqrt{s})=
\frac{
\exp\!\left[
-\dfrac{(\sqrt{s}-\sqrt{s}_c)^2}{2\sigma_{\sqrt{s}}^2}
\right]
}
{
\displaystyle
\int_{0}^{\infty}
\exp\!\left[
-\dfrac{(\sqrt{s'}-\sqrt{s}_c)^2}{2\sigma_{\sqrt{s}}^2}
\right] d\sqrt{s'}
} \quad,
\label{eq:BESnorm}
\end{equation}
where \(\sigma_{\sqrt{s}}\) is the standard deviation (width) of the distribution. By construction,

\begin{equation}
\int_0^\infty f(\sqrt{s})\,d\sqrt{s} = 1.
\end{equation}
%%%%%%%%%%%%%%%%%%%%%%%%%%%
%
The effective cross-section in the presence of BES is then obtained by convoluting the partonic cross-section with such distributions,
\begin{equation}
\sigma_{\rm eff}
:=
\int_{0}^{\infty}
f(\sqrt{s})\,
\sigma(s)\,
d\sqrt{s}.
\label{eq:sigmaeff}
\end{equation}

For our purpose, we define a peak energy window of width \( \Delta E = 5 \sigma_{\sqrt{s}} \) about the central value of the CM energy, \( \sqrt{s}_c \), which ensures a \( > 99\% \) containment of the CM energy under a Gaussian distribution.\footnote{
This captures all (resonant) contributions under the Gaussian envelope effectively. We, however, find that a narrower window with \( \Delta E = 3 \, \sigma_{\sqrt{s}} \) is sufficient to capture the dominant contribution to the signal, as demonstrated and discussed in detail in section~\ref{sec:resonant-exclusion}.}
In what follows, BES is parametrised as \( \sigma_{\sqrt{s}} / \sqrt{s}_c \), and is taken to be 0.1\%, which is a benchmark value commonly adopted in studies at the muon collider~\cite{Skoufaris:2023jnu}. For numerical purposes, the integration is thus restricted to the finite window of \( \Delta E = \mp 5 \sigma_{\sqrt{s}} \) about the central energy,

\begin{equation}
\sigma_{\rm eff}
\simeq
\int_{\sqrt{s}_c-5 \sigma_{\sqrt{s}}}^{\sqrt{s}_c+5 \sigma_{\sqrt{s}}}
f(\sqrt{s})\,
\sigma(s)\,
d\sqrt{s}.
\end{equation}
Such a convolution of the parton-level cross-section with the beam-energy profiles effectively smooths out narrow resonant peaks or threshold effects in \( \sigma(s) \). This helps ensure that theoretical predictions appropriately account for the realistic energy resolution of the collider. This treatment is thus especially important in the immediate vicinity of sharp peaks in invariant mass distributions characterising narrow resonances, where the interplay of finite mass resolution and finite beam spreads in the experiment could significantly affect observable event rates.
%
%%%%%%%
\subsubsection{Simulation details
\label{sec:mupmtomupm-technical}}
We retain KK modes up to the $n$-th level in a given tower with their masses \( M_n = (2n - 1) m_{\text{KK}} \) satisfying \( |M_n - \sqrt{s}| \lesssim \Delta E \).
This translates to the discrete mode range with
\begin{align}
n_{\min} = \left\lceil \frac{\sqrt{s} - \Delta E}{2 m_{\text{KK}}} + \frac{1}{2} \right\rceil, \quad
n_{\max} = \left\lfloor \frac{\sqrt{s} + \Delta E}{2 m_{\text{KK}}} + \frac{1}{2} \right\rfloor,
\end{align}
where \( \lceil x \rceil \) (the ceiling function) denotes the smallest integer greater than or equal to \( x \), and \( \lfloor x \rfloor \) (the floor function) denotes the largest integer less than or equal to \( x \).
\noindent
In this \textit{resonant regime}, we explicitly sum over the contributing modes using the full Breit-Wigner form for the propagators, i.e.,
\begin{align}
\sum_{n=1}^{\infty}\frac{f^{(n)}_{V}{(\tilde{y}_{\text{SM}})}^2}{s - M_n^2 + i M_n \Gamma_{n}} \simeq 
\sum_{n=n_{\min}}^{n_{\max}} \frac{f_{V}^{(n)}(\tilde{y}_{\text{SM}})^2}{s - M_n^2 + i M_n \Gamma_n} \, ,
\label{eqn:bwprop}
\end{align}
where \( f_{V}^{(n)}(\tilde{y}_{\text{SM}}) \) is the wavefunction profile of the KK modes evaluated at the SM brane location \( \tilde{y}_{\text{SM}} = \pi/2 \), yielding \( f_{V}^{(n)}(\pi/2) = \sqrt{2} \cos\left[(n - \tfrac{1}{2}) {\pi}/{2} \right] \).
Equation~\eqref{eqn:bwprop} incorporates the finite-width effects essential for capturing resonant enhancements near \( s \simeq M_n^2 \).

In contrast, for parameter regions where no KK mode lies within the peak energy window, i.e., \( |M_n - \sqrt{s}| > \Delta E \), for all \( n \), states from the tower contribute in the \textit{off-shell regime}. In this case, we approximate the sum by neglecting the small imaginary parts \( i M_n \Gamma_n \), which is valid when
\begin{align}
|s - M_n^2| \gg M_n \Gamma_n \quad \forall n \, ,
\end{align}
thus rendering the Breit-Wigner form of the propagator to a simple expression, viz.,
\begin{align*}
\sum_{n=1}^{\infty} \frac{f_{V}^n(\tilde{y}_{\text{SM}})^2}{s - M_n^2 + i M_n \Gamma_n}
 \, \leadsto \sum_{n=1}^{\infty} \frac{f_{V}^n(\tilde{y}_{\text{SM}})^2}{s - M_n^2} \, .
\end{align*}
\noindent
For the chosen brane location, this infinite sum admits an analytical continuation:
\begin{align}
\sum_{n=1}^{\infty} \frac{f_{V}^n(\pi/2)^2}{s - M_n^2} 
= -\frac{\pi}{4 \sqrt{s} m_{\text{KK}}} \tan\left( \frac{\pi \sqrt{s}}{2 m_{\text{KK}}} \right),\footnotemark
\end{align}
\footnotetext{To regulate the divergences in the KK-mode summation for \(s\)-channel exchange, where \(\tan x\) becomes singular near KK poles (\(x = {\pi}/{2} + n\pi\)), we replace \(\tan x\) with a deformed function \(\mathcal{T}(x, \Delta)\). This function flattens \(\tan x\) in narrow regions around each pole, with widths growing as \(n^2 \Delta\), where \(\Delta = m_{\text{KK}} \Gamma_{\text{KK}}\) with \( \Gamma_{\text{KK}} \) being the decay width of the lowest-lying KK mode. This stabilises the numerical evaluation while approximating the resonance widths.}
\noindent
which effectively encapsulates the off-resonance tail of the full KK tower contribution. In summary, our treatment distinguishes between \textit{on-shell} and \textit{off-shell} KK contributions, by incorporating the full Breit-Wigner form of the propagators for the resonant modes and an analytic approximation of the same otherwise. This hybrid prescription ensures both computational efficiency and physical accuracy of the results across the parameter space.

We now define the new effective couplings that enter the scattering amplitudes by following the conventions adopted in section~\ref{subsec:matrix-res}. For the benchmark brane location \( \tilde{y}_{\mathrm{SM}} = \pi/2 \), the \(s\)- and \(t\)-channel contributions are given by
\begin{equation}
\hat{c}_5\big|_{\tilde{y}_{\text{SM}} = \pi/2} \simeq 
\begin{cases}
\displaystyle -\sum_{n=n_{\text{min}}}^{n_{\text{max}}} \frac{g_D\, \big(f_{V}^{(n)}(\pi/2)\big)^2}{s - M_n^2 + i M_n \Gamma_n} \, , & \text{on-shell}, \\[6pt]
\displaystyle \frac{\pi g_D}{4 \sqrt{s} m_{\text{KK}}} \tan\left( \frac{\pi \sqrt{s}}{2 m_{\text{KK}}} \right), & \text{off-shell},
\end{cases}
\qquad
\hat{c}_6\big|_{\tilde{y}_{\text{SM}} = \pi/2} = -\frac{\pi g_D}{4 \sqrt{t} m_{\text{KK}}} \tan\left( \frac{\pi \sqrt{t}}{2 m_{\text{KK}}} \right).
\label{eq:coeff-yt=pi/2}
\end{equation}

For the alternative brane location \( \tilde{y}_{\text{SM}} = 0 \), the effective couplings are obtained explicitly as
\begin{equation}
\hat{c}_5\big|_{\tilde{y}_{\text{SM}} = 0} \simeq 
\begin{cases}
\displaystyle - \sum_{n=n_{\text{min}}}^{n_{\text{max}}} \frac{g_D\, \big(f_{V}^{(n)}(0)\big)^2}{s - M_n^2 + i M_n \Gamma_n} \, , & \text{on-shell}, \\[3pt]
\displaystyle \frac{\pi g_D}{2 \sqrt{s} m_{\text{KK}}} \tan\left( \frac{\pi \sqrt{s}}{2 m_{\text{KK}}} \right), & \text{off-shell},
\end{cases}
\qquad
\hat{c}_6\big|_{\tilde{y}_{\text{SM}} = 0} = -\frac{\pi g_D}{2 \sqrt{t} m_{\text{KK}}} \tan\left( \frac{\pi \sqrt{t}}{2 m_{\text{KK}}} \right).
\label{eq:coeff-yt=0}
\end{equation}
Note that the on-shell sum for \(\hat{c}_5\) must be computed numerically, as no expression in a closed form exists for the same. In the off-shell regime (for both $s$- and $t$-channels), as can be gleaned from equations~\eqref{eq:coeff-yt=pi/2} and \eqref{eq:coeff-yt=0}, the couplings for \(\tilde{y}_{\mathrm{SM}} = 0\) configuration are exactly twice the corresponding ones for the \(\tilde{y}_{\mathrm{SM}} = \pi/2\) case. Together, these expressions provide a unified description of the contributions from a given KK tower across both resonant and non-resonant regimes, and are to be used consistently in all numerical evaluations made in the present work (see Appendix~\ref{appendix-b} for details).

For small values of \( m_{\text{KK}} \sim \mathcal{O}(\text{MeV}) \), the narrow spacing between KK masses leads to a large number of resonant modes (\( \mathcal{O}(10^3) \)) falling within the collider’s energy spread. This results in a densely populated resonance structure, thus requiring a summation over a broad range of KK modes.  In contrast, for larger values of \( m_{\text{KK}} \), the mass splitting between adjacent KK modes becomes comparable to or larger than the width of the kinematically accessible peak energy window, so that only a few modes, often just one, lie within the peak energy window. For instance, at \( \sqrt{s} = 3~\text{TeV} \), a typical case with \( m_{\text{KK}} \sim \mathcal{O}(\text{TeV}) \) yields only a single on-shell KK mode contributing to the sum.

%%%%%%%%%%%%%%%%%%%%%%%%%%%%%%%%%%%%%%%%%%%%%%%%%%%%%%%%%%%%%%%%%%%%%%%%%%%%%%%%%%%%%%%
\subsection{Projected \texorpdfstring{$2\sigma$}{2 sigma} reach from resonant production
\label{sec:resonant-exclusion}}
\begin{figure}[!t]
    \centering
    % Row 1
    \begin{subfigure}[t]{0.48\linewidth}  
        \centering
        \includegraphics[width=\linewidth,height=6cm,keepaspectratio]{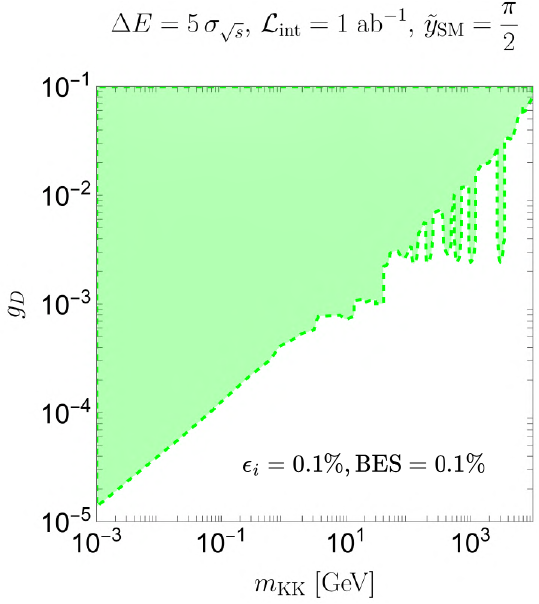}
        \caption{\hspace*{-4em}}
        \label{fig:MuonC_plot1a}
    \end{subfigure}
    \hspace{0.02\linewidth}
    \begin{subfigure}[t]{0.48\linewidth}
        \centering
        \includegraphics[width=\linewidth,height=6cm,keepaspectratio]{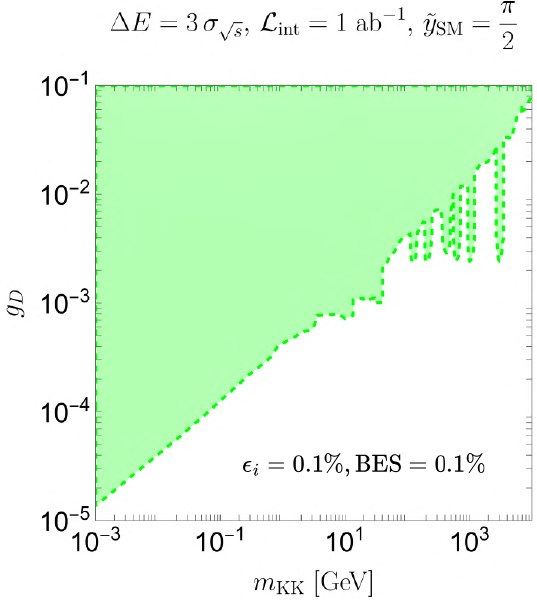}
        \caption{\hspace*{-4em}}
        \label{fig:MuonC_plot1b}
    \end{subfigure}

    \vspace{0.3cm} % Space between rows

    % Row 2
    \begin{subfigure}[t]{0.48\linewidth}
        \centering
        \includegraphics[width=\linewidth,height=6cm,keepaspectratio]{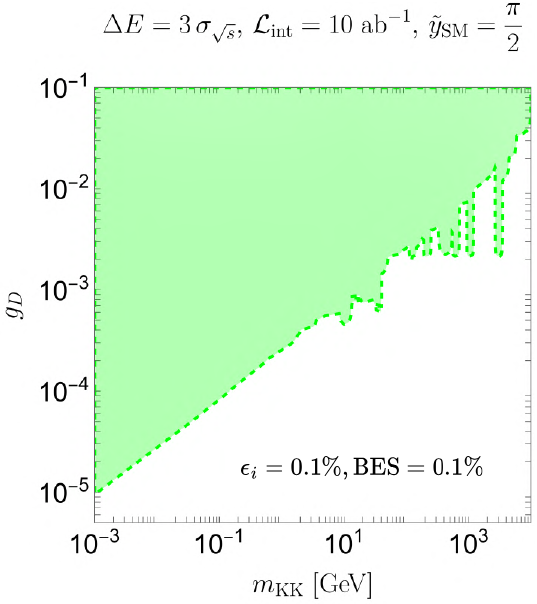}
        \caption{\hspace*{-4em}}
        \label{fig:MuonC_plot1c}
    \end{subfigure}
    \hspace{0.02\linewidth}
    \begin{subfigure}[t]{0.48\linewidth}
        \centering
        \includegraphics[width=\linewidth,height=6cm,keepaspectratio]{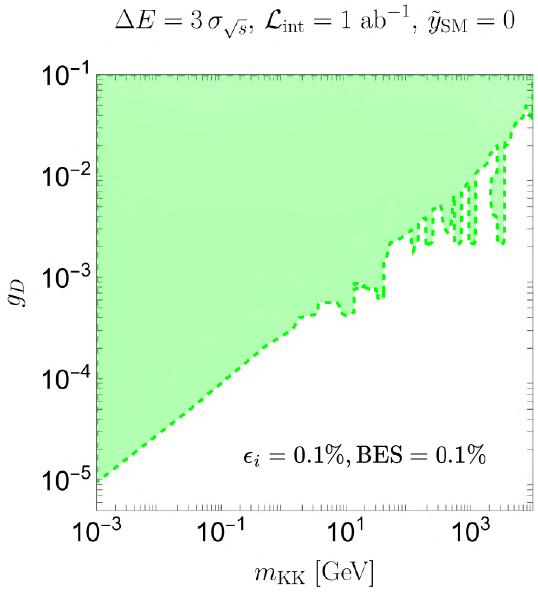}
        \caption{\hspace*{-4em}}
        \label{fig:MuonC_plot1d}
    \end{subfigure}
\caption{Projected \(2\sigma\) reach in the \(m_{\text{KK}}\)-\(g_D\) plane for a 3~TeV Muon Collider. Green regions yield a statistical significance exceeding \(2\sigma\).
%Regions above the curves yield a statistical significance exceeding \(2\sigma\).
All plots assume a BES value of \(0.1\%\) and a systematic uncertainty \(\epsilon_i = 0.1\%\). Each  plot corresponds to the indicated peak energy window \(\Delta E\), integrated luminosity \(\mathcal{L}_{\text{int}}\), and the value of $\tilde{y}_\text{SM}$.} 
\label{fig:MuonC_plot1}
\end{figure}

%%%%%%%%%%%%%%%%%%%%%%%%%%%%%%%%%%%%%%%%%%%%%%%%%%%%%%%%%%%%%%%%%%%%%%%%%%%%%%%%%%%%%%%%%%%
We investigate the sensitivity to BSM dynamics in the $\mu^- \mu^+ \rightarrow \mu^- \mu^+$ process employing a binned $\chi^2$ framework constructed from the effective differential cross-section. This approach incorporates a uniform binning of the angular distribution and accounts for both statistical and systematic uncertainties. Such a framework is particularly effective in preserving angular information, including forward-backward asymmetries, thereby enhancing the sensitivity of our analysis to BSM signatures.

We adopt a binned $\chi^2$ statistical analysis of the effective differential cross-section. While the statistical framework is identical to that described in section~\ref{sec:elastic-exclusion},  we briefly restate it here for completeness and clarity, given the modified treatment of the effective differential cross-section adopted in the present analysis.
%%
%%%%
%%
Here, we adopt the same framework that we have earlier used for the analysis of the elastic $\mu^+\mu^+$ process at the  $\mu$TRISTAN (discussed in section~\ref{sec:Mutristan}).
In our main results, we present the case with \( \epsilon_i = 0.1\% \), which reflects a realistic scenario for future high-precision $\mu^-\mu^+$ colliders.
%%%%%%%%%%%%%%%%%%%%%%%%%%%%%%%%%%%%%%%%%%%%%%%%%%%%%%%%%%%%%%%%%%%%%%%%%%%%%%%%%%%%%%%%%%%%%%%%

We discuss the projected $2\sigma$ reach arising from our analysis of resonant production of KK states at a future high-energy muon collider, for a few representative choices of theoretical and experimental parameters. These cases are illustrated in Fig.~\ref{fig:MuonC_plot1} where, for all the plots, we choose the reference plane to be that of \(m_{\text{KK}}\)--\(g_D\) and fix the CM energy to $\sqrt{s}=3$ TeV, with a flat BES and a universal systematic uncertainty ($\epsilon_i$), both set to 0.1\%. Furthermore, on the theoretical side, two locations of the SM brane, i.e., $\tilde{y}_\mathrm{SM}= \pi/2, 0$, are considered to contrast their effects. On the experimental side, two values of the peak energy window, viz., $\Delta E=5\sigma_{\sqrt{s}},3\sigma_{\sqrt{s}}$, and two values of integrated luminosity, viz., $\mathcal{L}_{\text{int}} = 1, 10$ ab$^{-1}$, are adopted to illustrate the roles of these parameters in determining the reach of the experiment.

Figures~\ref{fig:MuonC_plot1a} and \ref{fig:MuonC_plot1b} are both obtained for $\tilde{y}_\mathrm{SM}=\pi/2$ and $\mathcal{L}_{\text{int}}= 1$ ab$^{-1}$ but with different values of $\Delta E$: $5 \sigma_{\sqrt{s}}$ for the former, and $3 \sigma_{\sqrt{s}}$ for the latter. Note that although a much narrower peak-energy window is employed in Fig.~\ref{fig:MuonC_plot1b}, the overall $2\sigma$ reach remains nearly identical to that in \ref{fig:MuonC_plot1a}. This highlights that the dominant contribution comes from KK modes with masses near the nominal CM energy. Further, for KK modes with their masses in the MeV scale, the projected $2\sigma$ reach extends down to couplings as small as \( g_D \sim 10^{-5} \), which is close to the sensitivity obtained from elastic scattering processes at \(\mu\)TRISTAN, as discussed in section~\ref{sec:Mutristan}. 

Figs.~\ref{fig:MuonC_plot1c} and \ref{fig:MuonC_plot1d} present complementary scenarios that extend the primary analysis. Guided by our observations from Figs.~\ref{fig:MuonC_plot1a} and \ref{fig:MuonC_plot1b}, we now set $\Delta E=3 \sigma_{\sqrt{s}}$. Fig.~\ref{fig:MuonC_plot1c} is now obtained for $\mathcal{L}_{\text{int}} = 10$ ab$^{-1}$ and to be directly contrasted to Fig.~\ref{fig:MuonC_plot1b}, which was obtained for $\mathcal{L}_{\text{int}} = 1$ ab$^{-1}$.
As expected, the increased luminosity is found to improve the sensitivity and leads to tighter bounds in the  \(m_{\text{KK}} \)-\( g_D \) parameter space. Fig.~\ref{fig:MuonC_plot1d} can again be contrasted to Fig.~\ref{fig:MuonC_plot1b}, with a different choice for the brane-localisation of the SM brane which is now set to \( \tilde{y}_{\text{SM}} = 0 \), instead of \( \pi/2 \), as is the case with Fig.~\ref{fig:MuonC_plot1b}, other parameters remaining the same as in Fig.~\ref{fig:MuonC_plot1b}. The configuration with $\tilde{y}_\mathrm{SM}=0$ in Fig.~\ref{fig:MuonC_plot1d} presents us with marginally stronger $2\sigma$ reach thanks to the peaking nature of the KK wavefunctions in the vicinity of this location for the SM brane, thus leading to an enhanced overlap with their SM counterparts and hence a stronger mutual coupling. Some related discussions appear in sections~\ref{sec:elastic-exclusion} and~\ref{sec:MET channel} in the context of elastic scattering and semi-visible final state signals at the proposed \(\mu\)TRISTAN collider. These complementary results reinforce the robustness of our main conclusions and provide additional insight into how collider sensitivity depends on experimental and model-specific parameters.

The observations from the studies made in the current subsection can be summarised as follows.  For smaller values of \( m_{\text{KK}} \sim \mathcal{O}(\text{MeV}) \),  KK modes become densely packed so as to fall collectively within the Gaussian beam energy profile. This amounts to a large number of resonant modes contributing simultaneously, thus effectively enhancing the signal and hence leading to stronger bounds. While these light modes do not exactly coincide with the central or peak value of the CM energy distribution, owing to the finite precision and energy resolution of TeV-scale colliders, a significant fraction still lies in the vicinity of the central CM energy. Consequently, they are efficiently captured by the Gaussian tails, thus still contributing to the cross-section and retaining sensitivity in this regime. For intermediate values of \( m_{\text{KK}} \) (e.g., 200, 600, 1000, 3000 GeV), individual KK modes can closely match the CM energy, producing sharp peaks in the projected $2\sigma$ reach curves due to resonant amplification. However, in the low-\( m_{\text{KK}} \) regime, such peaks are not visually prominent due to the linear interpolation over a region densely populated by thousands of overlapping resonances, which collectively yield a smoother profile.
Furthermore, employing a narrower beam energy profile (i.e., smaller \( \Delta E \)) improves alignment with the resonant modes, thereby strengthening the $2\sigma$ reach. This highlights the significant potential of future muon colliders to probe such a 5D \( U(1)_{L_\mu - L_\tau} \) models not only for the weakly coupled KK states with smaller masses but also for their heavier cousins, when the coupling is strong enough. The resonant enhancement accessible through \(s\)-channel exchanges of such KK states at high energies offers a promising avenue for exploring such extra-dimensional gauge structures.

%%%%%%%%%%%%%%%%%%%%%%%%%%%%%%%%%%%%%%%%%%%%%%%%%%%%%%%%%%%%%%%%%%%%%%%%%%%%%%%%%%%%%%%%%%%%%%%%%%%%%%%%%%%%%%%%%%%%
\section{Complementarity of the processes studied
\label{sec:Summary}}
%%%%%%%%%%%%%%%%%%%%%%%%%%%%%%%%%%%%%%%%%%%%%%%%%%%%%%%%%%
%%%%%%%%%%%%%%%%%%%%%%%%%%%%%%%%%%%%%%%%%%%%%%%%%%%%%%%%%%

\begin{figure}[t]
    \centering
    \includegraphics[width=0.35\textwidth]{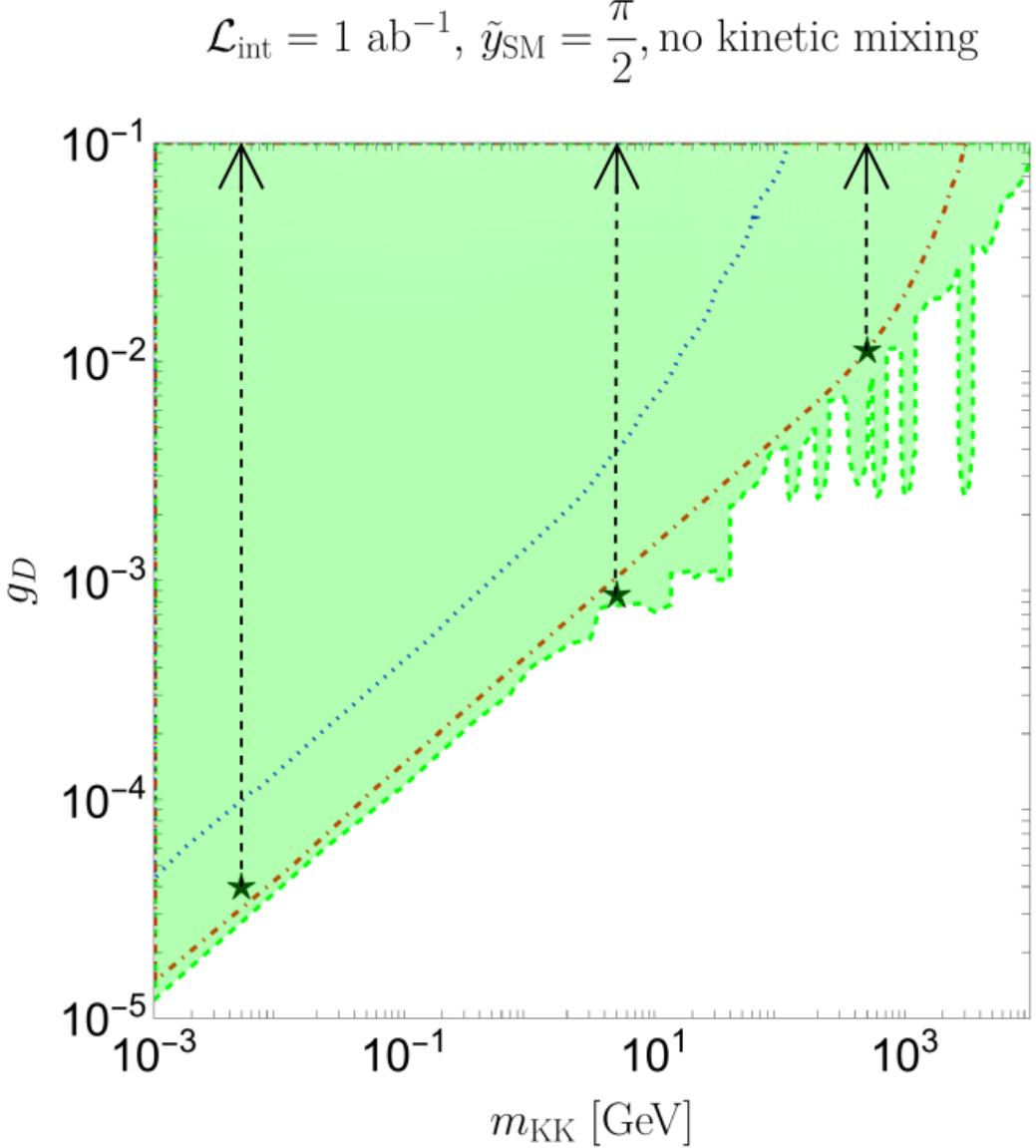}
    \hspace{1 em}
    \raisebox{5 em}{\includegraphics[width=0.50\textwidth]{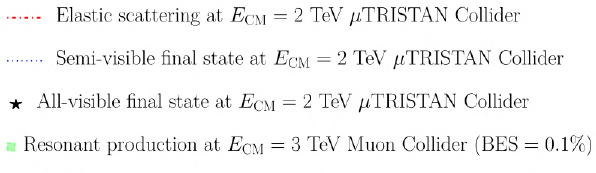}}
   \caption{Projected \(2 \sigma\) reach in the $m_\text{KK}$--$g_D$ plane via (i) elastic scattering at the future $\mu$TRISTAN collider with a vanishing systematic uncertainty ($\epsilon_i = 0\%$), (ii) semi-visible (SSDM + $\slashed{E}_T$) final state at the future $\mu$TRISTAN collider (see Eq.~\eqref{eq:toneutrino_METcut} for the kinematical cut), (iii) all-visible (four muon) final state at the future $\mu$TRISTAN collider (see table~\ref{tab:dimuon} for the kinematical cut), and (iv) resonant production of KK gauge boson(s) at a future $\mu^-\mu^+$ collider for a vanishing systematic uncertainty scenario ($\epsilon_i = 0\%$).
The parameter regions above each (boundary) curve or the dashed arrow lines above the filled asterisks indicate the projected significance beyond $2\sigma$ for representative benchmark parameters.} 
    \label{fig:combined_summary}
\end{figure}

\begin{figure}[t]
    \centering
    \includegraphics[width=0.35\textwidth]{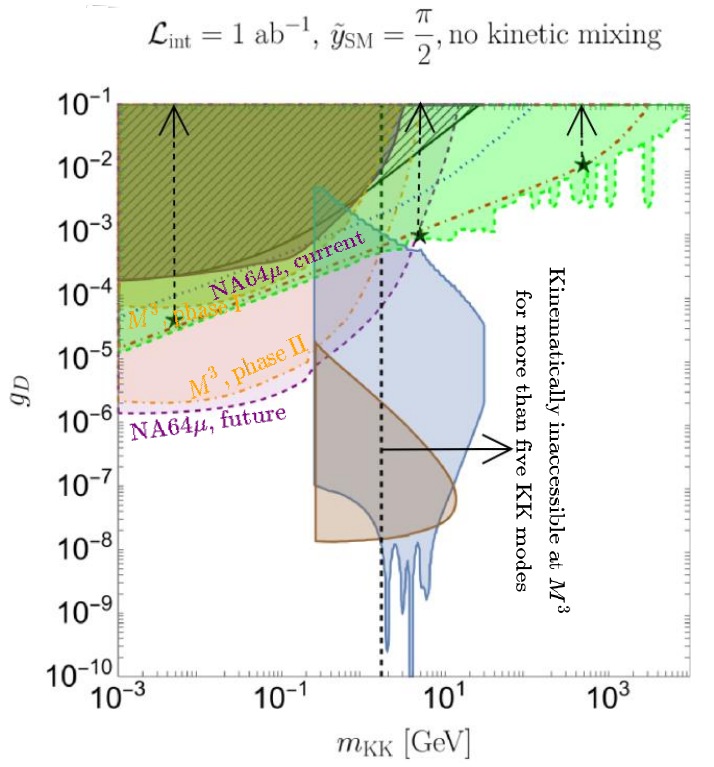}
    \hspace{1em}
    \begin{minipage}[c]{0.50\textwidth}
        \vspace{-5cm}
        \centering
        {\scriptsize {\scriptsize Future muon-based collider reach (this work)}}
        \vspace{0.01cm}

        \includegraphics[width=\textwidth]{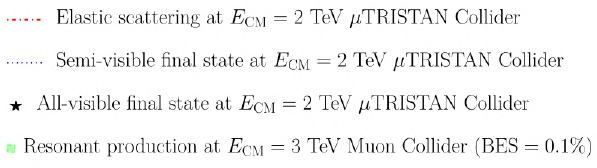}
        
        {\scriptsize Future muon beam-dump reach (Ref.~\cite{Chakraborty:2025jbd})}
        \vspace{0.01cm}
        
        \includegraphics[width=\textwidth]{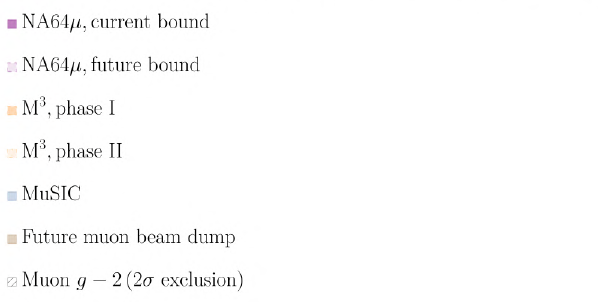}
    \end{minipage}
    \caption{Comparison of the projected muon-based collider reach and the reach obtained in our previous 5D muon beam-dump analysis, Ref.~\cite{Chakraborty:2025jbd}, in the $(m_{\rm KK}, g_D)$ plane. The beam-dump constraints include the current and projected NA64$\mu$ sensitivities (90\% CL), the projected M$^3$ sensitivity (90\% CL), the projected MuSIC sensitivity requiring $N_{\rm signal}>5$, and the projected future muon beam-dump sensitivity requiring $N_{\rm signal}>5$. The black hatched region is excluded by the measurement of the muon $(g-2)$, at the $2\sigma$ level. The muon-based collider extends the reach to larger KK masses and probes parameter space beyond the beam-dump coverage.}
    \label{fig:combined_all}
\end{figure}

In this section, we discuss the complementarity of the processes studied thus far. We have assessed the sensitivities of the proposed muon-based collider facilities in terms of their $2\sigma$ reach  in the relevant $m_\mathrm{KK}$--$g_D$ plane through four different final states:
\begin{itemize}
\item 
Elastic scattering at the future $\mu$TRISTAN \((\mu^+ \mu^+)\) collider (see section~\ref{sec:Mutristan}),
\item 
Semi-visible (SSDM + $\slashed{E}_T$) final state at the same facility (see section~\ref{sec:semi-visible}),
\item 
All-visible (four muon) final state at the same facility (for which three representative benchmark points in the $m_\mathrm{KK}$--$g_D$ plane are highlighted) (see section~\ref{sec:Dimuon Channel})
\item
Resonant productions of heavy KK modes at a future high-energy muon \((\mu^- \mu^+)\) collider (see section~\ref{sec:MuonC}).
\end{itemize}
Each of these channels probes distinct kinematic features and different regimes of KK mass and coupling of the scenario, thus collectively offering a broad and reinforced coverage of the parameter space.
%
%\cmag{(Broke the previous para and created a new one)}
To provide a comprehensive picture of the comparative and complementary sensitivities of the different search modes, we collect in Fig.~\ref{fig:combined_summary} the projected $2\sigma$ reach via each mode under realistic assumptions on the integrated luminosity and the detector performance. Some salient observations emerging from our studies of these final states and from Fig.~\ref{fig:combined_summary} are as follows.
\begin{itemize}
\item     
The elastic process (\(\mu^+ \mu^+ \to \mu^+ \mu^+\)), analysed via a binned \(\chi^2\) approach, reveals a strong sensitivity to the interference between the SM and the tower of massive KK gauge bosons. The resulting $2\sigma$ reach indicates that, for a benchmark luminosity of $\mathcal{L}_\text{int} = 1~\text{ab}^{-1}$ and a negligible systematic uncertainty, the analysis is sensitive to effective 4D gauge couplings as small as $g_D \sim 1.5 \times 10^{-5}$, with the greatest sensitivity achieved in the low-$m_{\text{KK}}$ regime due to dense stacking of light KK states.
\item
The semi-visible (SSDM + $\slashed{E}_T$) final state process (\(\mu^+ \mu^+ \to \mu^+ \mu^+ V^{(n)}, \, V^{(n)} \to \nu_\alpha \bar{\nu}_\alpha, \, n = 1,2,3,\dots, \, \alpha = \mu, \tau\)), arising from an on-shell KK gauge boson decaying to a neutrino pair, provides a complementary sensitivity in the small coupling regime. After implementing appropriate MET cuts and summing over the contributions from the KK states off the relevant KK tower, we find that the projected \(2\sigma\) reach in the \( m_{\text{KK}} \text{--} g_D\) plane extends down to \( g_D \sim 4 \times 10^{-5} \). Notably, in contrast to elastic scattering processes, where interference effects play an important role in the projected reach, the semi-visible final state is primarily controlled by on-shell KK production and decay kinematics, which yield a cleaner missing energy signature after MET selections.
\item
The all-visible (four muon) final state (\(\mu^+ \mu^+ \to \mu^+ \mu^+ V^{(n)}, \, V^{(n)} \to \mu^+ \mu^-, \, n = 1,2,3,\dots\)) further enhances the reach of the $\mu$TRISTAN collider.  By analysing three representative benchmark scenarios, we demonstrate that a projected $2\sigma$ reach can be comparable to that achieved in the search for the semi-visible final-state. In particular, for $m_{\text{KK}}$ in the MeV regime, this final state allows one to probe effective 4D gauge couplings as small as $g_D \sim 2 \times 10^{-5}$, thus demonstrating the all-visible mode to be a competitive probe alongside the elastic scattering process and semi-visible final state.
\item 
In the resonant production process (\(\mu^- \mu^+ \to V^{(n)} \to \mu^- \mu^+\)), the muon collider exploits an \(s\)-channel (resonant) enhancement of the signal to deliver an exceptional sensitivity. We have demonstrated that, even with a realistic  BES (\( = 0.1\% \)), multiple KK modes contribute resonantly when their masses fall within the peak energy window. For \( \sqrt{s} = 3\,\text{TeV} \), the projected $2\sigma$ reach touches a value as low as \( g_D \sim 10^{-5} \). Notably, the projected $2\sigma$ reach arising from the resonant production begins to dominate over the ones from other processes for \( m_{\text{KK}} \gtrsim 10\,\text{GeV} \).
This highlights the crucial role of the resonant processes in probing parameter regions involving higher KK masses. The process retains its sensitivity as a probe even within a narrower peak-energy window, corresponding to a smaller beam-energy spread around the central CM energy, thereby underscoring the need to achieve a precise control over the muon beam energies to maximise the overall sensitivity of the experiments to such scenarios.
\end{itemize}
To assess the complementarity between the two muon-based probes, in Fig.~\ref{fig:combined_all} we compare their projected reaches in the \((m_{\rm KK}, g_D)\) plane with that obtained in our previous muon beam-dump study~\cite{Chakraborty:2025jbd}.%
\footnote{We note that in the absence of kinetic mixing, as mentioned in section~\ref{sec:ModelSetup}, constraints that rely specifically on kinetic-mixing-induced couplings of the KK modes of the 5D $U(1)_{L_\mu-L_\tau}$ gauge field to electrons or quarks, such as those from elastic neutrino-electron scattering~\cite{CHARM-II:1993phx,CHARM-II:1994dzw}, electron-beam fixed-target searches~\cite{NA64:2018lsq}, and rare $\pi^0\to\gamma Z'$ decays~\cite{NA482:2015wmo}, are not applicable in our setup. Constraints arising from the direct couplings of these putative KK modes to muons, such as those from BaBar~\cite{BaBar:2016sci} and LHC multi-muon searches~\cite{ATLAS:2024uvu}, however, remain relevant.} The projected reaches obtained at the muon-based colliders extend substantially beyond the existing $\mathrm{NA64}\mu$ constraint, reaching out to significantly smaller values of $g_D$, for the masses of the KK-mode in the MeV--GeV range. This demonstrates the complementarity between low-energy fixed-target experiments and high-energy muon-based colliders in probing the 5D $U(1)_{L_\mu-L_\tau}$ framework. On the other hand, while the beam-dump setup is sensitive only to a restricted region of the parameter space of such a scenario, a muon-based collider has an extended reach to heavier KK modes. In particular, such a collider could probe KK modes having masses above $10~\mathrm{GeV}$, reaching up to $\mathcal{O}(10~\mathrm{TeV})$, and thereby turning sensitive to a new region of parameter space with $10^{-3}\lesssim g_D \lesssim 10^{-1}$, which is beyond the coverage of the beam-dump experiments. Thus, a significantly enlarged region of the $m_\mathrm{KK}-g_D$ plane of our scenario could be explored by the future muon-based colliders. 

We note, however, that the low $m_{\rm KK}$ region should be interpreted with some care. Cosmological constraints from \(N_{\rm eff}\) have been investigated in the minimal 4D \(U(1)_{L_\mu-L_\tau}\) model~\cite{Escudero:2019gzq}, where light mediators with masses below \(\sim \mathcal{O}(10\,\mathrm{MeV})\) are found to be strongly constrained, thus excluding couplings down to \(g'\sim10^{-8}\). A quantitative assessment of the corresponding constraints in the present 5D framework would require a dedicated cosmological analysis that should include the effects of the KK spectrum and the thermal history of the KK modes. Such an analysis is beyond the scope of our present work. Therefore, regions with low $m_{\rm KK}$ values shown in Figs.~12 and~13 are to be considered with this caveat in mind.

\begin{figure}[t]
    \centering
    \includegraphics[width=0.60\textwidth]{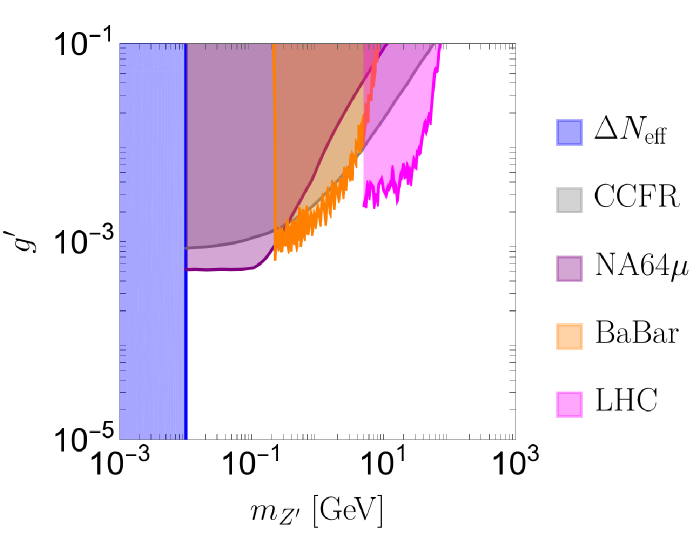}
    \caption{Summary of the current experimental and cosmological constraints on the minimal 4D $U(1)_{L_{\mu}-L_{\tau}}$ model in the $(m_{Z'},\,g')$ plane. The blue, grey, purple, orange, and magenta shaded regions are excluded by the cosmological $\Delta N_{\mathrm{eff}}>0.5$ criterion~\cite{Escudero:2019gzq}, CCFR (95\% CL)~\cite{Altmannshofer:2014pba}, NA64$\mu$ (90\% CL)~\cite{NA64:2024nwj}, BaBar (90\% CL)~\cite{BaBar:2016sci}, and LHC (95\% CL)~\cite{ATLAS:2024uvu} constraints, respectively.}
    \label{fig:Merged4D}
\end{figure}

We now briefly comment on some additional existing constraints. Neutrino trident production (e.g., the CCFR 
measurement)~\cite{Altmannshofer:2014pba},
supernova-cooling~\cite{Croon:2020lrf}, BaBar~\cite{BaBar:2016sci}, and LHC 
multi-muon analysis/search ~\cite{ATLAS:2024uvu} remain relevant even in the absence of 
kinetic mixing (see section~1). For completeness, we discuss the existing 
constraints on the minimal 4D $U(1)_{L_{\mu}-L_{\tau}}$ model, with the 
exclusion regions relevant to the mass and coupling ranges considered in the 
present work, as presented in 
Fig.~\ref{fig:Merged4D}. Around $m_{Z'}\simeq 0.01-0.2$~GeV, the NA64$\mu$ experiment provides the strongest constraint, probing couplings down to $g'\sim5\times10^{-4}$. In the mass range $m_{Z'}\simeq 0.2-5$~GeV, the BaBar constraint is the most stringent one, which excludes couplings in the approximate range of $g'\sim10^{-3}$--$10^{-2}$, while LHC multi-muon searches become sensitive for $m_{Z'}\gtrsim5$~GeV, and up to $\mathcal{O}(100)$~GeV, probing approximately $g'\sim10^{-3}$--$10^{-1}$. Across these mass ranges, the noted constraints are already stronger than the corresponding CCFR limits. 

Although the results discussed above provide useful phenomenological contexts, they should not be interpreted to derive direct constraints on our 5D framework. A consistent reinterpretation of these searches in our 5D scenario is non-trivial because the relevant amplitudes receive contributions from the entire KK tower, rather than from the exchange of a single gauge boson. In addition, the interference among diagrams mediated by different KK excitations modifies the signal prediction, preventing a straightforward recast of the existing 4D exclusion limits into the $(m_{\rm KK},g_D)$ parameter plane.\footnote{The following points should also be noted: (i) Since the BaBar statistical analysis is conducted under the assumption of a single $Z'$ boson, constructing a reliable likelihood function that reflects the experiment's internal data for cases involving multiple $Z'$ bosons presents significant difficulties. (ii) Regarding the multiple muon final state at the LHC, multiple $Z'$ bosons are involved as intermediate states in a multi-body final state; consequently, even if interference effects between scattering amplitudes are ignored, the analysis requires massive numerical computation, specifically, in generating a vast amount of $Z'$ signal data for each mass point individually and then appropriately combining them.}
Given these circumstances, simply reinterpreting the results of the previous 4D studies is entirely insufficient to obtain reliable results for our 5D model; instead, detailed calculations that take the aforementioned points into account are required.
Therefore, we conclude that these issues require a dedicated analysis, which is outside the scope of the present paper, and hence, we leave their details to future work.

The comparison shown in Fig.~\ref{fig:combined_all} is therefore restricted to constraints that have been explicitly studied within the 5D framework using the benchmark assumptions adopted in this work, such as the NA64$\mu$ bound obtained in our previous work~\cite{Chakraborty:2025jbd}. We also note that existing 4D studies~\cite{Huang:2021nkl} on a future muon collider can probe $g'$ values more than an order of magnitude smaller than those probed by current LHC multi-muon searches~\cite{ATLAS:2024uvu} for $m_{Z'}\sim100~\mathrm{GeV}$. The projected reach extends well into the TeV mass regime, covering a broad region of the $U(1)_{L_{\mu}-L_{\tau}}$ parameter space. Building upon this, we find that the projected reach of the muon-based colliders is further extended to smaller values of $g_D$ in our 5D framework, and an enhanced region of the parameter space becomes accessible thanks to the resonant enhancement of the signal with contributions coming from multiple KK modes.
%%%%%%%%%%%%%%%%%
%%%%%%%%%%%%%%%
%%%%%%%%%%%%%%%%%%%%%%%%%%%%%%%%%%%%%%%%%%%%%%%%%%%%%%%%%%%%%%%%%%%%%%%%%%

%%%%%%%%%%%%%%%%%%%%%%%%%%%%%%%%%%%
\section{Conclusions and outlook
\label{sec:conclusion}}
In this work, we have explored the collider phenomenology of a 5D 
\( U(1)_{L_\mu - L_\tau} \) gauge extension of the SM in the flat extra-dimensional setup, focusing on scenarios with vanishing kinetic mixing between the \( U(1)_{L_\mu - L_\tau} \) and \(U(1)_Y\) gauge groups. Our analysis considers a minimal realisation in which the corresponding 5D \( U(1)_{L_\mu - L_\tau} \) gauge boson can propagate in the bulk. Since these KK modes of the associated \( U(1)_{L_\mu - L_\tau} \)  gauge boson couple directly to muons and taus,  muon-based colliders, such as the proposed $\mu$TRISTAN facility and future high-energy muon colliders, are expected to provide ideal environments to test this framework. In contrast, a dedicated tau collider is not considered viable in the foreseeable future due to too short a lifetime for the $\tau$-lepton (${\cal O}$(10$^{-13}$ sec)) to have steady $\tau$-beams, thereby turning the focus on muon-based colliders. The phenomenology of the said scenario has been investigated at the muon beam-dump experiments \cite{Chakraborty:2025jbd}, which have only a limited reach for the masses of the KK excitations. In contrast, future high-energy muon colliders ($\mu^- \mu^+$) provide us with much more capable facilities to probe the higher (heavier) KK excitations, offering a complementary probe of this scenario.

Within the said theoretical setup, we investigated the signatures of the KK excitations via elastic scattering, semi-visible (SSDM + $\slashed{E}_T$) final state, and the all-visible (four muon) final state at a $\mu^+\mu^+$ collider like the $\mu$TRISTAN, together with the resonant productions of these states at a future $\mu^- \mu^+$ collider by taking into account realistic spreads in the beam energy and systematic uncertainties. Collectively, these analyses reveal a significant {competition} in the reach in the relevant $m_\mathrm{KK}$--$g_D$ plane for some parameter choices, when compared to our earlier study~\cite{Chakraborty:2024xxc}, which considered low-energy electron-neutrino scattering experiments. %Notably, no direct bound, either from the LEP or the LHC, currently exists on this class of extra-dimensional realisations. A complementary study of the regime involving light and weakly coupled KK modes, focusing on the prospects of future muon beam-dump experiments, has been presented in Ref.~\cite{Chakraborty:2025jbd}.

These results demonstrate that the upcoming muon-based colliders would provide sensitive and complementary probes of the 5D $U(1)_{L_\mu-L_\tau}$ scenario considered in this work, extending the accessible parameter space into the small-coupling and high-mass regimes where traditional low-energy experiments, such as fixed-target and beam-dump searches, lose sensitivity. The existing 4D constraints discussed in section~\ref{sec:Summary} are generally weaker than the projected collider reaches over most of the parameter space considered here, although the low-$m_{\rm KK}$ region below $\mathcal{O}(10\,\mathrm{MeV})$ requires particular caution in view of cosmological constraints. These constraints should, however, be regarded only to provide some phenomenological contexts rather than as direct constraints on the present 5D scenario. The extents of the projected $2\sigma$ reach, derived from muon elastic scattering that includes the interference effects between the KK modes and the SM gauge bosons, the semi-visible (SSDM + $\slashed{E}_T$) channel, the all-visible four-muon final state, and the resonantly enhanced KK production with interference effects taken into account, underscore the complementarity of searches in different final states in probing extended gauge structures beyond the SM.

Future directions in such studies include extending the theoretical framework to warped geometries and investigating its implications for the dark matter phenomenology and its cosmological consistency, in the form of its compliance with the constraints from the observed relic abundance, and with observations pertaining to cosmic microwave background (CMB), and formation of the large-scale structures in the Universe~\cite{Foldenauer:2018zrz,Lee:2025lko,Krnjaic:2019rsv,Kamada:2018zxi}.
A comprehensive analysis that addresses these aspects would help produce a more complete assessment of the viable parameter space for this class of extra-dimensional gauge extensions of the SM.
%
%%%%%%%%
\section*{Acknowledgements}
D.C.\ thanks Saiyad Ashanujjaman for helpful discussions on collider physics tools, particularly \texttt{ROOT}, during the ML4HEP workshop at IOP, Bhubaneswar. D.C.\ also thanks CSIR, India, for financial support through the Senior Research Fellowship (Direct) (File No: 09/1128(23623)/2025-EMR-I). We acknowledge the Shiv Nadar Institution of Eminence for providing the computational infrastructure utilised in this work.
%
%%%%%%%%%
\appendix
\section*{Appendix}
%%%%%%%%%%%%%%%%%%%%%%%%%%%%%%%%%%%%%%%%%%%%%%%%%%%%%%%%%%%%%%%%%%%%%%%%%
\section{Coefficients of $s^2$, $t^2$, and $u^2$ for elastic scattering case
\label{appendix-a}}

In this appendix, we present the explicit expressions for the coefficients of the square of the Mandelstam variables \(s\), \(t\), and \(u\) that appear in the squared, spin-averaged matrix element discussed in section~\ref{sec:sig-mutristan}.

\subsection*{Coefficient of $s^2$: $X$}

\begin{align}
X ={}& 2 \Bigg[ (\tilde c_1^2 + \tilde d_1^2)c_1^2
- 2 \tilde d_1(\tilde c_2 d_2 + c_2 \tilde d_2 - \tilde c_3 d_3 - c_3 \tilde d_3 + \tilde c_4 d_4 + c_4 \tilde d_4 - \tilde c_5 d_5 - c_5 \tilde d_5 + \tilde c_6 d_6 + c_6 \tilde d_6)c_1 \nonumber\\
&- 2 \tilde c_1(c_2 \tilde c_2 - c_3 \tilde c_3 + c_4 \tilde c_4 - c_5 \tilde c_5 + c_6 \tilde c_6 - 2 d_1 \tilde d_1 + d_2 \tilde d_2 - d_3 \tilde d_3 + d_4 \tilde d_4 - d_5 \tilde d_5 + d_6 \tilde d_6)c_1 \nonumber \\[4pt]
&+ c_3^2 \tilde c_3^2 + c_4^2 \tilde c_4^2 + c_5^2 \tilde c_5^2 + c_6^2 \tilde c_6^2 + \tilde c_1^2 d_1^2 + d_1^2 \tilde d_1^2 + \tilde c_2^2 d_2^2 + d_2^2 \tilde d_2^2 + \tilde c_3^2 d_3^2 + c_3^2 \tilde d_3^2 + d_3^2 \tilde d_3^2 \nonumber \\[4pt] 
&+ \tilde c_4^2 d_4^2 + c_4^2 \tilde d_4^2 + d_4^2 \tilde d_4^2 + \tilde c_5^2 d_5^2 + c_5^2 \tilde d_5^2 + d_5^2 \tilde d_5^2 + \tilde c_6^2 d_6^2 + c_6^2 \tilde d_6^2 + d_6^2 \tilde d_6^2 \nonumber \\[4pt] 
&- 2 c_3 \tilde c_3 c_4 \tilde c_4 + 2 c_3 \tilde c_3 c_5 \tilde c_5 - 2 c_4 \tilde c_4 c_5 \tilde c_5 - 2 c_3 \tilde c_3 c_6 \tilde c_6 + 2 c_4 \tilde c_4 c_6 \tilde c_6 - 2 c_5 \tilde c_5 c_6 \tilde c_6 \nonumber \\[4pt] 
&+ 2 c_3 \tilde c_3 d_1 \tilde d_1 - 2 c_4 \tilde c_4 d_1 \tilde d_1 + 2 c_5 \tilde c_5 d_1 \tilde d_1 - 2 c_6 \tilde c_6 d_1 \tilde d_1 - 2 \tilde c_1 \tilde c_2 d_1 d_2 - 2 c_3 \tilde c_3 d_2 \tilde d_2 \nonumber \\[4pt] 
&+ 2 c_4 \tilde c_4 d_2 \tilde d_2 - 2 c_5 \tilde c_5 d_2 \tilde d_2 + 2 c_6 \tilde c_6 d_2 \tilde d_2 - 2 d_1 \tilde d_1 d_2 \tilde d_2 + c_2^2(\tilde c_2^2 + \tilde d_2^2) + 2 \tilde c_1 \tilde c_3 d_1 d_3 \nonumber \\[4pt] 
&- 2 \tilde c_2 \tilde c_3 d_2 d_3 + 2 \tilde c_1 c_3 d_1 \tilde d_3 - 2 \tilde c_2 c_3 d_2 \tilde d_3 + 4 c_3 \tilde c_3 d_3 \tilde d_3 - 2 c_4 \tilde c_4 d_3 \tilde d_3 + 2 c_5 \tilde c_5 d_3 \tilde d_3 \nonumber \\[4pt] 
&- 2 c_6 \tilde c_6 d_3 \tilde d_3 + 2 d_1 \tilde d_1 d_3 \tilde d_3 - 2 d_2 \tilde d_2 d_3 \tilde d_3 - 2 \tilde c_1 \tilde c_4 d_1 d_4 + 2 \tilde c_2 \tilde c_4 d_2 d_4 - 2 \tilde c_3 \tilde c_4 d_3 d_4 \nonumber \\[4pt] 
&- 2 c_3 \tilde c_4 \tilde d_3 d_4 - 2 \tilde c_1 c_4 d_1 \tilde d_4 + 2 \tilde c_2 c_4 d_2 \tilde d_4 - 2 \tilde c_3 c_4 d_3 \tilde d_4 - 2 c_3 c_4 \tilde d_3 \tilde d_4 - 2 c_3 \tilde c_3 d_4 \tilde d_4 \nonumber \\[4pt] 
&+ 4 c_4 \tilde c_4 d_4 \tilde d_4 - 2 c_5 \tilde c_5 d_4 \tilde d_4 + 2 c_6 \tilde c_6 d_4 \tilde d_4 - 2 d_1 \tilde d_1 d_4 \tilde d_4 + 2 d_2 \tilde d_2 d_4 \tilde d_4 - 2 d_3 \tilde d_3 d_4 \tilde d_4 \nonumber \\[4pt] 
&+ 2 \tilde c_1 \tilde c_5 d_1 d_5 - 2 \tilde c_2 \tilde c_5 d_2 d_5 + 2 \tilde c_3 \tilde c_5 d_3 d_5 + 2 c_3 \tilde c_5 \tilde d_3 d_5 - 2 \tilde c_4 \tilde c_5 d_4 d_5 - 2 c_4 \tilde c_5 \tilde d_4 d_5 \nonumber \\[4pt] 
&+ 2 \tilde c_1 c_5 d_1 \tilde d_5 - 2 \tilde c_2 c_5 d_2 \tilde d_5 + 2 \tilde c_3 c_5 d_3 \tilde d_5 + 2 c_3 c_5 \tilde d_3 \tilde d_5 - 2 \tilde c_4 c_5 d_4 \tilde d_5 - 2 c_4 c_5 \tilde d_4 \tilde d_5 \nonumber \\[4pt] 
&+ 2 c_3 \tilde c_3 d_5 \tilde d_5 - 2 c_4 \tilde c_4 d_5 \tilde d_5 + 4 c_5 \tilde c_5 d_5 \tilde d_5 - 2 c_6 \tilde c_6 d_5 \tilde d_5 + 2 d_1 \tilde d_1 d_5 \tilde d_5 - 2 d_2 \tilde d_2 d_5 \tilde d_5 \nonumber \\[4pt] 
&+ 2 d_3 \tilde d_3 d_5 \tilde d_5 - 2 d_4 \tilde d_4 d_5 \tilde d_5 - 2 \tilde c_2\big\{ \tilde d_2(\tilde c_1 d_1 + \tilde c_3 d_3 + c_3 \tilde d_3 - \tilde c_4 d_4 - c_4 \tilde d_4 + \tilde c_5 d_5 + c_5 \tilde d_5 - \tilde c_6 d_6 - c_6 \tilde d_6) \nonumber \\[4pt] 
&\quad + \tilde c_2(c_3 \tilde c_3 - c_4 \tilde c_4 + c_5 \tilde c_5 - c_6 \tilde c_6 + d_1 \tilde d_1 - 2 d_2 \tilde d_2 + d_3 \tilde d_3 - d_4 \tilde d_4 + d_5 \tilde d_5 - d_6 \tilde d_6) \big\} \Bigg].
\end{align}

\subsection*{Coefficient of $t^2$: $Y$}
\begin{align}
Y ={}& 
2 \Bigg[
(\tilde c_2^2 + \tilde d_2^2)c_2^2
- 2\big\{\tilde d_2(\tilde c_4 d_4 - c_4 \tilde d_4 + \tilde c_6 d_6 - c_6 \tilde d_6)
+ \tilde c_2(-c_4 \tilde c_4 - c_6 \tilde c_6 + 2 d_2 \tilde d_2 + d_4 \tilde d_4 + d_6 \tilde d_6)\big\}c_2 \nonumber\\[4pt]
&+ c_4^2 \tilde c_4^2 + c_6^2 \tilde c_6^2
+ \tilde c_2^2 d_2^2 + d_2^2 \tilde d_2^2
+ \tilde c_4^2 d_4^2 + c_4^2 \tilde d_4^2 + d_4^2 \tilde d_4^2 \nonumber \\[4pt] 
&+ \tilde c_6^2 d_6^2 + c_6^2 \tilde d_6^2 + d_6^2 \tilde d_6^2
+ 2 c_4 \tilde c_4 c_6 \tilde c_6
- 2 c_4 \tilde c_4 d_2 \tilde d_2
- 2 c_6 \tilde c_6 d_2 \tilde d_2 \nonumber \\[4pt] 
&+ 2 \tilde c_2 \tilde c_4 d_2 d_4
- 2 \tilde c_2 c_4 d_2 \tilde d_4
- 4 c_4 \tilde c_4 d_4 \tilde d_4
- 2 c_6 \tilde c_6 d_4 \tilde d_4 \nonumber \\[4pt] 
&+ 2 d_2 \tilde d_2 d_4 \tilde d_4
+ 2 \tilde c_2 \tilde c_6 d_2 d_6
+ 2 \tilde c_4 \tilde c_6 d_4 d_6
- 2 c_4 \tilde c_6 \tilde d_4 d_6 \nonumber \\[4pt] 
&- 2 \tilde c_2 c_6 d_2 \tilde d_6
- 2 \tilde c_4 c_6 d_4 \tilde d_6
+ 2 c_4 c_6 \tilde d_4 \tilde d_6
- 2 c_4 \tilde c_4 d_6 \tilde d_6 \nonumber \\[4pt] 
&- 4 c_6 \tilde c_6 d_6 \tilde d_6
+ 2 d_2 \tilde d_2 d_6 \tilde d_6
+ 2 d_4 \tilde d_4 d_6 \tilde d_6 \Bigg].
\end{align}
%
%%%%%%%%%
\subsection*{Coefficient of $u^2$: $Z$}
\begin{align}
Z ={}& 2 \Bigg[
(\tilde c_1^2 + \tilde d_1^2)c_1^2
- 2 \tilde d_1(\tilde c_3 d_3 - c_3 \tilde d_3 + \tilde c_5 d_5 - c_5 \tilde d_5)c_1 \nonumber\\[4pt]
&- 2 \tilde c_1(-c_3 \tilde c_3 - c_5 \tilde c_5 + 2 d_1 \tilde d_1 + d_3 \tilde d_3 + d_5 \tilde d_5)c_1 \nonumber \\[4pt] 
&+ \tilde c_1^2 d_1^2 + d_1^2 \tilde d_1^2
+ \tilde c_3^2 d_3^2 + d_3^2 \tilde d_3^2
+ \tilde c_5^2 d_5^2 + d_5^2 \tilde d_5^2 \nonumber \\[4pt] 
&+ 2 \tilde c_1 \tilde c_3 d_1 d_3
+ 2 d_1 \tilde d_1 d_3 \tilde d_3
+ c_3^2(\tilde c_3^2 + \tilde d_3^2) \nonumber \\[4pt] 
&+ 2 \tilde c_1 \tilde c_5 d_1 d_5
+ 2 \tilde c_3 \tilde c_5 d_3 d_5
+ 2(d_1 \tilde d_1 + d_3 \tilde d_3)d_5 \tilde d_5 \nonumber \\[4pt] 
&- 2 c_5\big(\tilde c_5 d_1 \tilde d_1
+ \tilde c_5 d_3 \tilde d_3
+ \tilde c_1 d_1 \tilde d_5
+ \tilde c_3 d_3 \tilde d_5
+ 2 \tilde c_5 d_5 \tilde d_5 \big) + c_5^2(\tilde c_5^2 + \tilde d_5^2) \nonumber \\[4pt] 
&- 2 c_3\big\{ \tilde d_3(\tilde c_1 d_1 + \tilde c_5 d_5 - c_5 \tilde d_5)
 + \tilde c_3(-c_5 \tilde c_5 + d_1 \tilde d_1 + 2 d_3 \tilde d_3 + d_5 \tilde d_5) \big\}
\Bigg].
\end{align}
%
%%%%%%%%%%%%%%%%%%%%%%%%%%%%%%%%%%%%%%%%%%%%%%%%%%%%%%%%%%%%%%%%%%%%%%%%%%%%%%%%%%%%%%%%%%%%%%%%%%%%%%%%%%%%%%%%%%%%%%%%%%%%%%%%%%%%%%%%%%%%%%%%%%%%
\section{Infinite KK tower sums and decay widths}
\label{appendix-b}
In this appendix, we provide a detailed presentation of the KK mode functions, the analytic infinite sums over the KK tower, including the general dependence on the SM brane position, and the decay widths of the KK modes relevant for phenomenology.
%
%%%%%%%%%%%%%%
\subsubsection*{KK mode functions}
The profile of the \( n \)-th KK gauge boson mode along the extra dimension \(\tilde{y}_{\text{SM}}\) is given by
\begin{equation}
f_V^{(n)}(\tilde{y}_{\text{SM}}) = \sqrt{2} \cos\left[\left(n - \frac{1}{2}\right) \tilde{y}_{\text{SM}} \right], \quad n=1,2,3,\ldots \,
\end{equation}
where \(\tilde{y}_{\text{SM}}\) is the coordinate of the SM brane along the extra dimension. The KK masses are quantised as
\begin{equation}
M_n = \left(2 n - 1\right) m_{\text{KK}} \, ,
\end{equation}
with \( m_{\text{KK}} \) setting the KK scale.
%
%%%%%%%%%%%
\subsection*{Infinite sum over KK modes: general brane position}
The infinite sum over KK propagators, weighted by the squared mode functions at an arbitrary brane location \(\tilde{y}_{\mathrm{SM}}\), is defined as
\begin{equation}
S(a, \tilde{y}_{\mathrm{SM}}) \equiv \mathlarger{\sum_{n=1}^\infty} 
\frac{\left[f_V^{(n)}(\tilde{y}_{\mathrm{SM}})\right]^2}{a - M_n^2} \, ,
\end{equation}
where `\(a\)' denotes the squared four-momentum flowing through the propagator and may take values of the Mandelstam variables \(s\), \(t\), or \(u\). Unless otherwise stated, we restrict ourselves to the off-shell regime, where this summation is well-defined. This sum can be evaluated analytically and expressed in terms of the Gauss hypergeometric function \({}_2F_1(a,b;c;z)\) as
\begin{align}
S(a, \tilde{y}_{\mathrm{SM}}) &= \frac{e^{-i \tilde{y}_{\mathrm{SM}}}}{4 \sqrt{a} m_{\mathrm{KK}} (a - m_{\mathrm{KK}}^2)} \Bigg[
\sqrt{a} m_{\mathrm{KK}} \, \Bigg\{
{}_2F_1\!\left(1, \frac{1}{2} - \frac{\sqrt{a}}{2 m_{\mathrm{KK}}}; \frac{3}{2} - \frac{\sqrt{a}}{2 m_{\mathrm{KK}}}; e^{-2 i \tilde{y}_{\mathrm{SM}}}\right) \nonumber\\
&\quad + e^{2 i \tilde{y}_{\mathrm{SM}}} \, {}_2F_1\!\left(1, \frac{1}{2} - \frac{\sqrt{a}}{2 m_{\mathrm{KK}}}; \frac{3}{2} - \frac{\sqrt{a}}{2 m_{\mathrm{KK}}}; e^{2 i \tilde{y}_{\mathrm{SM}}}\right) \Bigg\} \nonumber \\
&\quad + m_{\mathrm{KK}}^2 \, \Bigg\{
{}_2F_1\!\left(1, \frac{1}{2} - \frac{\sqrt{a}}{2 m_{\mathrm{KK}}}; \frac{3}{2} - \frac{\sqrt{a}}{2 m_{\mathrm{KK}}}; e^{-2 i \tilde{y}_{\mathrm{SM}}}\right) \nonumber \\
&\quad + e^{2 i \tilde{y}_{\mathrm{SM}}} \, {}_2F_1\!\left(1, \frac{1}{2} - \frac{\sqrt{a}}{2 m_{\mathrm{KK}}}; \frac{3}{2} - \frac{\sqrt{a}}{2 m_{\mathrm{KK}}}; e^{2 i \tilde{y}_{\mathrm{SM}}}\right) \Bigg\} \nonumber \\
&\quad + \sqrt{a} m_{\mathrm{KK}} \, \Bigg\{
{}_2F_1\!\left(1, \frac{1}{2} + \frac{\sqrt{a}}{2 m_{\mathrm{KK}}}; \frac{3}{2} + \frac{\sqrt{a}}{2 m_{\mathrm{KK}}}; e^{-2 i \tilde{y}_{\mathrm{SM}}}\right) \nonumber \\
&\quad + e^{2 i \tilde{y}_{\mathrm{SM}}} \, {}_2F_1\!\left(1, \frac{1}{2} + \frac{\sqrt{a}}{2 m_{\mathrm{KK}}}; \frac{3}{2} + \frac{\sqrt{a}}{2 m_{\mathrm{KK}}}; e^{2 i \tilde{y}_{\mathrm{SM}}}\right) \Bigg\} \nonumber \\
&\quad - m_{\mathrm{KK}}^2 \, \Bigg\{
{}_2F_1\!\left(1, \frac{1}{2} + \frac{\sqrt{a}}{2 m_{\mathrm{KK}}}; \frac{3}{2} + \frac{\sqrt{a}}{2 m_{\mathrm{KK}}}; e^{-2 i \tilde{y}_{\mathrm{SM}}}\right) \nonumber \\
&\quad + e^{2 i \tilde{y}_{\mathrm{SM}}} \, {}_2F_1\!\left(1, \frac{1}{2} + \frac{\sqrt{a}}{2 m_{\mathrm{KK}}}; \frac{3}{2} + \frac{\sqrt{a}}{2 m_{\mathrm{KK}}}; e^{2 i \tilde{y}_{\mathrm{SM}}}\right) \Bigg\} \Bigg] \nonumber \\
&\quad - a e^{i \tilde{y}_{\mathrm{SM}}} \pi \tan\left( \frac{\pi \sqrt{a}}{2 m_{\mathrm{KK}}} \right) + e^{i \tilde{y}_{\mathrm{SM}}} m_{\mathrm{KK}}^2 \pi \tan\left( \frac{\pi \sqrt{a}}{2 m_{\mathrm{KK}}} \right) \,.
\end{align}
This general result reduces to closed-form expressions for specific brane locations.  
For instance, at $\tilde{y}_{\mathrm{SM}} = \pi/2$, one finds
\begin{equation}
S\!\left(a, \frac{\pi}{2}\right) 
= - \frac{\pi}{4\, m_{\text{KK}} \sqrt{a}} 
\tan\!\left( \frac{\pi \sqrt{a}}{2\, m_{\mathrm{KK}}} \right),
\end{equation}
whereas, for $\tilde{y}_{\mathrm{SM}} = 0$, the sum takes the form
\begin{equation}
S(a,0) 
= - \frac{\pi}{2\, m_{\text{KK}} \sqrt{a}} 
\tan\!\left( \frac{\pi \sqrt{a}}{2\, m_{\text{KK}}} \right).
\end{equation}

Although the KK propagator is written as an infinite sum, the 5D theory is understood as an effective theory with a finite cutoff. Since the 5D gauge coupling $g_{5D}$ has a mass dimension of $-\frac{1}{2}$, an estimate based on naive dimensional analysis gives
\begin{equation}
\Lambda_{5D}\sim \frac{24\pi^3}{g_{5D}^2} \; .
\end{equation}
Using the relations between $g_{5D}$, $g_D$, $R$, and $m_{\rm KK}$ presented in section~\ref{sec:ModelSetup}, this can be expressed as
\begin{equation}
\frac{\Lambda_{5D}}{m_{\rm KK}}
\sim \frac{48\pi^2}{g_D^2} \; .
\end{equation}
Thus, the number of KK modes below the perturbative cutoff is parametrically large over the parameter space explored in this work. We have explicitly verified that truncating the KK tower at the EFT cutoff does not appreciably modify the relevant results over the parameter region considered in our collider analysis.
%%%%%%%%%%%%%%%%%%%%%%%%%%%%%%%%%%%%%%%%%%%%%%%%%%%%%%%%%%%%%%%%%
\subsection*{KK mode decay widths}
The partial widths for the decays of the \(n\)-th KK mode into neutrinos and charged leptons are given by
\begin{align}
\Gamma_{\nu_\alpha}^{(n)} &= \frac{g_D^2 M_n}{24 \pi} \left[f_V^{(n)}(\tilde{y}_{\mathrm{SM}})\right]^2, & \alpha = \mu, \tau, \\
\Gamma_\ell^{(n)} &= \frac{g_D^2 M_n}{12 \pi} \left[f_V^{(n)}(\tilde{y}_{\mathrm{SM}})\right]^2 
\left(1 + \frac{2 m_\ell^2}{M_n^2}\right) \sqrt{1 - \frac{4 m_\ell^2}{M_n^2}} \, , & \ell = \mu, \tau.
\end{align}
The total decay width can then be written in a compact way as
\begin{equation}
\Gamma_n \equiv \Gamma^{(n)}_{\text{Total}} 
= \sum_{\alpha = \mu, \tau} \Gamma_{\nu_\alpha}^{(n)} 
+ \sum_{\ell = \mu, \tau} \Theta(M_n - 2 m_\ell)\, \Gamma_\ell^{(n)},
\end{equation}
where \(\Theta\) is the Heaviside step function that ensures relevant kinematic thresholds are respected.  

%%%%%%%%%%%%%%%%%%%%%%%%%%%%%%%%%%%%%%%%%%%%%%%%%%%%%%

\section{Validation of the significance approximation
\label{appendix-c}}

\begin{figure}[t]
\centering
\includegraphics[width=0.50\linewidth]{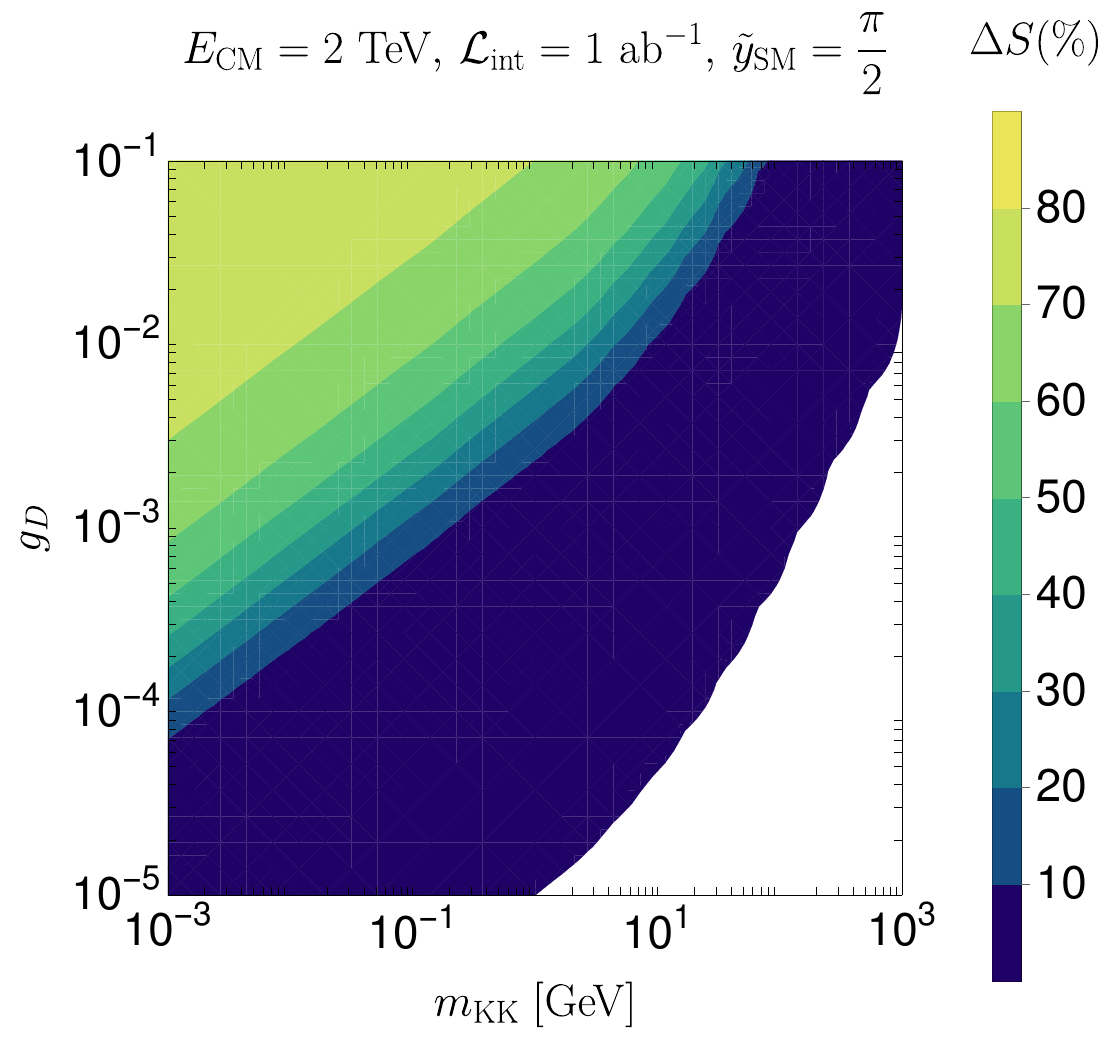}
\caption{
Percentage deviation between the naive Gaussian significance
$S_{\text{Gauss}} = s/\sqrt{s+b}$, used in the main text, and the log-likelihood significance
$S_{\text{Asimov}} = \sqrt{2[(s+b)\ln(1+s/b)-s]}$, shown over the $(m_\tx{KK}, g_D)$ parameter space
for the semi-visible (SSDM + $\slashed{E}_T$) final state arising from
$\mu^+ \mu^+ \to \mu^+ \mu^+ V^{(n)}$ with $V^{(n)} \to \nu_\alpha \bar{\nu}_\alpha$ ($\alpha = \mu, \tau$).
The deviation is evaluated wherever $S_{\text{Asimov}} > 10^{-4}$. Regions with vanishing sensitivity, where the relative deviation is ill-defined, are excluded. This comparison demonstrates that the use of the simple Gaussian significance in the main analysis does not introduce any appreciable bias.}
\label{fig:appendix-deviation}
\end{figure}

In the main analysis, the projected $2\sigma$ reach for the semi-visible final state (SSDM + $\slashed{E}_T$) arising from
\begin{equation}
\mu^+ \mu^+ \to \mu^+ \mu^+ V^{(n)}, \qquad 
V^{(n)} \to \nu_\alpha \bar{\nu}_\alpha,
\end{equation}
with $n = 1,2,3,\dots$ and $\alpha = \mu, \tau$, are obtained using the widely employed Gaussian estimator for statistical significance,
\begin{equation}
S_{\text{Gauss}} = \frac{s}{\sqrt{s+b}},
\end{equation}
where $s$ and $b$ denote the expected numbers of signal and background events, respectively, after the application of all selection criteria. This estimator provides a reliable approximation in the asymptotic regime where both $s$ and $b$ are sufficiently large, such that Gaussian statistics may well approximate Poisson fluctuations.

To assess the validity of the Gaussian significance approximation for the said final state over the parameter space explored in this work, we perform an explicit comparison of the same with the log-likelihood-based significance given by
\begin{equation}
S_{\text{Asimov}} = \sqrt{2\left[(s+b)\ln\left(1+\frac{s}{b}\right)-s\right]},
\end{equation}
which corresponds to the Asimov significance~\cite{Cowan:2010js} for a single counting experiment and remains applicable beyond the Gaussian regime. This definition is widely employed in precision sensitivity studies, particularly in scenarios with moderate event yields.

To quantify the agreement between the two significance estimators, we define the relative deviation as
\begin{equation}
\Delta S(\%) = \frac{S_{\text{Asimov}} - S_{\text{Gauss}}}{S_{\text{Asimov}}} \times 100 \, ,
\end{equation}
and is presented in the $m_\text{KK}$--$g_D$ plane in Fig.~\ref{fig:appendix-deviation} shows that, in the regions of parameter space not already excluded in the analysis (see Figs.~\ref{fig:6} and \ref{fig:combined_summary}), the relative deviation $\Delta S$ remains small. This indicates that using the log-likelihood-based significance would not result in any appreciable modification of the projected $2\sigma$ reach presented in the main text (as log-log plots).
%%%%%%%%%%%%%%%%%%%%%%%%%%%%%%%%%%%%%%%%%%%%%%%%%%%%%%%%%%%%%%%%%%%%%%%%%%%%%%%%%%%%%%%%%%%%%%%
\section{Dependence of the projected reach on the MET selection
\label{appendix-d}}

\begin{figure}[t]
    \centering
    % First plot: current yt = pi/2 case
    \begin{subfigure}[b]{0.35\textwidth}
        \centering
        \includegraphics[width=\textwidth]{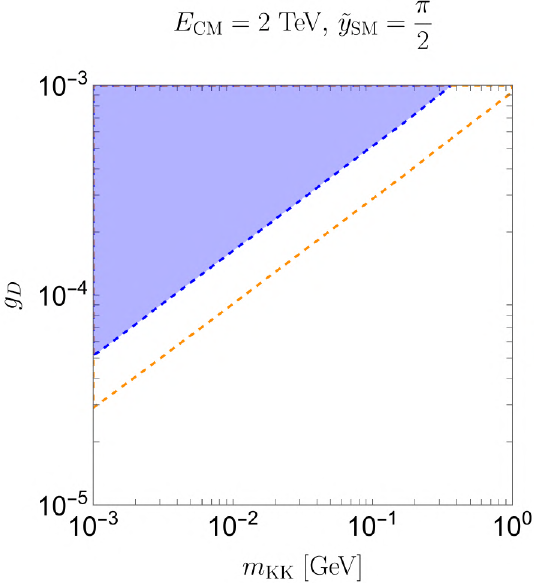}
        \caption{\hspace*{-3em}}
        \label{fig:met-50}
    \end{subfigure}
    \hfill
    % Second plot: yt = 0 case
    \begin{subfigure}[b]{0.35\textwidth}
        \centering
        \includegraphics[width=\textwidth]{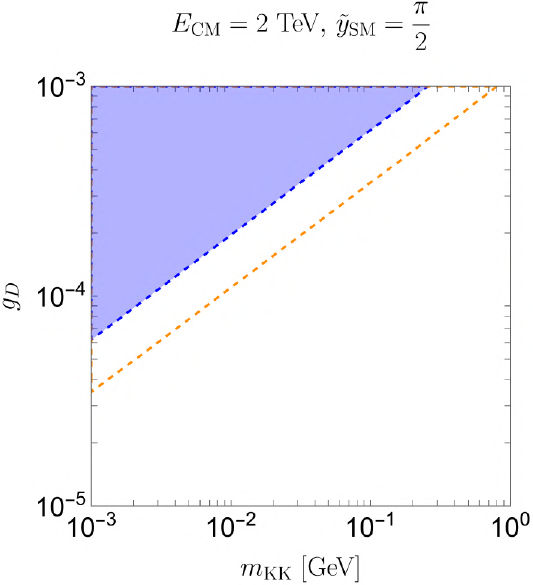}
       \caption{\hspace*{-3em}}
        \label{fig:met-100}
    \end{subfigure}
     \begin{subfigure}[b]{0.18\textwidth}
        \centering
        \raisebox{3cm}
        {
        \includegraphics[width=\textwidth]{met_legends.pdf} }
    \end{subfigure}

\caption{Projected $2\sigma$ reach for the MET windows
(a) $1~\mathrm{GeV}\leq\slashed{E}_T\leq50~\mathrm{GeV}$ and
(b) $1~\mathrm{GeV}\leq\slashed{E}_T\leq100~\mathrm{GeV}$.}
\label{app:met50-100}
\end{figure}
The benchmark analysis presented in the section~\ref{sec:MET channel} employs the MET window
\begin{align}
1~\mathrm{GeV}\leq \slashed{E}_T \leq 10~\mathrm{GeV},
\end{align}
which is motivated by the characteristic low-MET spectrum of the signal and the resulting suppression of the SM background. To examine the dependence of the projected reach on this choice, we repeat the analysis using two broader MET windows:
\begin{align}
1~\mathrm{GeV}\leq \slashed{E}_T \leq 50~\mathrm{GeV},
\qquad
1~\mathrm{GeV}\leq \slashed{E}_T \leq 100~\mathrm{GeV}.
\end{align}
%%%%%%%%%%

Fig.~\ref{app:met50-100} illustrates the dependence of the projected $2\sigma$ reach on the choice of the MET window. As the upper MET threshold is progressively increased, the projected reach becomes weaker. Quantitatively, for the largest integrated luminosity considered, $10~\mathrm{ab}^{-1}$, extending the upper limit of the MET window from $10~\mathrm{GeV}$ to $50~\mathrm{GeV}$ and $100~\mathrm{GeV}$ weakens the projected reach in $g_D$ by factors of approximately $1.5$ and $2$, respectively.

%%%%%%%%%%%%%%%%%%%%%%%%%%%%%%%%%%%%%%%%%%%%%%%%%%%%%%%%%%%%%%%%%%%%%%%%%%%%%%%%%%%%%%%%%%%%
\section{Coefficients of $s^2$, $t^2$, and $u^2$ for resonant production case}
\label{appendix-e}
This appendix collects the explicit expressions for the coefficients of the square of the Mandelstam variables 
\(s\), \(t\), and \(u\) appearing in the squared, spin-averaged matrix element defined in section~\ref{subsec:matrix-res}.
Note that $\Re\fn{a}$ represents the real part of $a$.

\subsection*{Coefficient of $s^2$: $P$}
\begin{align}
P \;=\; 2 \Bigg[
& (\hat c_2^{\,2} + \hat d_2^{\,2})\, c_2^{\,2}
+ 2 \hat d_2 \big( -\hat c_4 d_4 + c_4 \hat d_4 - \hat c_6 d_6 + c_6 \hat d_6 \big) c_2
\nonumber \\[4pt]
& + 2 \hat c_2 \big( c_4 \hat c_4 + c_6 \hat c_6 - 2 d_2 \hat d_2 - d_4 \hat d_4 - d_6 \hat d_6 \big) c_2
\nonumber \\[4pt]
& + c_6^{\,2} \hat c_6^{\,2}
+ \hat c_2^{\,2} d_2^{\,2}
+ d_2^{\,2} \hat d_2^{\,2}
+ \hat c_4^{\,2} d_4^{\,2}
+ d_4^{\,2} \hat d_4^{\,2}
+ \hat c_6^{\,2} d_6^{\,2}
+ c_6^{\,2} \hat d_6^{\,2}
+ d_6^{\,2} \hat d_6^{\,2}
\nonumber \\[4pt]
& - 2 c_6 \hat c_6 \, d_2 \hat d_2
+ 2 \hat c_2 \hat c_4 \, d_2 d_4
- 2 c_6 \hat c_6 \, d_4 \hat d_4
+ 2 d_2 \hat d_2 \, d_4 \hat d_4
\nonumber \\[4pt]
& + c_4^{\,2} (\hat c_4^{\,2} + \hat d_4^{\,2})
+ 2 \hat c_2 \hat c_6 \, d_2 d_6
+ 2 \hat c_4 \hat c_6 \, d_4 d_6
\nonumber \\[4pt]
& - 2 \hat c_2 c_6 \, d_2 \hat d_6
- 2 \hat c_4 c_6 \, d_4 \hat d_6
- 4 c_6 \hat c_6 \, d_6 \hat d_6
\nonumber \\[4pt]
& + 2 d_2 \hat d_2 \, d_6 \hat d_6
+ 2 d_4 \hat d_4 \, d_6 \hat d_6
\nonumber \\[4pt]
& - 2 c_4 \big\{
\hat d_4 \big( \hat c_2 d_2 + \hat c_6 d_6 - c_6 \hat d_6 \big)
+ \hat c_4 \big( - c_6 \hat c_6 + d_2 \hat d_2 + 2 d_4 \hat d_4 + d_6 \hat d_6 \big)
\big\}
\Bigg] .
\end{align}

\subsection*{Coefficient of $t^2$: $Q$}
\begin{align}
Q \;=\; 2 \Bigg[
& (\hat c_1^{\,2} + \hat d_1^{\,2})\, c_1^{\,2}
- 4 \hat c_1 d_1 \hat d_1 \, c_1
+ 2 \hat c_1 c_3 \Re(\hat c_3)\, c_1
- 2 \hat d_1 d_3 \Re(\hat c_3)\, c_1
 \nonumber \\[4pt]
& + 2 \hat c_1 c_5 \Re(\hat c_5)\, c_1
- 2 \hat d_1 d_5 \Re(\hat c_5)\, c_1
+ d_1^{\,2} (\hat c_1^{\,2} + \hat d_1^{\,2})
\nonumber \\[4pt]
& + \big\{
\hat c_3 c_3^{\,2}
+ (c_5 \hat c_5 - 2 d_3 \hat d_3 - d_5 \hat d_5) c_3
+ d_3 (\hat c_3 d_3 + \hat c_5 d_5 - c_5 \hat d_5)
\big\} \hat c_3^{*}
\nonumber \\[4pt]
& + c_3 \hat c_3 c_5 \hat c_5^{*}
+ c_5^{\,2} \hat c_5 \hat c_5^{*}
- c_3 \hat d_3 d_5 \hat c_5^{*}
+ d_5 (\hat c_3 d_3 + \hat c_5 d_5) \hat c_5^{*}
\nonumber \\[4pt]
& - c_5 (d_3 \hat d_3 + 2 d_5 \hat d_5) \hat c_5^{*}
+ \big\{
\hat d_3 c_3^{\,2}
+ (-2 \hat c_3 d_3 - \hat c_5 d_5 + c_5 \hat d_5) c_3
\nonumber \\[2pt]
& \hspace{1.6cm}
+ d_3 (- c_5 \hat c_5 + d_3 \hat d_3 + d_5 \hat d_5)
\big\} \hat d_3^{*}
\nonumber \\[4pt]
& + \big\{
c_3 c_5 \hat d_3
+ d_3 d_5 \hat d_3
- 2 c_5 \hat c_5 d_5
- \hat c_3 (c_5 d_3 + c_3 d_5)
+ (c_5^{\,2} + d_5^{\,2}) \hat d_5
\big\} \hat d_5^{*}
\nonumber \\[4pt]
& - 2 c_3 d_1 \hat d_1 \Re(\hat c_3)
+ 2 \hat c_1 d_1 d_3 \Re(\hat c_3)
- 2 c_5 d_1 \hat d_1 \Re(\hat c_5)
+ 2 \hat c_1 d_1 d_5 \Re(\hat c_5)
\nonumber \\[4pt]
& - 2 \hat c_1 (c_3 d_1 + c_1 d_3) \Re(\hat d_3)
+ 2 \hat d_1 (c_1 c_3 + d_1 d_3) \Re(\hat d_3)
\nonumber \\[4pt]
& - 2 \hat c_1 (c_5 d_1 + c_1 d_5) \Re(\hat d_5)
+ 2 \hat d_1 (c_1 c_5 + d_1 d_5) \Re(\hat d_5) 
\Bigg] .
\end{align}
%%%

\subsection*{Coefficient of $u^2$: $R$}
\begin{align}
R \;=\; 2 \Bigg[
& c_1^{\,2} \hat c_1^{\,2}
+ d_1^{\,2} \hat c_1^{\,2}
- 2 \hat c_2 d_1 d_2 \hat c_1
- 2 \hat c_4 d_1 d_4 \hat c_1
- 2 \hat c_6 d_1 d_6 \hat c_1
\nonumber \\[4pt]
& + 2 c_1 c_3 \Re(\hat c_3)\, \hat c_1
+ 2 d_1 d_3 \Re(\hat c_3)\, \hat c_1
+ 2 c_1 c_5 \Re(\hat c_5)\, \hat c_1
+ 2 d_1 d_5 \Re(\hat c_5)\, \hat c_1
\nonumber \\[4pt]
& + 2 (c_3 d_1 + c_1 d_3) \Re(\hat d_3)\, \hat c_1
+ 2 (c_5 d_1 + c_1 d_5) \Re(\hat d_5)\, \hat c_1
\nonumber \\[4pt]
& + c_1^{\,2} \hat d_1^{\,2}
+ d_1^{\,2} \hat d_1^{\,2}
+ \hat c_2^{\,2} d_2^{\,2}
+ d_2^{\,2} \hat d_2^{\,2}
+ \hat c_4^{\,2} d_4^{\,2}
+ d_4^{\,2} \hat d_4^{\,2}
\nonumber \\[4pt]
& + \hat c_6^{\,2} d_6^{\,2}
+ d_6^{\,2} \hat d_6^{\,2}
- 2 d_1 \hat d_1 d_2 \hat d_2
+ c_2^{\,2} (\hat c_2^{\,2} + \hat d_2^{\,2})
\nonumber \\[4pt]
& + 2 \hat c_2 \hat c_4 d_2 d_4
- 2 d_1 \hat d_1 d_4 \hat d_4
+ 2 d_2 \hat d_2 d_4 \hat d_4
+ c_4^{\,2} (\hat c_4^{\,2} + \hat d_4^{\,2})
\nonumber \\[4pt]
& + 2 \hat c_2 \hat c_6 d_2 d_6
+ 2 \hat c_4 \hat c_6 d_4 d_6
- 2 d_1 \hat d_1 d_6 \hat d_6
+ 2 d_2 \hat d_2 d_6 \hat d_6
+ 2 d_4 \hat d_4 d_6 \hat d_6
\nonumber \\[4pt]
& + 2 c_2 \hat d_2 \big(
- \hat c_1 d_1 - c_1 \hat d_1
+ \hat c_4 d_4 + c_4 \hat d_4
+ \hat c_6 d_6 + c_6 \hat d_6
\big)
\nonumber \\[4pt]
& + 2 c_2 \hat c_2 \big(
- c_1 \hat c_1 + c_4 \hat c_4 + c_6 \hat c_6
- d_1 \hat d_1 + 2 d_2 \hat d_2 + d_4 \hat d_4 + d_6 \hat d_6
\big) + c_6^{\,2} (\hat c_6^{\,2} + \hat d_6^{\,2})
\nonumber \\[4pt]
& - 2 c_1 \big\{
\hat d_1 ( \hat c_2 d_2 + \hat c_4 d_4 + \hat c_6 d_6 + c_6 \hat d_6 )
+ \hat c_1 ( c_6 \hat c_6 - 2 d_1 \hat d_1 + d_2 \hat d_2 + d_4 \hat d_4 + d_6 \hat d_6 )
\big\}
\nonumber \\[4pt]
& + 2 c_4 \big\{
- c_1 (\hat c_1 \hat c_4 + \hat d_1 \hat d_4)
+ \hat d_4 ( - \hat c_1 d_1 + \hat c_2 d_2 + \hat c_6 d_6 + c_6 \hat d_6 )
\nonumber \\[4pt]
& \hspace{1.6cm}
+ \hat c_4 ( c_6 \hat c_6 - d_1 \hat d_1 + d_2 \hat d_2 + 2 d_4 \hat d_4 + d_6 \hat d_6 )
\big\}
\nonumber \\[4pt]
& + 2 c_6 \big\{
( - \hat c_1 d_1 + \hat c_2 d_2 + \hat c_4 d_4 ) \hat d_6
+ \hat c_6 ( - d_1 \hat d_1 + d_2 \hat d_2 + d_4 \hat d_4 + 2 d_6 \hat d_6 )
\big\}
\nonumber \\[4pt]
& + \big\{
\hat c_3 c_3^{\,2}
+ ( c_5 \hat c_5 + 2 d_3 \hat d_3 + d_5 \hat d_5 ) c_3
+ d_3 ( \hat c_3 d_3 + \hat c_5 d_5 + c_5 \hat d_5 )
\big\} \hat c_3^{*}
\nonumber \\[4pt]
& + c_3 \hat c_3 c_5 \hat c_5^{*}
+ c_5^{\,2} \hat c_5 \hat c_5^{*}
+ c_3 \hat d_3 d_5 \hat c_5^{*}
+ d_5 ( \hat c_3 d_3 + \hat c_5 d_5 ) \hat c_5^{*} + c_5 ( d_3 \hat d_3 + 2 d_5 \hat d_5 ) \hat c_5^{*}
\nonumber \\[4pt]
& + \big\{
\hat d_3 c_3^{\,2}
+ ( 2 \hat c_3 d_3 + \hat c_5 d_5 + c_5 \hat d_5 ) c_3
+ d_3 ( c_5 \hat c_5 + d_3 \hat d_3 + d_5 \hat d_5 )
\big\} \hat d_3^{*}
\nonumber \\[4pt]
& + \big\{
\hat c_3 c_5 d_3 + \hat d_3 d_5 d_3 + c_3 c_5 \hat d_3
+ c_3 \hat c_3 d_5 + 2 c_5 \hat c_5 d_5
+ ( c_5^{\,2} + d_5^{\,2} ) \hat d_5
\big\} \hat d_5^{*}
\nonumber \\[4pt]
& + 2 c_3 d_1 \hat d_1 \Re(\hat c_3)
+ 2 c_1 \hat d_1 d_3 \Re(\hat c_3)
\nonumber \\[4pt]
& - 2 d_3 ( \hat c_2 d_2 + c_2 \hat d_2 + \hat c_4 d_4 + c_4 \hat d_4 + \hat c_6 d_6 + c_6 \hat d_6 ) \Re(\hat c_3)
\nonumber \\[4pt]
& - 2 c_3 ( c_2 \hat c_2 + c_4 \hat c_4 + c_6 \hat c_6 + d_2 \hat d_2 + d_4 \hat d_4 + d_6 \hat d_6 ) \Re(\hat c_3)
\nonumber \\[4pt]
& + 2 c_5 d_1 \hat d_1 \Re(\hat c_5)
+ 2 c_1 \hat d_1 d_5 \Re(\hat c_5)
\nonumber \\[4pt]
& - 2 d_5 ( \hat c_2 d_2 + c_2 \hat d_2 + \hat c_4 d_4 + c_4 \hat d_4 + \hat c_6 d_6 + c_6 \hat d_6 ) \Re(\hat c_5)
\nonumber \\[4pt]
& - 2 c_5 ( c_2 \hat c_2 + c_4 \hat c_4 + c_6 \hat c_6 + d_2 \hat d_2 + d_4 \hat d_4 + d_6 \hat d_6 ) \Re(\hat c_5)
\nonumber \\[4pt]
& - 2 \big(
- c_1 c_3 \hat d_1 - d_1 d_3 \hat d_1
+ \hat c_2 c_3 d_2 + c_2 c_3 \hat d_2 + c_2 \hat c_2 d_3
+ c_4 \hat c_4 d_3 + c_6 \hat c_6 d_3
\nonumber \\[4pt]
& \hspace{1.6cm}
+ d_2 \hat d_2 d_3
+ c_3 \hat c_4 d_4 + c_3 c_4 \hat d_4 + d_3 d_4 \hat d_4
+ c_3 \hat c_6 d_6 + c_3 c_6 \hat d_6 + d_3 d_6 \hat d_6
\big) \Re(\hat d_3)
\nonumber \\[4pt]
& - 2 \big(
- c_1 c_5 \hat d_1 - d_1 d_5 \hat d_1
+ \hat c_2 c_5 d_2 + c_2 c_5 \hat d_2
+ c_5 \hat c_4 d_4 + c_4 c_5 \hat d_4
+ c_2 \hat c_2 d_5 + c_4 \hat c_4 d_5 + c_6 \hat c_6 d_5
\nonumber \\[4pt]
& \hspace{1.6cm}
+ d_2 \hat d_2 d_5 + d_4 \hat d_4 d_5
+ c_5 \hat c_6 d_6 + c_5 c_6 \hat d_6 + d_5 d_6 \hat d_6
\big) \Re(\hat d_5) 
\Bigg].
\end{align}
%%%%%%%%%%%%%
%%%%%%%%%%%%%%%%%%%%%%%%%%%%%%%%%%%%%%%%%%%%%%%%%%%%%%%%%%%%%%%%%%%%%%%%%%%%%%%%%%%%%%%%%%%%%%%%%%%%%%%%%%%%%%%%%%%%%%%%%%%%%%
\bibliographystyle{utphys}
\bibliography{LmuLtau-ref, Exdim-Zprime, updated-ref, Collider-ref}
\end{document}